\documentclass[useAMS,usenatbib,usegraphicx,usedcolumn]{mn2e}

\usepackage[T1]{fontenc}
\usepackage{ae,aecompl}

\usepackage{lscape}

\usepackage{longtable}

\usepackage{booktabs}

\usepackage{multirow}

\usepackage{subfigure}

\usepackage{rotating}

\usepackage{colortbl}

\usepackage{relsize}

\usepackage{caption}
\usepackage{natbib}

\newcommand{\Msun}{M$_{\odot}$}
\newcommand{\JK}{$\mbox{\em J}-\mbox{\em K}_{\rm s}$}
\newcommand{\JH}{$\mbox{\em J}-\mbox{\em H}$\,}
\newcommand{\HK}{$\mbox{\em H}-\mbox{\em K}_{\rm s}$}
\newcommand{\JHK}{{\em JHK}$_{s}$}
\newcommand{\RK}{$\mbox{\em R}_{\rm c}-\mbox{\em K}_{\rm s}$}
\newcommand{\IK}{$\mbox{\em I}_{\rm c}-\mbox{\em K}_{\rm s}$}
\newcommand{\ImJ}{$\mbox{\em I}_{\rm c}-\mbox{\em J}$\,}
\newcommand{\BK}{$\mbox{\em B}_{\rm J}-\mbox{\em K}_{\rm s}$}
\newcommand{\RI}{$\mbox{\em (R}-\mbox{\em I)}$\,$_{\rm c}$}
\newcommand{\Ks}{{\em K}$_{\rm s}$}
\newcommand{\Bj}{{\em B}$_{\rm J}$}
\newcommand{\Rone}{{\em R}$_{\rm 63F}$}
\newcommand{\Rtwo}{{\em R}$_{\rm 59F}$}
\newcommand{\Rf}{{\em R}$_{\rm F}$-bands}
\newcommand{\Rc}{{\em R}$_{\rm c}$}
\newcommand{\Ic}{{\em I}$_{\rm c}$}
\newcommand{\In}{{\em I}$_{\rm N}$}
\newcommand{\Bb}{{\em B}-band}
\newcommand{\Rb}{{\em R}-band}
\newcommand{\Ib}{{\em I}-band}
\newcommand{\Galb}{$|\mbox{{\em b}}|$}
\newcommand{\Gall}{\ell}
\newcommand{\Kcmd}{{\em K}$_{\rm s}$/($\mbox{\em J}-\mbox{\em K}_{\rm s}$)}

\newcommand{\Mj}{{\em M}$_{\rm J}$}
\newcommand{\Mk}{{\em M}$_{K_{\rm s}}$}
\newcommand{\asec}{~arcsec~}
\newcommand{\peryr}{yr$^{-1}$~}
\newcommand{\rpm}{{\em H}$_{(K)}$}
\newcommand{\microns}{$\mu\mbox{m}$}

\newcommand{\Vtan}{{\em V}$_{\rm tan}$}
\newcommand{\methane}{CH$_4$}

\newcommand{\kms}{\,km\,s$^{-1}$}
\newcommand{\catnum}{246}
\newcommand{\catarea}{3223}

\title[Identifying Ultra-Cool Dwarfs at Low Galactic Latitudes: A Southern Candidate Catalogue]{Identifying Ultra-Cool Dwarfs at Low Galactic Latitudes: A Southern Candidate Catalogue}
\author[S. L. Folkes et al.]{S. L. Folkes$^{1,2}$\thanks{E-mail: s.l.1.folkes@herts.ac.uk}, D. J. Pinfield$^{2}$, H. R. A. Jones$^{2}$, R. Kurtev$^{1}$, Z. Zhang$^{2}$
\newauthor
M. C. G\'{a}lvez-Ortiz$^{3,2}$, F. Marocco$^{2}$, A. C. Day-Jones$^{4,2}$, and J. R. A. Clarke$^{2}$\\
$^{1}$Departamento de F\'isica y Astronom\'ia, Facultad de Ciencias, Universidad de Valpara\'iso, Ave. Gran Breta\~na 1111, Playa Ancha,\\ Casilla 53, Valpara\'iso, Chile.\\
$^{2}$Centre for Astrophysics Research, University of Hertfordshire, College Lane, Hatfield, AL10 9AB, United Kingdom.\\
$^{3}$Centro de Astrobiolog\'ia (CSIC-INTA). Crta, Ajalvil km 4. E-28850 Torrej\'on de Ardoz, Madrid, Spain\\
$^{4}$Departamento de Astronom\'ia, Universidad de Chile, Camino el Observatorio 1515, Santiago,Chile.\\}
\begin{document}

\date{Accepted April 2012. Received April 2012; in original form December 2011}

\pagerange{\pageref{firstpage}--\pageref{lastpage}} \pubyear{2012}


\maketitle

\label{firstpage}

\bibliographystyle{aa}
\begin{abstract}
We present an Ultra-Cool Dwarf (UCD) catalogue compiled from low southern Galactic latitudes and mid-plane, from a cross-correlation of the 2MASS and SuperCOSMOS surveys. The catalogue contains \catnum~members identified from 5042~deg$^2$ within $220^{\circ}\leqslant\ell\leqslant360^{\circ}$ and $0^{\circ}<\ell\leqslant30^{\circ}$, for $|\mbox{{\em b}}|\leqslant15^{\circ}$. Sixteen candidates are spectroscopically confirmed in the near-IR as UCDs with spectral types from M7.5V to L9, the latest being the unusual blue L dwarf 2MASS\,J11263991-5003550. Our catalogue selection method is presented enabling UCDs from $\sim$M8V to the L--T transition to be selected down to a 2MASS limiting magnitude of $\mbox{\Ks}\simeq14.5\,\mbox{\rm mag}$ (for $\mbox{SNR}\geqslant10$). This method does not require candidates to have optical detections for catalogue inclusion. An optimal set of optical/near-IR and reduced proper-motion selection criteria have been defined that includes: an {\em R}$_F$~and \In~photometric surface gravity test, a dual {\em R}$_F$-band variability check, and an additional photometric classification scheme to selectively limit numbers of potential contaminants in regions of severe overcrowding. We identify four candidates as possible companions to nearby \textit{Hipparcos} stars -- observations are needed to identify these as potential benchmark UCD companions. We also identify twelve UCDs within a possible distance 20\,pc, three are previously unknown of which two are estimated within 10\,pc, complimenting the nearby volume-limited census of UCDs. An analysis of the catalogue spatial completeness provides estimates for distance completeness over three UCD \Mj~ranges, while Monte-Carlo simulations provide an estimate of catalogue areal completeness at the 75 per cent level. We estimate a UCD space density of $\rho_{\rm (total)}=(6.41\pm3.01)\times10^{-3}\,{\rm pc}^{-3}$ over the range of $10.5\leqslant\mbox{\Mj}\la 14.9$, similar to values measured at higher Galactic latitudes ($|\mbox{{\em b}}|\ga 10^{\circ}$) in the field population and obtained from more robust spectroscopically confirmed UCD samples.
\end{abstract}

\begin{keywords}
stars: low mass, brown dwarfs - star: kinematics - infrared: stars - surveys: techniques - photometric - techniques - spectroscopic
\end{keywords}


\section{Introduction} \label{intro}
For about four decades before the advent of large area near-IR sensitive detectors, the M dwarfs marked the observational limit at the lower luminosity (and lower mass) end of the main sequence. Early searches for Very Low-Mass Stars (VLMS) and brown dwarfs were limited to all-sky surveys based on optical photographic plates, which are not sensitive to the very red spectral energy distributions of these intrinsically faint objects. The first photographic plates to be used were the blue (103aO) and red (103aE) sensitive emulsions in the northern first Palomar Observatory Sky Survey (POSS-I: 1949-1957). Later in the 1970s, the UK/AAO and ESO (40inch Chile) Schmidt telescopes began surveys in the southern hemisphere using blue-green IIIaJ, and far-red IIIaF/\In~plate emulsions respectively. Second epoch photographic surveys were carried out in the northern sky by the POSS-II in the late 1980s, which added an \In~sensitive emulsion, and also in the south using the UK/AAO (blue and red plates) in the late 1990s.

The first all-sky search exploiting these initial photographic surveys for nearby stars was by Luyten, who published two proper motion catalogues: the Luyten Half-Second catalogue \citep[LHS:][]{Luyten79LHS}, and the New Luyten Two-Tenths catalogue \citep[NLTT:][]{Luyten80NLTTv1,Luyten80NLTTv2,Luyten80NLTTv3,Luyten80NLTTv4}. However, the low sensitivity of the photographic surveys to VLMS meant that only a limited number of the nearest examples were identified. Among these are the well known and well studied examples Wolf\,359 (M5.5V), VB\,8 (M7V), VB\,10 (M8V), LHS\,2924 (M9V), and LHS\,2065 (M9V). The bottom of the main sequence was observationally defined by VB\,10 for nearly 40 years, until in 1983 LHS\,2924 was identified as an unusual low luminosity late-M dwarf \citep{LHS2924_discov83}. The mass of LHS\,2924 appears to be close to that of the stellar hydrogen-burning mass limit (HBML), but is probably not a sub-stellar brown dwarf due to weak H$\alpha$ emission and no detectable Li\,{\sc i} absorption \citep[inferring older age: ][]{Martin94}. Current theoretical models give the HBML as M$_{\rm HBML}\simeq0.075\mbox{\,M}_{\sun}$ \citep[e.g.,][ for Solar metallicity]{dusty00_Ch_model,dusty02_Ba_model}.

More recently digitised automated scanning of these photographic plates has opened up the possibility of discovering many more nearby VLMS, and even brown dwarfs, either directly or by cross-correlation with near-IR surveys. Digitised catalogues that are currently available and containing photometric and proper motion data are: SuperCOSMOS Sky Survey and SuperCOSMOS Science Archive \citep[SSS and SSA respectively:][]{ssa_p1a,ssa_p2b,ssa_p3c}, USNO-B1.0 \citep{USNO_B1_Monet03}, GSC-2 \citep{gsc_cat90,gsc2p3_cat08new}, and the APS archive \citep[][POSS-I only]{maps_cat03}. One automated digitised scanning process in particular (which uses specialised software called SUPERBLINK) was developed specifically to search for new high proper-motion stars in both the northern hemisphere \citep[][$\mu>0.15^{\prime\prime}\mbox{yr}^{-1}$]{lepine_north_cat05}, and southern hemisphere \citep[][$0.45^{\prime\prime}\mbox{yr}^{-1}<\mu<2.0^{\prime\prime}\mbox{yr}^{-1}$]{lepine_south_m30deg_cat05}.

The term `Ultra-Cool Dwarf' (UCD) has recently become synonymous with low-mass late-type dwarfs of spectral types $\geqslant$M7 (both VLMS and brown dwarfs). The term UCD was first deployed in \cite{kirkp}, and then began to be used more extensively  \citep[e.g.,][]{martin}. The justification for M7 in defining the earliest spectral sub-type is based on two observed physical characteristics of dwarf late-M stars: the first is that the sub-stellar boundary is located at about the spectral type of M6.5 in the Pleiades Galactic cluster (having an age of $\sim125$\,Myrs), as inferred by the Lithium depletion edge test \citep{Martin96,Stauffer98_Li}. The second is the observed dramatic decrease in the 0.65 -- 0.75$\,\mu$m TiO band strength, also located at around M6.5 for field dwarfs \citep{jones97_TiO}.

At the time of writing there were just over 600 L dwarfs, and nearly 200 T dwarfs known (from the M, L, and T dwarf compendium housed at: {\sc http://dwarfarchives.org})\footnote{The {\sc http://dwarfarchives.org} compendium is maintained by Chris Gelino, Davy Kirkpatrick, and Adam Burgasser.}, which had been mostly discovered photometrically in the red-optical and near-IR: DENIS \citep[DEep Near Infrared Survey of the southern sky:][]{denis99}, 2MASS \citep[Two Micron All-Sky Survey:][]{Skrutskie97_2M}, SDSS \citep[Sloan Digital Sky Survey:][]{sdss}, and UKIDSS \citep[UKIRT Infrared Deep Sky Survey:][]{ukidss_laurence}. Using these more recent surveys, early searches for L and T dwarfs have largely avoided the Galactic plane (i.e., $\mbox{\Galb}\la15^{\circ}$), with the first attempt to systematically search the `zone of avoidance' by \citet{Reid03_GP_search}, who identified two UCDs: one M8V and an L1.5. Both of these objects had been previously discovered by the SUPERBLINK survey of \citet{lepine02_PM_cat_north}. The search by \citet{Reid03_GP_search} used a near-IR selection criteria requiring $\mbox{\Ks} \geqslant 10.5\,\mbox{\rm mag}$. More recently, a search of the Galactic plane conducted by \citet{phan_bao_gp07} published results for 26 confirmed UCDs from DENIS -- 15 of which have been independently identified here, and previously in \cite{my_paper}.

The Galactic plane represents a significant area of sky ($\sim26$ per cent within $\mbox{\Galb}\la15^{\circ}$), offering considerable latent potential for new discoveries of UCDs, from deeper searches over a wider range of spectral types. It is possible that such discoveries may contain highly interesting, unusual, and nearby/bright examples. Such a search will also compliment those made at higher Galactic latitudes, in particular by adding to the statistics of the 10\,pc and 20\,pc volume-limited census (e.g., RECONS\footnote{See http://www.recons.org/ for details of the research consortium.} and 2MU2, \citet{reid_cool_N10_08, cruz07} respectively), which aim to produce complete luminosity and field mass functions of the stellar/sub-stellar population within the Solar neighbourhood.

However, the Galactic plane is referred to as the `zone of avoidance' for very good reasons as it contains the highest stellar densities down to faint limiting magnitudes. One also has to contend with regions of dark molecular clouds, nebulosity, and regions of current star formation. It is interesting to note that $\sim75$ per cent of all the objects contained in the 2MASS Point Source Catalogue (PSC) lie within $\mbox{\Galb}\la15^{\circ}$. The Galactic plane is also host to several types of objects that cause confusion and contamination in a search to identify UCDs, e.g., `O'-rich and `C'-rich Long Period Variable (LPV) AGB stars, distant highly reddened luminous early-type main sequence and giant branch stars, Young Stellar Objects (YSOs) -- primarily CTTS (Classical T-Tauri Stars), and Herbig AeBe stars. All these sources of contamination can mimic the near-IR, and in some cases optical-NIR, colours of UCDs.

Here, we present a method which uses a comprehensive set of photometric and proper motion criteria, to create a catalogue of field UCDs at low Galactic latitudes in the southern hemisphere. Our method has been tailored to discriminate UCDs from sources of contamination, that one is most likely to encounter in the crowed regions of the Galactic plane.


\section{Defining the Selection Method} \label{meth}

In considering the choice of photometric survey data to use for the primary near-IR query, the 2MASS all-sky release and the DENIS catalogue could have been utilised. DENIS has been used to great effect in discovering many UCDs \citep{first_denis_LDs_97,martin,Phan-Bao_NNV_03,Crifo2005,phan_bao_gp07,martin2010}, and offers {\em J},{\em K}, and \Ib~(Gunn-{\em i}) photometry. The (Gunn-{\em i}) passband is effectively contemporaneous with its near-IR bands and can be used to impose a \ImJ~colour constraint to good effect for selection of very red candidates \citep[e.g.,][]{phan_bao_gp07}. However, DENIS lacks {\em H}-band photometry, preventing the use of the \JH/\HK~two-colour plane and \JK~colour indices for a primary near-IR selection. Therefore, the requirement to have an \Ib~detection significantly limits the depth to which one can identify very red objects, thus probing a smaller space volume. The 2MASS Point Source Catalogue (PSC) has contains data for the {\em J},{\em H}, and \Ks~bands and has extensive quality flag information for the catalogue entries, and has also been extensively mined in searches for UCDs at higher Galactic latitudes since its first data release \citep[e.g.,][]{kirkp,Kirkp00,Gizis02_2mass,Cruz03,Wilson03_2mass,Kendall03_2mass,knapp,cruz07,Kirkpatrick07_2m_prop_mot}. Thus, we chose 2MASS for the primary near-IR photometric data to enable a more flexible search method, while allowing a self consistent comparison to the pervious UCD searches at higher Galactic latitude. The 2MASS photometry can then be combined with multi-band photometry from an optical survey of complimentary depth. 

For optical photometry there were two choices available at the beginning of our search, which could facilitate a cross-correlation with the 2MASS data: the SuperCOSMOS Science Archive (SSA) and the USNO-B1.0 catalogue -- both of which contain digitised photographic \Bj, \Rone, \Rtwo~and \In~plate scans covering the whole sky. The SSA and SSS resources were chosen due to the availability of excellent database mining tools and the extensive enhanced quality information for each detection.

While the near-IR photometric data from the three 2MASS \JHK~bands are contemporaneous, the optical photometric data from the individual bands of the SSA will not be, as the original photographic plates range in epoch from $\sim1950$ (POSS-I) to the late-nineties (\In). A significant proportion of the contaminant objects in the Galactic plane are mostly in the form of distant AGB Long Period Variable (LPV) stars which can exhibit a large amplitude in variability of up to five magnitudes in the {\em V}-band, and typically have periods of 200 to 500 days \citep{Aller91_book}. Thus, non-contemporaneous colours of LPV stars resulting from combining the near-IR and optical photometry are expected to exhibit a large range in scatter. Due to this behaviour in the optical/near-IR colours it was necessary to construct selection criteria which utilise the full range of photometric bands ({\em J, H}, \Ks, \Bj, \Rone, \Rtwo~and \In) available, to remove as much of the contamination as possible. 

Although intrinsic differences exist in the loci of optical/near-IR colours between UCDs and LPVs, variability induced scatter in the colour ranges of these latter objects spreads them through the UCD loci. The more optical/near-IR two-colour planes one has, the greater the chance of 'catching' these contaminants outside the selection criteria defined for UCDs. In defining this method we have characterised seven colours, as well as a reduced proper-motion criterion, over the UCD spectral range of interest. These have been utilised in both our primary near-IR candidate selection, and secondary optical/near-IR selection process.

The characterisation of these selection criteria was achieved using a sample of 186 known bright reference UCDs (mostly L dwarfs but with some late-M dwarfs) with reliable near-IR 2MASS photometry ($\mbox{SNR}\geqslant20$), which were obtained from {\sc http://dwarfarchives.org}. SSA \Rone, \Rtwo~and \In~photometry was also obtained for these reference UCDs.

In total we searched an area of sky of 5042\,deg$^2$ at low southern Galactic latitudes ($220^{\circ}\leqslant\ell\leqslant360^{\circ}$ and $0^{\circ}<\ell\leqslant30^{\circ}$), partitioned into smaller regions we termed 'sky tiles', to facilitate a more manageable number of objects to be checked in each instance. A sky tile area typically of 200\,deg$^2$ was found adequate for this purpose, especially through the Galactic mid-plane ($\mbox{\Galb}\la5^{\circ}$).

\subsection{The Near-IR Candidate Selection}\label{nir:sel}
The primary near-IR candidate selections were obtained via the 2MASS Point Source Catalogue (PSC) {\sc gator}\footnote{Available at http://irsa.ipac.caltech.edu/applications/Gator/} tool. A number of query parameters were set to govern the spatial and photometric constraints for a specific sky tile query. The following parameters and quality flags have been used consistently throughout:

\begin{description}
\item []{\bf cc\_flg:} The \JHK~Artefact/confusion/contamination flag: set to '000'.
\item []{\bf mp\_flg=0:} Minor planet identification flag: is always set to '0'.
\item []{\bf gal\_flg=0:} Flag indicating if object is contaminated by an extended source -- always set to '0'. 
\item []{\bf prox:} Nearest neighbour distance in the PSC. Always set to prox $\geqslant6$\asec.
\item []{\bf ph\_qual:} Photometric quality flag: set to $\geqslant\mbox{'CCC'}$.

\end{description}

We also use a combination of the {\em ph\_qual} flag value and the 2MASS source uncertainty (\(\sigma_{\mbox{\JHK}}(\rm mag)\)) for each candidate to retain those candidates with an implied $\mbox{SNR}\geqslant 10$. Using this SNR limit should allow limiting magnitudes of $\mbox{\em J}\simeq16.5\,\mbox{\rm mag}$, $\mbox{\em H}\simeq15.3\,\mbox{\rm mag}$, and $\mbox{\Ks}\simeq14.9\,\mbox{\rm mag}$ to be reached (see the 2MASS users guide documentation\footnote{see http://www.ipac.caltech.edu/2mass/releases/allsky/doc/sec2\_2.html.}). However, these PSC limiting magnitudes are for high Galactic latitudes ({\em b}$\ >+75^{\circ}$) and the true limiting magnitudes in the Galactic plane due to confusion noise will be brighter at about $\mbox{\em J}\simeq16.0\,\mbox{\rm mag}$, $\mbox{\em H}\simeq15.0\,\mbox{\rm mag}$, and $\mbox{\Ks}\simeq14.5\,\mbox{\rm mag}$. In \textsection\,\ref{meth:class} we define a candidate selection and classification criteria which has this photometric limit of $\mbox{\Ks}\simeq14.5$ as a baseline, and thus compliments this.

The primary near-IR candidate selection is achieved by passing the 2MASS photometry through a custom set of \JHK~two-colour, and \JK~colour criteria. These were devised by carefully characterising the loci of known LPV Carbon and `O'-rich AGB stars (S-type Zirconium group AGB stars also exist, but are not considered here as they have a much lower observed frequency), in relation to the 186 known bright reference L dwarfs. Photometric data for these known LPV stars were obtained from catalogues available through VizieR \citep{Mennessier01,c_star_cat01}. The slopes defining the \JH/\HK~selection region were carefully crafted such that they were kept as tight as possible to the L dwarf locus, avoiding as best as possible the main bulk of the LPV locus along the \JH~red end of the region.

Three of the reference L dwarfs fell outside of the two-colour selection cut, but this was felt acceptable as adjusting the criteria to accommodate them would let through an unacceptable amount of contaminant objects. In addition to this reason two of the three L dwarfs have unusal properties which may explain their bluer \HK~colours: LSR 1610-0040 (2MASS\,J16102900-0040530), an older metal-poor early-type L subdwarf \citep[sdL:][]{LSR1610_0040}; 2MASS\,J00145575-4844171, a young field L dwarf with a slightly peculiar spectrum, perhaps indicative of lower metallicity \citep[L2.5pec:][]{Kirkp08_young_Ldwfs}. The third object is 2MASS\,J21373742+0808463, an L5 not reported as having unusal properties \citep{reid_cool_N10_08} which lies close to the redder \JH~colour selection boundary.

It was also noted that there is an overlap in \JK~between late-M dwarfs and early-L dwarfs. Under further investigation, four of the 186 reference L dwarfs were found to be bluer than expected, and to accommodate these we carefully adjusted the lower blue-ward limit to $\mbox{\JK}\geqslant1.075\,\mbox{\rm mag}$ whilst still allowing field M dwarf contamination to be reduced, compared to the range in this colour expected for late-M and L-dwarfs in the literature \citep[e.g.,][]{leggett92,leggett01,dahn02,Cruz03,Phan-Bao_NNV_03,vrba}. The final \JH, \HK, and \JK~colour cuts allow objects in the spectral range from $\sim$M8V to the L--T transition to be selected.

\begin{figure*}
 \begin{center}
  \begin{minipage}{0.75\textwidth}
  \begin{center}
  \includegraphics[width=\textwidth,angle=0]{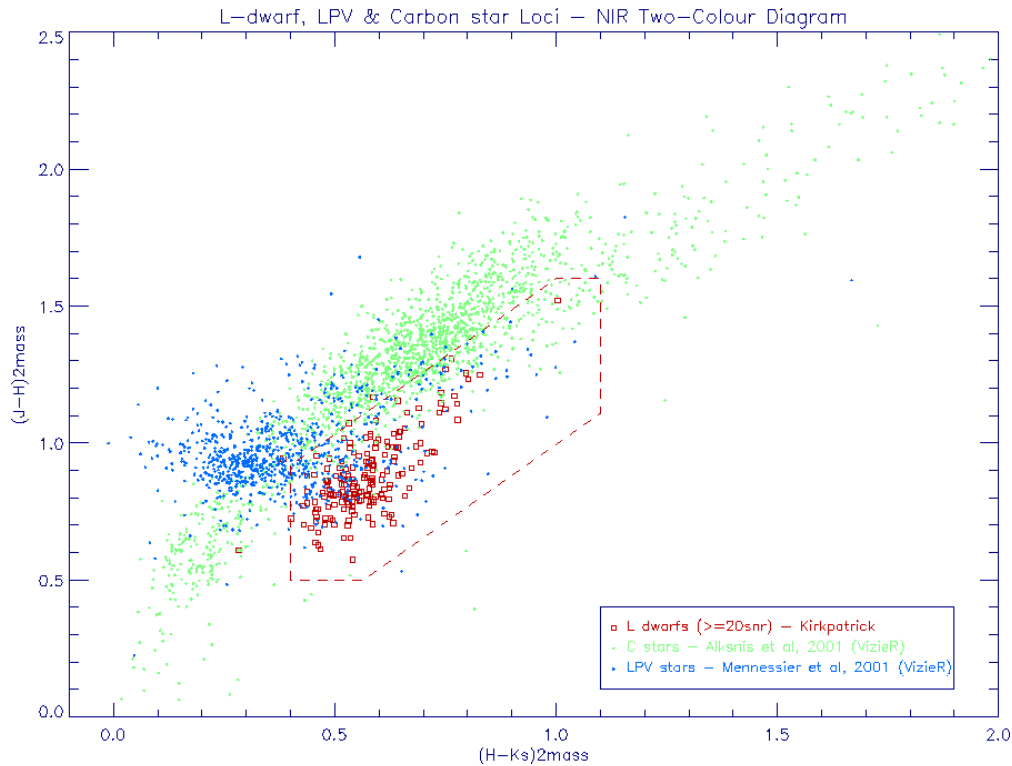}
  \caption{The \JH/\HK~near-IR two-colour diagram showing the locations of LPV Carbon (green dots) and `O'-rich (blue dots) AGB stars in relation to known reference L dwarfs with good photometry $\mbox{SNR}\geqslant 20$ (red squares). The colour-cut criteria (outlined as the red dashed line) shows the selection criteria avoiding the main locus of the LPV stars while still allowing through the full spectral range of $\sim$M8V to the L--T transition.}
  \label{nir_crit}
  \end{center}
  \end{minipage}
 \end{center}
\end{figure*}

The adopted \JH/\HK~two-colour UCD selection criteria is shown in Fig.\,\ref{nir_crit}, which also plots the reference L dwarfs with the high signal-to-noise 2MASS photometry, along with the LPV `O'-rich and carbon stars. The \JHK~selection criteria are defined as follows:   

\begin{displaymath}
\left. \begin{array}{rl}
0.4\leqslant \mbox{\HK} \leqslant 1.1 \\
0.5\leqslant \mbox{\JH} \leqslant 1.6 \\
\mbox{\JH} \geqslant 1.1367\times \mbox{\HK} - 0.142 \\
\mbox{\JH} \leqslant 1.137\times \mbox{\HK} + 0.463
\end{array} \right\} 1.075\leqslant \mbox{\JK} \leqslant 2.8
\end{displaymath}
In this \JHK~two-colour plane the \JK~$\geqslant 1.075$ criterion results in the following:

\begin{itemize}
 \centering
  \item []\(\mbox{\JH}\geqslant 1.075 - \mbox{(\HK)}\)
\end{itemize}

\subsection{The Optical SSA Cross-Correlation} \label{opt:ssa_query}

The SSA {\em R}-band magnitudes are obtained from digitised scans of \Rone~and \Rtwo~photographic plates. These photographic {\em R}$_F$ passbands have non-standard profiles that differ quite considerably from the standard Kron-Cousins \Rc~profile, which has a long declining transmission tail that extends nearly 1500\AA~further into the red. As a result this can lead to significant differences when comparing uncorrected magnitudes between the two systems for very red objects \citep[e.g., see][for details on the passband profiles and system transformations]{bessell86}. However, the situation appears much better between the \In-band and \Ic~magnitudes, with only very small offsets being required. Where possible, all the \Rone, \Rtwo~and \In~photometry obtained from the SSA are transformed onto the standard Kron-Cousins system (\Rc~and \Ic) using the transformations defined by \citet{bessell86}, to produce consistent magnitudes across the whole dataset which are also more readily comparable to those in the literature, such as those used in defining the selection criteria.

For the SSA optical photometry query we return the magnitudes calibrated for point sources ({\em sCorMag[band]}), and determine the quality of each detection via the bitwise quality flags ({\em qual[band]}). Photometric detections of each band which have quality flags $\ga2048$ are rejected in line with the advice given on the SSA on-line data overview.

Due to the significantly increased source density in the Galactic plane, identifying the correct optical counterparts to 2MASS detections will be more problematic compared to regions at higher Galactic latitude. SSA query cross-matching search radii of 2\asec~and 5\asec~were tested on the same 2MASS dataset for a region of the Galactic plane; it was found that the 2\asec~radius let through significantly more objects inhabiting regions of a \Kcmd~CMD and \JH/\HK~two-colour diagram that are consistent with being mid- to late-M dwarfs. This is most likely due to the larger proper-motions of nearby late-type dwarf stars, whose motion over the average 2MASS--SSA epoch difference ($\sim16$ years: see \textsection\,\ref{ssa_xmatch}) resulted in the 2\asec~angular radius being exceeded. Thus, these objects passed the primary near-IR colour selection, but were not subject to the full set of criteria that would have rejected them. Therefore, throughout our search we used a 5\asec~cross-matching radius, noting the possibility that some nearby high proper-motion objects may not be cross-matched with an SSA detection. However, as optical SSA detections are not a requirement for candidate selection, any genuine nearby (possibly bright) high proper-motion UCD should still be identified from our full range of selection criteria.

\subsection{Optical Data Preparation} \label{meth:opt_prep}
\subsubsection{Gravity Skewed Photographic-to-Cousins Conversion} \label{prep:conv}
If either, or both, of the SSA \Rf~are available then the {\em R}$_F$-band(s) and \In-band are transformed onto the standard Cousins system. These {\em R}$_F$-band transformations are valid to a red limit of $\mbox{\RI}\simeq2.8\,\mbox{\rm mag}$, which is the same as the red limit defined in our \RI~colour criterion (see \textsection\,\ref{opt_nir:ri_jk}). However, due to the spectral differences between the late-M and L dwarfs, it is uncertain how valid these transformations are for the later UCD spectral types of interest here. To limit the possible effects of such differences, the optical colour selection criteria were all defined around the reference L dwarfs (as mentioned above), as well as LPV stars with well constrained photometry, using \Rc~and \Ic~magnitudes transformed from the SSA photometry.

An interesting and potentially useful aspect of one of these transformations is that they appear to be dependant on the stellar luminosity class, i.e., on their surface gravity. The \Rone~magnitudes show differences of up to $\simeq 0.2\,\mbox{\rm mag}$ between K- and M-type giants compared to dwarfs of the same spectral types \citep{bessell86}. Use is made of this property by only applying the dwarf transformation to all the candidates which have an \Rone~and an \In~detection. Thus, if a potential candidate has a low surface gravity -- i.e., is of luminosity class I-III -- the converted \Rc~magnitude will be offset compared to the \Rc~magnitude derived from the \Rtwo~detection (unaffected by gravity). This gravity induced offset between the two \Rc~derived magnitudes will assist in the rejection of giant and AGB contaminant stars during the check for variability as discussed next in \textsection\,\ref{prep:dual}, if both {\em R}$_F$-bands are available.
The transformations are carried out according to the following procedure:
\begin{enumerate}
\item If one, or both, of the \Rf and the \In-band are present then all are transformed to the standard (Cousins) system.
\item If only one or both of the \Rf~are present, but there is no \In-band, then an offset is applied to the {\em R}$_F$-band assuming \(\left(\mbox{{\em R}}_F-\mbox{\In}\right)=2\) (typical for an L dwarf).
\item When only an \In~detection is available no correction is made, as the match to \Ic~is good enough to be acceptable.
\end{enumerate}

\subsubsection{Dual Epoch {R}$_F$-Band Variability} \label{prep:dual}
The availability of two \Rb~detections at different epochs from the SSA gives the ability to test for brightness variability for each candidate. As mentioned previously one of the main types of contaminant object which share similar areas in the near-IR colour planes are the LPV stars, of both Oxygen-rich, and Carbon-rich types. These LPV stars typically have periods of between 200 and 500 days, and brightness variability of up to five magnitudes (visual). This makes them easy to distinguish as long as their minimum apparent \Rb~magnitudes are bright enough to accommodate a realistic chance of a detection, at two widely spaced epochs.

In contrast, optical variability in L dwarfs has only been observed at a very low level (in a small number of cases): \citet{ldw_Iband_vari} showed RMS \Ib~variations to be no more than $0.083\,\mbox{\rm mag}$, while \citet{ucd_weather_ultrcam} find evidence variability induced by dust-cloud weather in the L dwarf 2MASS J1300+1912 (L1) at the 5 per cent level, using the fast cadence photometer ULTRACAM \citep[also see][]{martin96_cs12,tinney99_vari,martin01_weather}. Therefore, {\em R}- and {\em I}-band variability in UCDs is not considered to be significant compared to variations due to the larger photometric uncertainties. However, flare activity in can lead to significantly increased H$\alpha$ emission in Late-M dwarfs \citep[e.g.,][]{Martin01_M_flares}, potentially causing a rejection in the \Rb~variability test here. However, the occurrence of such flares appears to be low at 3 per cent \citep[for late-M subtypes at low Galactic latitudes:][]{Hilton10_M_flare}, therefore, such flares are not expected to cause a significant problem here. 

To apply the variability test in a consistent way requires a variability threshold to be defined in relation to a control sample: our reference UCD sample with well constrained near-IR photometry, for which both \Rone~and \Rtwo~detections are also available (a total of 65 objects). This threshold value was defined as the modulus of the maximum difference between the \Rone~and \Rtwo~transformed \Rc~magnitudes, which was found to be \(\Delta m=0.312\,\mbox{\rm mag}\). Thus, any candidate displaying a variability of \[\mid\mbox{\Rc}_{\rm (R63F)}-\mbox{\Rc}_{\rm (R59F)}\mid\ >\ 0.312\,\mbox{\rm mag}\] is rejected.

\subsubsection{{R}$_{\rm F}$-band Brightness Bias} \label{prep:bias}
For cases where candidates have both \Rf~available a choice has to be made which transformed \Rc~magnitude will be used for the optical selection criteria. Rather than just choosing the \Rtwo~detection for all candidates, which generally has a better correction and SNR than \Rone, and also reaches to a fainter limiting magnitude by $\sim0.8\,\mbox{\rm mag}$ \citep{ssa_p2b}, the brightest of the two detections is chosen instead. For the more unlikely cases where the two magnitudes are equal the \Rtwo~is selected in preference. 

The problematic carbon stars lie very close to, and slightly overlap with, the L dwarf loci and two-colour plane selection criteria shown in Fig.\,\ref{rminusk} and Fig.\,\ref{rminusi}. Using a positional argument based on the overwhelming number of these contaminant objects, even small differences between the \Rone~and \Rtwo~transformed \Rc~magnitudes (\(\Delta {\rm mag}<0.312\)) can be exploited by choosing the brightest \Rc~magnitude, to bias in favour of a non-selection. While this potentially applies to a genuine UCD candidates, it is more likely that this bias will aid in the rejection of LPV stars.

\subsection{Near-IR/Optical Selection Criteria} \label{meth:opt_nir}
As might be expected from looking at the overlap between the AGB stellar locus with that of the UCDs in the $(\mbox{\JH})/(\mbox{\HK})$ two-colour plane of Fig.\,\ref{nir_crit}, that 2MASS \JHK~data alone are not adequate for a selection to be made. The sheer number of intrinsically bright contaminant objects detected to large distances, and thus large space volumes, swamps any near-IR selection made close to the Galactic plane. To alleviate this contamination problem it was therefore necessary to combine the 2MASS near-IR photometry with optical photometry and proper-motion data from the SSA. We thus defined a set of well characterised optical-NIR and RPM based selection criteria, by investigating the photometric and kinematic properties of the distinct contaminant and non-contaminant populations.

\subsubsection{The $\mbox{B}_{\rm J}-\mbox{K}_{\rm s}$~Criterion}\label{opt_nir:bk_jk}

Although UCDs are not usually detected in the \Bb, very nearby examples could be. Furthermore, we expect there to be many bright contaminants with detections in the \Bb, such as `O'-rich LPVs in which the dominant molecular absorber is titanium oxide that blankets more of the red end of the spectrum compared with carbon stars. Thus it is important to constrain the candidate sample using this colour.

\begin{figure*}
 \begin{center}
  \begin{minipage}{0.75\textwidth}
  \begin{center}
  \includegraphics[width=\textwidth,angle=0]{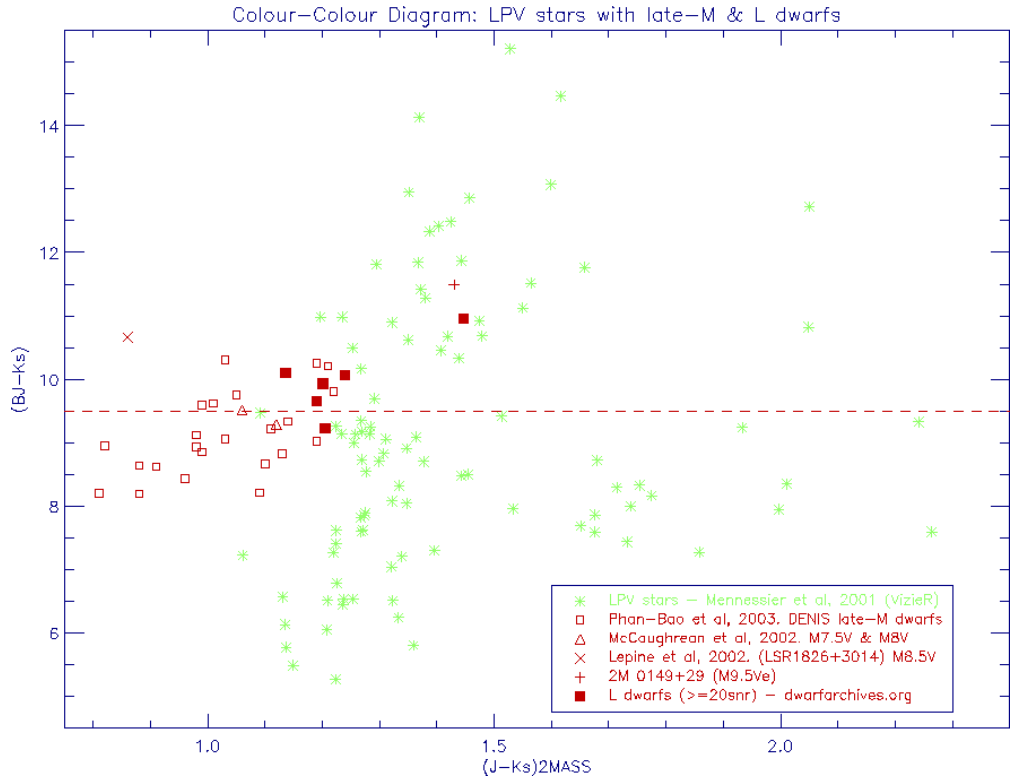}
  \caption{The $(\mbox{\BK})/(\mbox{\JK})$ colour cut criterion showing the end of the M dwarf sequence, six L dwarfs, and 96 `O'-rich LPV stars. The cut is made at $\mbox{\BK}=9.5\,\mbox{\rm mag}$ and is constant throughout \JK. The average \BK~colour for these LPV stars is $\simeq9.1\,\mbox{\rm mag}$ indicating that the majority of them should fall below this limit.}
 \label{bminusk}
  \end{center}
  \end{minipage}
 \end{center}
\end{figure*}

To define a \BK~colour cut, SSA \Bj~photometry was extracted for 96 of the LPVs from the catalogue of \citet{Mennessier01}, as well as for a selection of late-M dwarfs with good \Bb~photometry available from the literature: LSR1826+3014 \citep{LSR1826_02}; LHS2065 \citep{leggett92}; 2M0149+2956 \citep{2M0149_flare99}; LP775-31 and LP655-48 \citep{McCaughrean_2_MDws,Phan-Bao_NNV_03}. We aso obtained \Bj~magnitudes from the SSA for six further UCDs listed on {\sc http://dwarfarchives.org}; two late-M dwarfs and four L dwarfs from our high $\mbox{S/N}$~reference sample (photometry for these six UCDs are presented in Table.\,\ref{Bmags_Ldwfs}), which together, allowed a boundary between the M- and L-type dwarfs to be defined in relation to the LPV objects. The resulting $(\mbox{\BK})/(\mbox{\JK})$ two-colour diagram is shown in Fig.\,\ref{bminusk}, in which a trend appears to be visible extending from the M dwarf to the L dwarf spectral types as the \BK~colour increases with an increasing \JK index. A range in variability spanning $\sim10$ magnitudes in \BK~was seen to exist for the `O'-rich LPV stars, with an average \BK~colour of $\simeq 9.1\,\mbox{\rm mag}$. A cut in this colour plane was made such that objects with $(\mbox{\BK})\geqslant9.5\,\mbox{\rm mag}$ are accepted, corresponding to a value of $(\mbox{\JK}) = 1.075\,\mbox{\rm mag}$ (our\JK~lower limit) along the visible trend, and therefore an approximate spectral type of M8V.

\begin{table*}
\begin{minipage}{0.8\textwidth}
 \begin{center}
 \caption[]{2MASS and SSA photometry for the six Late-M and L dwarfs with \Bj~magnitudes.}
 \label{Bmags_Ldwfs}
 \begin{tabular}{lcccc}
  \toprule
     & (SSA) & (2MASS) &     & (Optical/Near-IR)\\
Name & \Bj   &   \BK   & \JK & SpT\\
  \midrule
2MASS\,J23453903+0055137    & 22.238 &  9.66 & 1.190 & M9V/\textemdash~$^a$\\
DENIS\,J220002.05-303832.9  & 22.261 & 10.06 & 1.240 & \textemdash/L0~$^b$\\
2MASS\,J10220489+0200477    & 22.829 &  9.93 & 1.201 & L0\,pec/\textemdash~$^c$\\
2MASSI\,J0523382-140302     & 22.594 & 10.96 & 1.446 & L2.5/L5~$^d$\\
SDSS\,J161928.31+005011.9   & 22.414 &  9.23 & 1.205 & L2/\textemdash~$^e$\\
SSSPM\,J0109-5101           & 21.193 & 10.10 & 1.136 & M8.5V/L2~$^f$\\
 \bottomrule
 \end{tabular}
 \end{center}
{($^a$) Discovery and optical spectral type by \citet{reid_cool_N10_08}.\\
($^b$) Discovered by \citet{Kendall03_denis}. Near-IR spectral type from \citet{Burgy2006_M9_L0_bin}.\\
($^c$) Discovered by \citet{reid_cool_N10_08} and optical spectral type from \citet{Kirkp08_young_Ldwfs}.\\
($^d$) Discovery and optical spectral type by \citet{Cruz03}. Near-IR spectral type from \citet{Wilson03_2mass}.\\
($^e$) Discovery and optical spectral type by \citet{hawley}.\\
($^f$) Discovery and optical spectral type by \citet{Lodieu2002}. Near-IR spectral type from \citet{SSSPM0829_05}.}
\end{minipage}
\end{table*}

\subsubsection{The $\mbox{R}_{\rm c}-\mbox{K}_{\rm s}$~and $\mbox{I}_{\rm c}-\mbox{K}_{\rm s}$~Criteria}\label{opt_nir:rk_ik}
\begin{figure*}
 \begin{center}
  \begin{minipage}{0.75\textwidth}
  \begin{center}
  \includegraphics[width=\textwidth,angle=0]{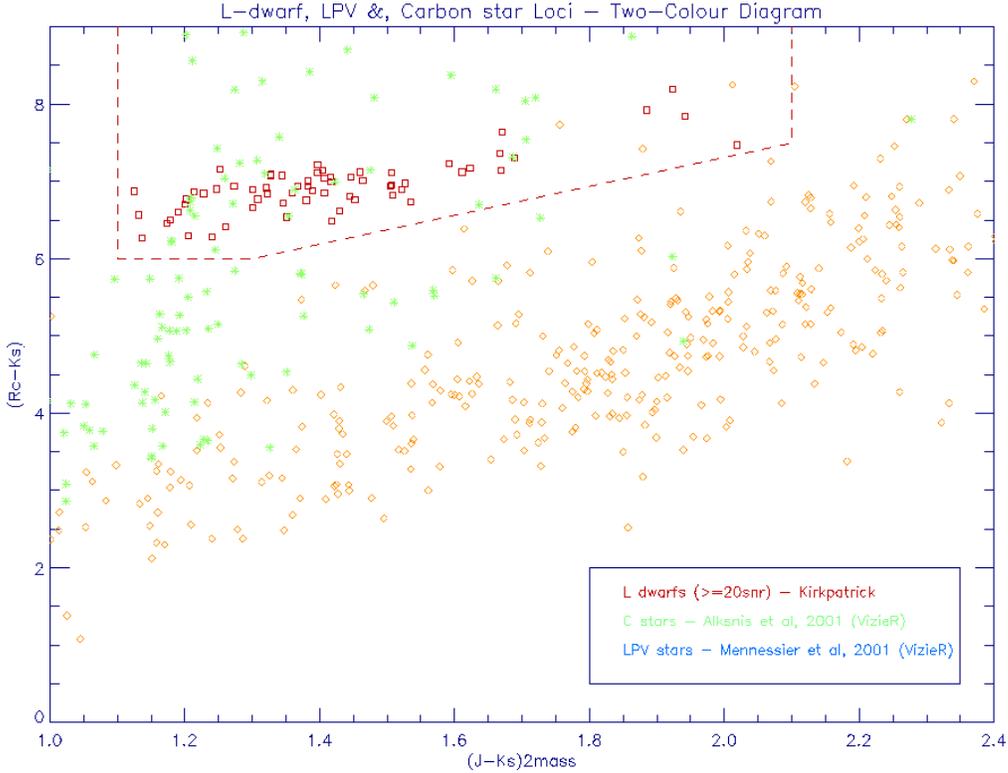}
  \caption{The \RK/\JK~colour cut criteria is the dashed red line (upper limit of $(\mbox{\RK})=9.0\,\mbox{\rm mag}$), showing a good separation between the L dwarfs and carbon stars. However, for redder \JK~values they become more of a potential problem. The `O'-rich LPV stars are seen bisecting the L dwarf locus nearly vertically due to their large amplitude of variability.}
  \label{rminusk}
  \end{center}
  \end{minipage}
 \end{center}
\end{figure*}

\begin{figure*}
 \begin{center}
  \begin{minipage}{0.75\textwidth}
  \begin{center}
  \includegraphics[width=\textwidth,angle=0]{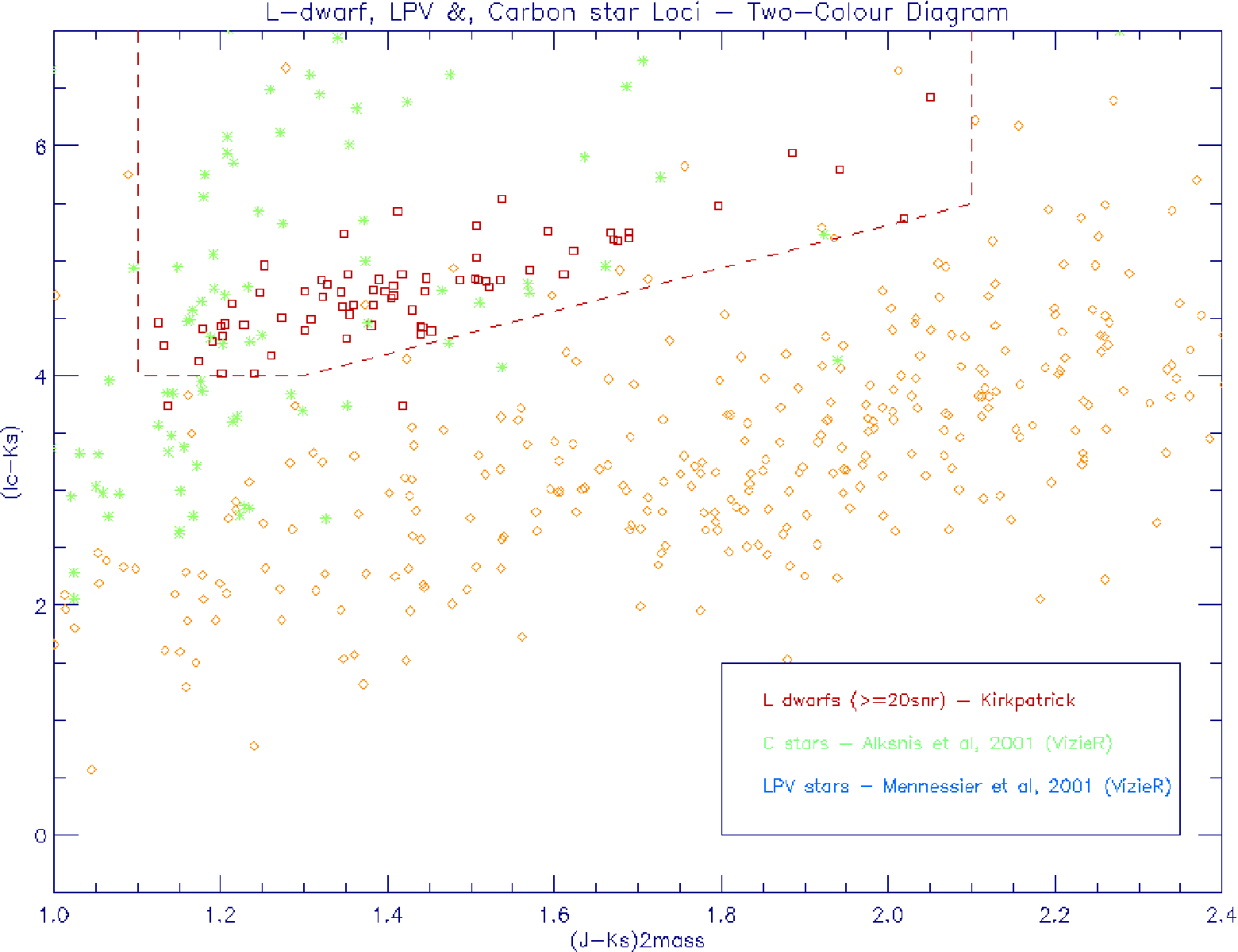}
  \caption{The (\IK)/(\JK)~colour cut criteria shown as the dashed red line (upper limit of $(\mbox{\IK})=7.0\,\mbox{\rm mag}$). The carbon stars are less well separated from the L dwarf locus than in \RK~over the same \JK~range. The `O'-rich LPV stars also bisect the L dwarf range as seen for the \RK~colour.}
  \label{iminusk}
  \end{center}
  \end{minipage}
 \end{center}
\end{figure*}

To maximise the efficiency of the contaminant rejection these two selection criteria were defined as slopes in the two-colour planes against \JK. As well as using our reference set of L dwarfs (65 with both \Rb~and \Ib~data), scatter in the colours of L dwarfs was taken into account from the literature (see \textsection\,\ref{nir:sel}) in defining these slopes.
Although the \ImJ~colour generally gives a better monotonic relation for UCDs \citep{dahn02} extending down to the latest T-dwarfs (as the {\em I}- and {\em J}-bands are relatively unaffected by \methane, but are heavily dependant on the K{\sc\,i} absorption \citep{dahn02}), we decided to use the \IK~colour in preference because it gives a better baseline separation in the two-colour $(\mbox{\IK})/(\mbox{\JK})$~selection plane of $\sim2$ magnitudes, instead of $\sim1.5$ magnitudes. As can be seen in Fig.\,\ref{rminusk} and Fig.\,\ref{iminusk}, the locus of the L dwarfs is separated better from the carbon stars in \RK~than in \IK, but appear to overlap more in both at the extreme red end of the L dwarf range $(\mbox{\JK})\approx2\,\mbox{\rm mag}$. The `O'-rich LPVs, on the other hand, extend right through the L dwarf locus upwards to redder \RK~and \IK~colours, and in general have much bluer \JK~colours than the carbon stars ($1\,\mbox{\rm mag}\leqslant(\mbox{\JK})\leqslant 2\,\mbox{\rm mag}$). Thus, these two-colour selection criteria together should provide a good rejection efficiency against carbon star contamination. Our adopted $(\mbox{\RK})/(\mbox{\JK})$~and $(\mbox{\IK})/(\mbox{\JK})$~selection criteria are given in Exp.\,\ref{rk_defin} and Exp.\,\ref{ik_defin} below, and are also outlined as dashed lines on the two-colour planes shown in Fig.\,\ref{rminusk}, and Fig.\,\ref{iminusk}.
\begin{eqnarray}
\mbox{The (\RK)/(\JK) selection plane;}\nonumber \\
                          6.0 \leqslant (\mbox{\RK}) \leqslant 9.0\nonumber \\
      \mbox{for  } 1.075\leqslant (\mbox{\JK}) \leqslant 1.3\nonumber \\
\mbox{and,}\nonumber \\
 (1.875\times(\mbox{\JK }))+3.563 \leqslant (\mbox{\RK}) \leqslant 9.0\nonumber \\
     \mbox{for  } 1.3 < (\mbox{\JK}) \leqslant 2.8\label{rk_defin}\\
\nonumber \\
\mbox{The (\IK)/(\JK)~selection plane;} \nonumber \\
                          4.0 \leqslant (\mbox{\IK}) \leqslant 7.0\nonumber \\
      \mbox{for  } 1.075\leqslant (\mbox{\JK}) \leqslant 1.3\nonumber \\
\mbox{and,}\nonumber \\
 (1.875\times(\mbox{\JK }))+1.563 \leqslant (\mbox{\IK}) \leqslant 7.0\nonumber \\
       \mbox{for  } 1.3 < (\mbox{\JK}) \leqslant 2.8\label{ik_defin}
\end{eqnarray}

As shown in Fig.\,\ref{iminusk} a couple of the known reference L dwarfs would be rejected due to their \IK~values being just under the lower limit of $(\mbox{\IK})=4.0\,\mbox{\rm mag}$. We do not adjust the lower limit to include them as the \IK~values of mid-to-late M dwarfs appear to be very sensitive to small changes in \JK~in the range of $1.0\,\mbox{\rm mag}\leqslant(\mbox{\JK})\leqslant 1.2\,\mbox{\rm mag}$, due at least in part to variations in dust, cloud coverage, and metallicity \citep[e.g., see][for details]{leggett92,Liebert06}, and show scatter in this colour. Thus, lowering the \IK~limit could potentially let through many early- to mid-M dwarfs.

\subsubsection{The $\mbox{(R}-\mbox{I)}$\,$_{\rm c}$ Criterion}\label{opt_nir:ri_jk}

\begin{figure*}
 \begin{center}
  \begin{minipage}{0.75\textwidth}
  \begin{center}
  \includegraphics[width=\textwidth,angle=0]{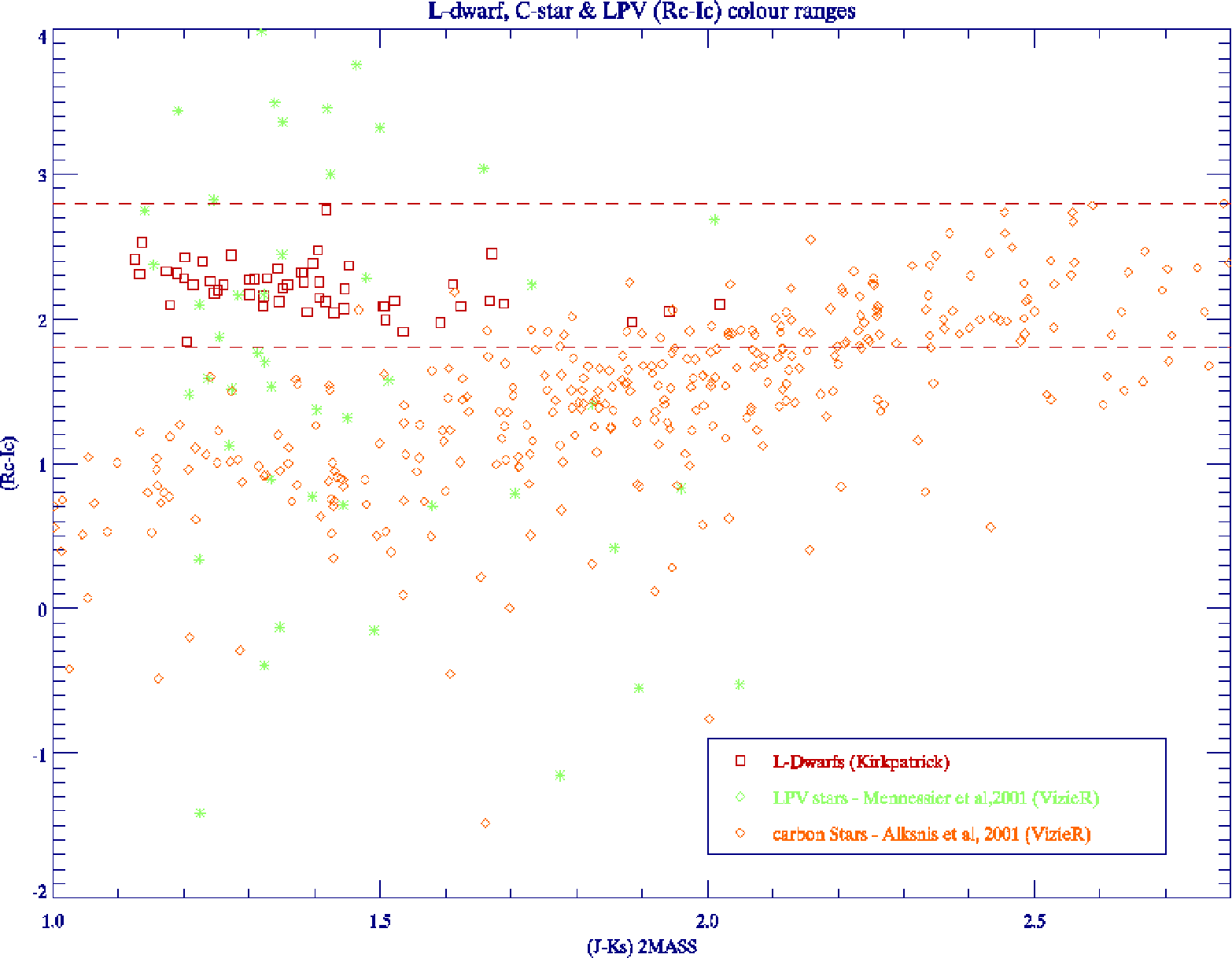}
  \caption{The $\mbox{\RI}/(\mbox{\JK})$ two-colour plane criteria. Despite the separate \RK~and \IK~cuts the \RI~contributes significantly to removing many LPV stars. Notice that the L dwarfs occupy a narrow range of $\approx 1\,\mbox{\rm mag}$ in \RI, and how the carbon stars have an overlap with L dwarfs for $(\mbox{\JK})>1.7\,\mbox{\rm mag}$.}
 \label{rminusi}
  \end{center}
  \end{minipage}
 \end{center}
\end{figure*}

Initially, no selection was made using this colour as this was implied from both the previous \RK~and \IK~criteria (i.e., $\mbox{\RI}\leqslant 2$). However, we subsequently included this colour criterion as some bright ($\mbox{\Ks}\leqslant8\,\mbox{\rm mag}$) variables, which had passed both the previous \RK~and \IK~two-colour criteria, had \RI~colours outside that expected for L dwarfs due to variability between the non-contemporaneous \Rb~and \Ib epochs. Therefore, an $\mbox{\RI}/(\mbox{\JK})$ criterion was developed around our reference set of L dwarfs that was found to occupy a flat locus with increasing \JK, having a narrow range of $\approx 1\,\mbox{\rm mag}$. The flat relation is not surprising as they display similar slopes in their loci for both the \RK~and \IK~colours against \JK. The $\mbox{\RI}/(\mbox{\JK})$~selection criterion has been defined in Exp.\,\ref{ri_defin}, and is shown in Fig.\,\ref{rminusi};
\begin{equation}\label{ri_defin}
1.8\leqslant\mbox{\RI}\leqslant 2.8;\quad \mbox{for  } 1.075\leqslant(\mbox{\JK})\leqslant 2.8
\end{equation}

The \RI~criterion contributes significantly to removing many bright contaminant stars. The narrow $\Delta\mbox{\RI}=1\,\mbox{\rm mag}$ selection range in this colour acts to constrain the possible range of \RK~and \IK~colours in a way that behaves as a second pseudo variability constraint, for objects which either pass the dual \Rb~variability test (see \textsection\,\ref{prep:dual}), or those that have just a single \Rb~detection along with an \Ib. Our selection limits in this colour match very well with the range of values obtained by \citet{Liebert06} for spectral types greater than mid-M to L--T transition.

\subsection{Reduced Proper-Motion Criterion}\label{opt_nir:rpm}

Many of the photometric selected candidates were found to be bright compared to that expected for typical UCDs (i.e., $\mbox{\Ks}\leqslant10\,\mbox{\rm mag}$), with the number increasing dramatically toward Galactic longitudes of $\Gall>260^{\circ}$ and latitudes of $|b|\la 10^{\circ}$. Therefore, these are clearly dominated by contaminant objects that would be expected to have little or no discernible proper motion. Nearby UCDs and contaminant distant luminous giants and supergiants represent kinematically distinct stellar populations within the Galaxy, allowing proper motions and apparent magnitudes to be combined to create a reduced proper-motion selection criterion. When reduced proper-motion is plotted against a suitable colour to create a Reduced Proper-Motion Diagram (RPMD), it becomes a powerful discriminant in segregating stellar (and sub-stellar) population types \citep{luyten78}. Here we utilise the 2MASS \Ks-band apparent magnitude as our UCD candidates are brightest in the near-IR spectral region.

We determine reduced proper-motion as follows;
\begin{equation} \label{H}
\mbox{\rpm}=\mbox{\em K}_{\rm s}+5+5\log\mu
\end{equation}

where the proper motion, $\mu$, is in units of arcseconds \peryr.

 The \rpm~selection criterion we developed was chosen by accessing the same `O'- and `C'-rich LPV star catalogues, and reference L dwarfs used previously, for which reliable proper-motions could be obtained from the SSA. For `O'-rich types there were 221, for carbon stars 546, while 65 L dwarfs were used (from {\sc http://dwarfarchives.org}). The accuracy of the SSA derived proper-motions \citep[see][for details]{ssa_p3c} are primarily dependant on the number of bands which have a detections (maximum of four), the separations in epoch of the scanned plates, and the faintness of individual detections. Typical uncertainties are \(\pm10<\mu<\pm50\)mas \peryr for magnitudes between \(\sim18\leqslant\mbox{{\em R}$_F$}\leqslant 21\), and become larger for short epoch baselines with faint detections.

\begin{figure}
 \begin{center}
  \includegraphics[width=0.48\textwidth,angle=0]{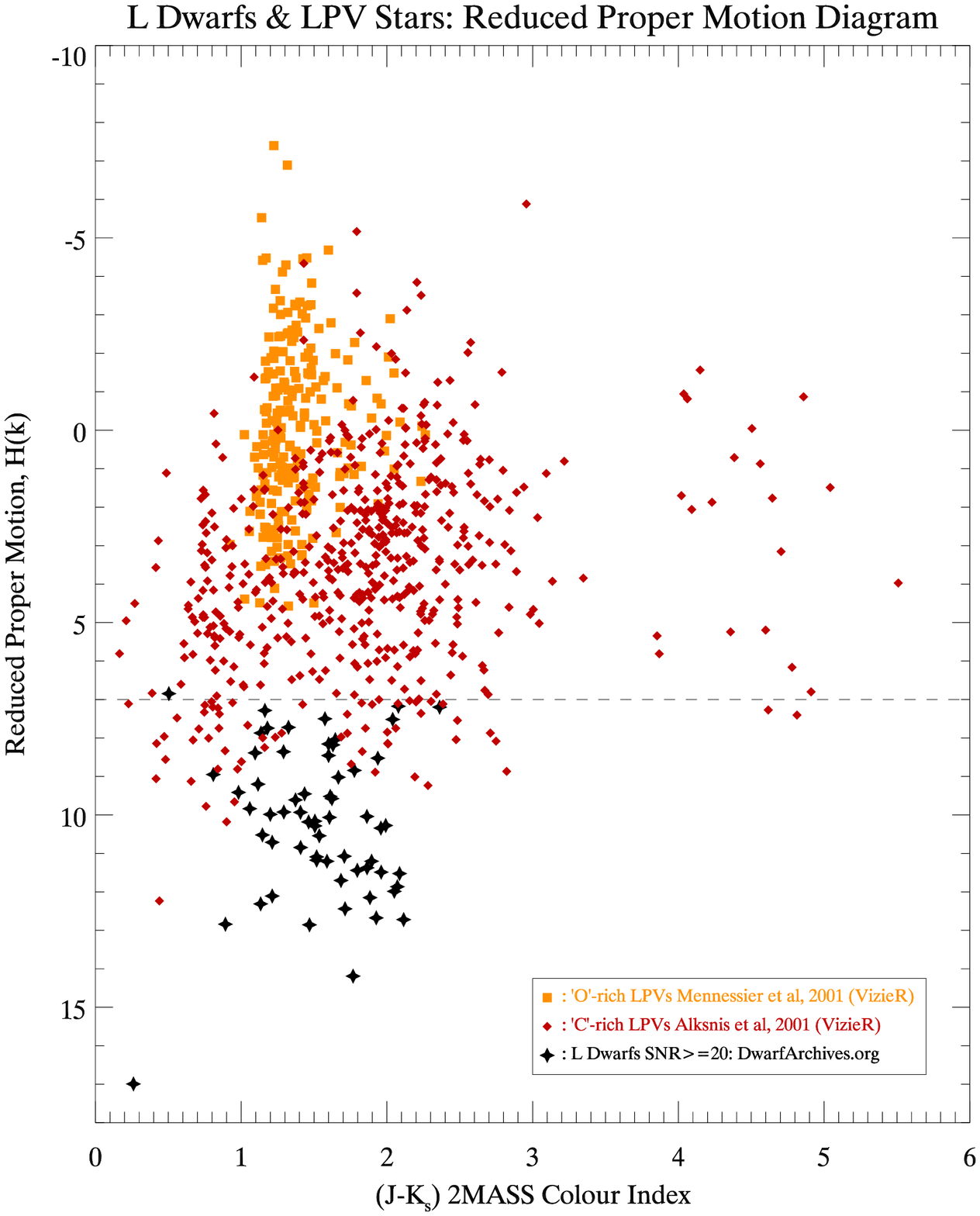}
  \caption{The RPMD showing the loci of the contaminant LPV stars and the L dwarf population. Our candidate selection is made at $\mbox{\rpm}\geqslant7$, indicated by the dashed line. The `O'- and `C'-rich LPV stars are denoted separately by the squares (orange), and diamonds (red) respectively. L dwarfs are denoted as filled star symbols (black).}
 \label{rpm_fig}
 \end{center}
\end{figure}

Our RPMD can be seen in Fig.\,\ref{rpm_fig}, which shows that the reference L dwarfs can be effectively segregated from the `O'-rich LPV stars. The carbon stars are also reasonably well segregated from the L dwarfs, but do appear to have some overlap. This may be due to the \citet{c_star_cat01} carbon star catalogue being contaminated by the recently identified population of dwarf carbon stars. These carbon dwarfs share the same disk kinematics of nearby M dwarfs, and are potentially more numerous than carbon giants (e.g., \citealt{dC_Lowrance03}; \citealt{dC_Mauron04}).

The \rpm~selection criterion was defined to reject objects with $\mbox{\rpm}\geqslant7$. This limit was found to work well in removing many of the bright candidates that we know are dominated by contamination. In Fig.\,\ref{rpm_fig} one reference L dwarf is seen as rejected (representing 1.5 per cent of the reference sample), and a couple of other L dwarfs lie close to this limit. However, given the number of potential contaminant objects at lower \rpm~values we consider it wise to stay with this conservative limit.

\begin{figure}
 \begin{center}
  \includegraphics[width=0.48\textwidth,angle=0]{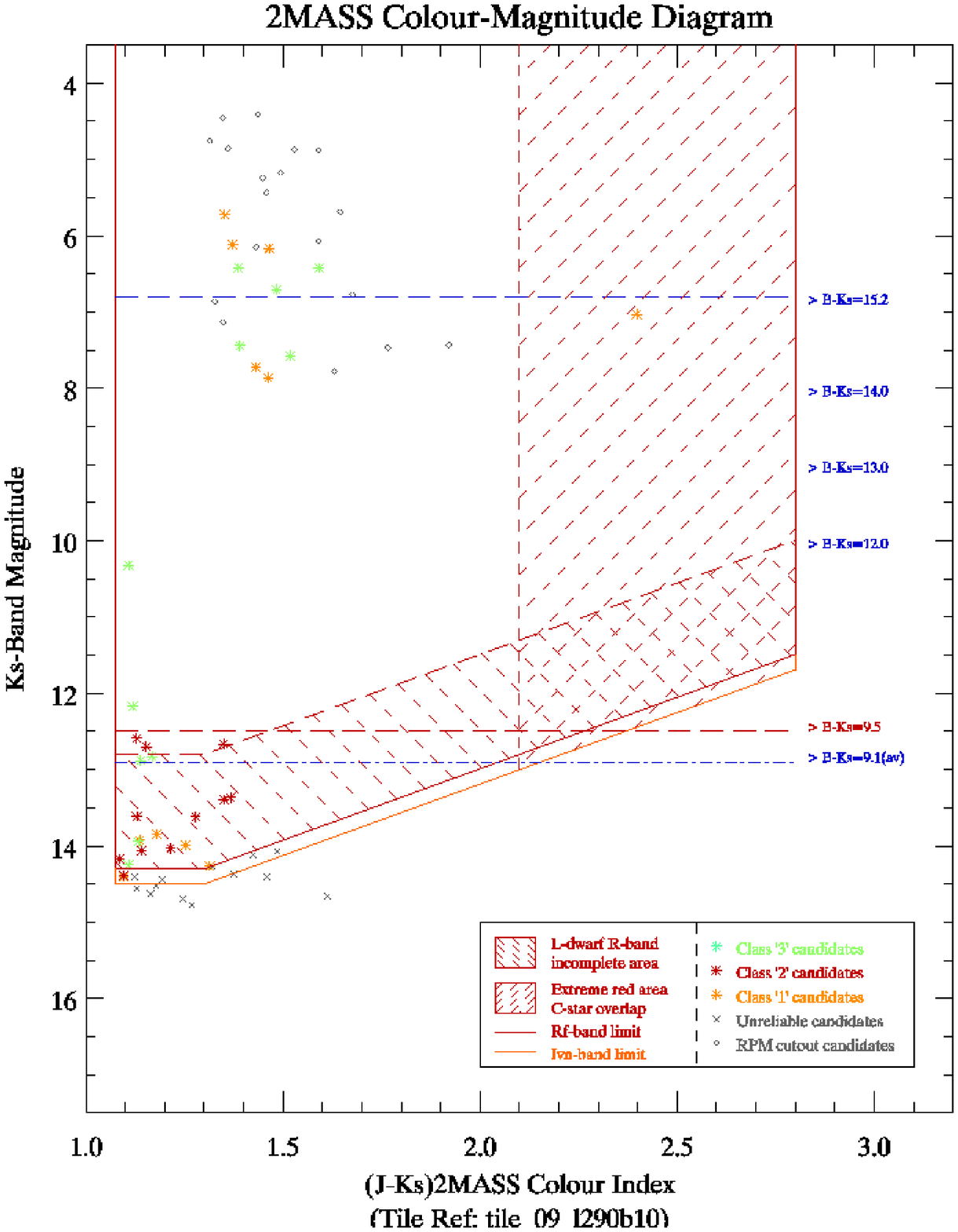}
  \caption{A \Kcmd~colour-magnitude diagram from a sky tile centred on $\Gall=290^{\circ}$ and $b=10^{\circ}$ (200 deg$^2$) showing the effectiveness of the \rpm~cut in reducing contamination. The grey circles denote 18 candidates removed after the $\mbox{\rpm}\geqslant7$ selection criterion was applied.}
 \label{cmd_rpm}
 \end{center}
\end{figure}

The effectiveness of the \rpm~cut can be seen in the \Kcmd~colour-magnitude diagram of Fig.\,\ref{cmd_rpm}, taken from one of the sky tiles (200 deg$^2$) centred on $\Gall=290^{\circ}$ and $b=10^{\circ}$. The 18 grey circles in this diagram are candidates that were removed by the \rpm~cut, which amounts to 62 per cent of the bright objects with $\mbox{\Ks}<8\,\mbox{\rm mag}$.

\subsection{Candidate Classification}\label{meth:class}
As our search progressed eastwards to Galactic longitudes of $\Gall\approx260^{\circ}$ the number of candidates per typical sky tile increased significantly to several hundred (number dependent on exact $\Gall$~and \Galb). Confronted with such large numbers of candidates a classification scheme was developed to help identify the `quality' of the candidates, in terms of their likelihood of being genuine UCDs.

 This was achieved by defining a quality parameter which is defined in two parts, both in relation to a \Kcmd~CMD. The first part indicates whether there is a detection in each optical {\em BRI}-band, and whether that detection falls inside or outside the photometric completeness limit defined for that band. The photometric completeness limits of the optical bands are derived in relation to the \Ks-band limiting magnitude using the following information;

\begin{enumerate}
\item  The limiting magnitude that corresponds to the SuperCOSMOS Sky Survey plate detection limits at an $\sim$80 per cent confidence level of an object being detected as a star \citep[see][for details]{ssa_p2b}. These limits for each band are: $\mbox{\Bj}\simeq22\,\mbox{\rm mag}$; $\mbox{\Rone}\simeq19.5\,\mbox{\rm mag}$; $\mbox{\Rtwo}\simeq20.3\,\mbox{\rm mag}$; $\mbox{\In}\simeq18.5\,\mbox{\rm mag}$. Given the two \Rf~we adopt a limit of $\mbox{\Rc}\simeq20.3$.
\item  How the slopes of the \RK/\JK~and \IK/\JK~selection criteria, coupled with the above magnitude limits, translate onto the \Kcmd~CMD.
\end{enumerate}

For an \Rc~magnitude limit case this translates onto the \Kcmd~CMD as a faint photometric completeness limit defined as;
\begin{displaymath}
1.075\leqslant \mbox{\JK}\leqslant 1.3\left\{ \begin{array}{l}
\mbox{\RK~lower limit}=6.0\\
\mbox{\Ks~limit}=(20.3-6.0)=14.3\,\mbox{\rm mag} \end{array} \right.
\end{displaymath}
\begin{displaymath}
1.3 < \mbox{\JK}\leqslant 2.8\left\{ \begin{array}{l}
\mbox{\RK~lower limit}=\\(1.875\times\mbox{\JK })+3.5625.\\
\mbox{\Ks~limit}=20.3-\mbox{(\RK)} \end{array} \right.
\end{displaymath}

For an \In~magnitude limit$\simeq18.5\,\mbox{\rm mag}$, this translates onto the \Kcmd~CMD as a faint photometric completeness limit defined as;
\begin{displaymath}
1.075\leqslant \mbox{\JK}\leqslant 1.3\left\{ \begin{array}{l}
\mbox{\IK~lower limit}=4.0\\
\mbox{\Ks~limit}=(18.5-4.0)=14.5\,\mbox{\rm mag} \end{array} \right.
\end{displaymath}
\begin{displaymath}
1.3 < \mbox{\JK}\leqslant 2.8\left\{ \begin{array}{l}
\mbox{\IK~lower limit}=\\(1.875\times\mbox{\JK })+1.5625.\\
\mbox{\Ks~limit}=18.5-\mbox{(\IK)} \end{array} \right.
\end{displaymath}

The \Bj~magnitude limit is $\simeq22.0\,\mbox{\rm mag}$, which translates onto the \Kcmd~CMD as a single faint photometric completeness limit defined by;
\begin{displaymath}
1.075\leqslant\mbox{\JK}\leqslant 2.8\left\{ \begin{array}{l}
\mbox{\BK~single limit}=9.5\\
\mbox{\Ks~limit}=(22.0-9.5)=12.5 \end{array} \right.
\end{displaymath}

This part of the parameter contains a set of three character indicators, one for each of the {\em BRI} bands, which have been given one of the following attributes:
\begin{description}
\item []{{\bf `-'}:} for a non-detection which would fall outside the completeness limit of that band, given its \Ks~magnitude.
\item []{{\bf `N'}:} for a non-detection that lies within the completeness limit of that band, so is theoretically bright enough to be detected as an UCD.
\item []{{\bf `W'}:} for a detection in that band which lies within the photometric completeness limit.
\item []{{\bf `O'}:} for a detection in that band which lies outside the photometric completeness limit (but has passed the error flag check).
\end{description}
So far this parameter can therefore contain combinations of the three individual band indicators such as `NOW' or `-NW', which in this example have the meaning in the first case of: has no detection in \Bj, but is within the expected photometric limits, has an {\em R}$_F$ magnitude outside of the detection limits, and has an \In~magnitude which falls within the \Ic~limit. In this example the {\em R}$_F$-band detection indicator was said to be outside the completeness limits; for some of the {\em R}$_F$ and \In~plate scans the detection limits have been found to be fainter than the $\simeq20.3\,\mbox{\rm mag}$ and $\simeq18.5\,\mbox{\rm mag}$ limits expected, but still have SSA quality flags set to $\leqslant2048$, therefore these objects have been accepted as candidates if they pass all the selection criteria.

The full quality parameter is now defined based on the first part, and is designed to prioritise the candidates into four classes to denote their `quality' as an aid to the visual checking process. This parameter (which we have termed OP - an `Optimistic Parameter' to denote optimism in finding a potential UCD) classifies the candidates according to their position on the \Kcmd~CMD, relative to the photometric completeness limits, and according to the combination of indicators in the first parameter. Thus, encoded in these indicators are potentially useful information such as the likelihood of a candidate being a bright high proper-motion nearby object -- based on how many, and which, optical detections they have. Therefore, a bright $\mbox{\Ks}\approx10\,\mbox{\rm mag}$ candidate having an indicator attribute combination of `NNN' would be very interesting as it might be nearby and have a high proper motion, explaining why it has no optical detections in the merged SSA -- when it should have. Conversely, an object with a combination of `-~-~O' has only an \Ib~detection outside of the completeness limit, and is most likely to be a faint, distant, and reddened contaminant object.

This classification scheme has been defined such that the differentiation in class indicate the likelihood of a true UCD detection -- {\em Class `3'} being the best. Candidates whose position falls outside the \Rc- and \Ic-band limits on the CMD, with or without actual optical detections are classified as {\em Class `0'}. These are most likely highly reddened distant giants and/or luminous main sequence stars.

The definition of each of the OP `{\em classes}' were carefully chosen from different combinations of attributes from the first quality parameter, such that they are likely to indicate the most promising candidates in priority of each `{\em class}'. These combinations are listed below:
\begin{description}
\item []{{\bf{\em `Class 3'}}:} Potentially the most interesting with possible large proper-motions and being bright enough in the \Ks-band to have at least two optical detections (actual or possible). Have the attributes of: `NNN', `NWN", `NON', `-NN', `-WN', and `NWW'.
\item []{{\bf{\em `Class 2'}}:} Good candidates with a mixture of one or two optical detections for either brighter or fainter candidates. Have the attributes of: `NOW', `NNO', `NOO', `OWW', `-OW', `-WW', `-NW', `-ON', and `-~-W'.
\item []{{\bf{\em `Class 1'}}:} Are possible candidates, but may also be contaminants, and include attribute combinations not given in the previous {\em `Class 3'} and {\em `Class 2'} cases, and also for{\em `Class 0'} (see below). Comprise of the fainter candidates near the photometric limits.
\item []{{\bf{\em `Class 0'}}:} Objects in this class lie below the photometric limits with either one, or no, optical detections, and are not considered reliable candidates. Have the attributes of: `-~-~-', `-~-O', and `-O-".
\end{description}
All the {\em `Class 3'} to {\em `Class 1'} candidates are retained for the final visual checking stage.

\begin{figure}
 \begin{center}
  \includegraphics[width=0.48\textwidth,angle=0]{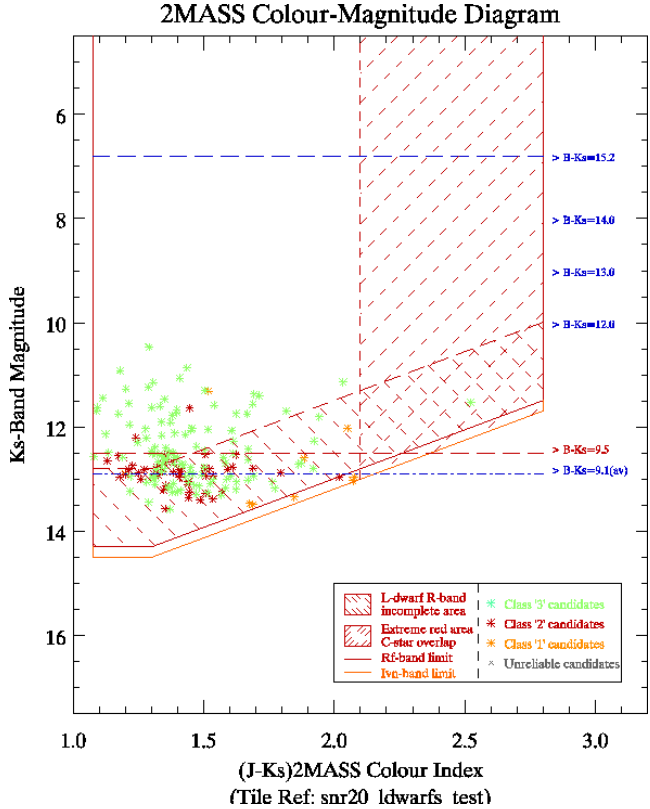}
  \caption{The candidate classification scheme as applied to the $\mbox{SNR}\geqslant20$ reference L dwarf sample, and showing their locations relative to the {\em BRI}-band photometric completeness limits. The candidate `class' is denoted by colour: best {\em `Class 3'} are green; good {\em Class `2'} in red, and the {\em `Class 1'} candidates as orange. The \Rc- and \Ic-band completeness limits are shown as red and orange solid lines respectively (bottom of the enclosed region). The \Bj~completeness limit is the horizontal red dashed line at $\mbox{\Ks}=12.5\,\mbox{\rm mag}$. Other \Bj~limits are also shown to denote the variability range of the LPV stars in \BK.}
 \label{cmd_lds_class}
 \end{center}
\end{figure}

The validity of the classification scheme was tested by comparing the attribute combinations of the first quality parameter for the reference L dwarfs, relative to their positions on the same CMD to check that these differing classes do reflect their potential to indicate nearby/interesting UCDs. This reference \Kcmd~CMD is shown in Fig.\,\ref{cmd_lds_class}, and it may be seen that the classification scheme works well with the brightest L dwarfs being given the best rating (green symbols), and the L dwarfs near the lower edge of the limits having the least rating (orange).

\subsection{Additional selection requirements in Highly Overcrowded Regions}\label{meth:overcrow}

For Galactic longitudes of $260\geqslant\Gall\leqslant340^{\circ}$ and $b\geqslant 5^{\circ}$, and eastwards of $\Gall=340^{\circ}$ for $b\leqslant 15^{\circ}$ the colour and \rpm~selection alone returned up to several hundred UCD candidates for a typical 200\,deg$^2$ region, due to problems of overcrowding and reddening. The majority of the candidates saturating the colour and \rpm~selection method were of the fainter {\em `class 1'} candidates (largely without proper-motion measurements).

An additional approach was developed to address this issue that also retains the best chance of detecting bright/nearby UCD candidates, and is summarised thus: ({\sc\,i}) A limit of one {\em class [321]} candidate per square degree (averaged over an individual sky tile) was chosen to facilitate a sensible number to be checked by visual means.({\sc\,ii}) To impose a magnitude limit of $\mbox{\Ks}\leqslant12.5\,\mbox{\rm mag}$.

A favourable aspect of this approach is that it can be used to bias the candidate selection away from pockets of highly clumped near-IR only detections due to regions of high reddening, to one that is from a more homogeneous distribution. However, this additional overcrowding and photometric `clipping' criteria would only be activated when the surface density of {\em `class [321]'} candidates exceeded $\Sigma_{\rm Cands}\geqslant\langle1\rangle\mbox{\,deg}^{-2}$, and was implemented as part of our automated procedure with details summarised as follows:

\begin{enumerate}
\item Breaks each sky tile into a 2-dimensional grid of $1^{\circ}\times1^{\circ}$ elements in $\ell$ and {\em b}.
\item Candidates are selected from these elements begining with elements containing the least number, but which have been sorted such that the best quality candidates ({\em class [321]}) are selected first in a stepwise fashion.
\item The selection continues until the maximum number of candidates (equal to the whole number of square degrees in the sky tile) is reached, or would be exceeded, given the number in the next element.   
\item Candidates are then selected from the remaining elements until the maximum permitted number is reached, except that this time only Candidates with magnitudes of $\mbox{\Ks}\leqslant12.5\,\mbox{\rm mag}$ are selected. If not enough bright candidates are found to reach the maximum limit, then different elements containing successively greater number densities are checked in ascending order using the $\mbox{\Ks}\leqslant12.5\,\mbox{\rm mag}$ criterion, until it is reached, or no more bright candidates are found. 
\end{enumerate}
The advantage of this approach is that most of the candidates in the elements are selected in the same way to regions where the overcrowding procedure does not apply (i.e., $\Sigma_{\rm Cands}<\langle1\rangle\mbox{\,deg}^{-2}$), except that they are biased towards the best candidate {\em class 3} objects. In practice it was found that typically only a few photometrically `clipped' candidates of either {\em `class'} brighter than $\mbox{\Ks}=12.5\,\mbox{\rm mag}$ were selected from a typical region.

\begin{figure*}
 \begin{center}
  \begin{minipage}{1.0\textwidth}
  \begin{center}
  \includegraphics[width=0.95\textwidth,angle=90]{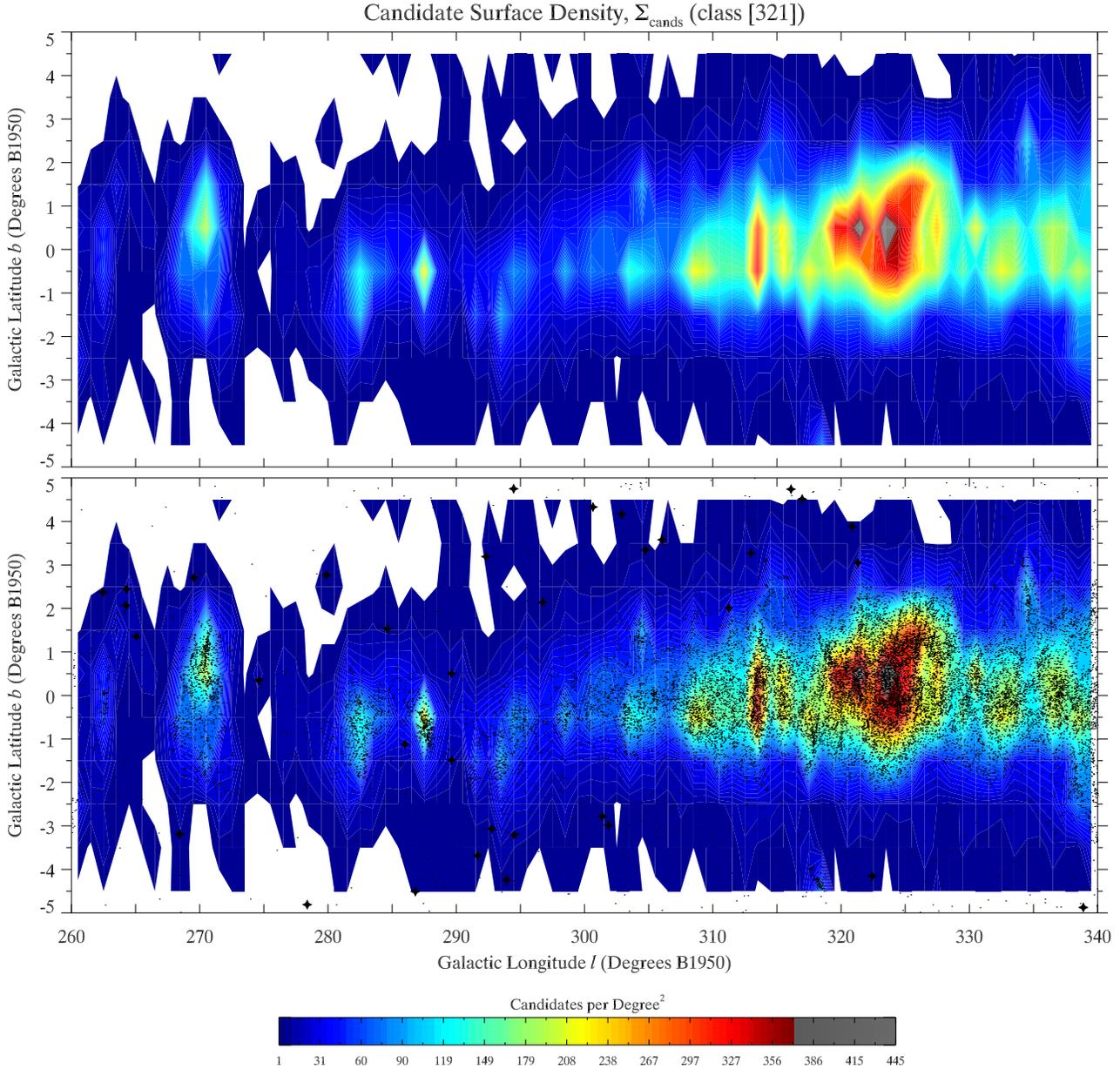}
  \caption{A colour-coded surface density contour map ($\Sigma_{\rm Cands}$) of all {\em class [321]} candidates before application of the `overcrowding criteria'. Surface densities are obtained from $1^{\circ}\times1^{\circ}$ resolution elements in Galactic $\ell$ and {\em b} from the region: $260^{\circ}\leqslant\Gall\leqslant340^{\circ}$ and $|\mbox{{\em b}}|\leqslant5^{\circ}$. The region of high density correlates with the position of the Norma dark molecular cloud. The lower plot shows the actual candidates from the initial selection indicating the degree of clumping on small spatial scales. The base level colour is taken as one candidate per $1^{\circ}\times1^{\circ}$ resolution element the same value as used to invoke the overcrowding criteria. Note that the axis are not plotted isotropically in both panels, elongating the contours in Galactic latitude.}
  \label{surf_dens_t7to16}
  \end{center}
  \end{minipage}
 \end{center}
\end{figure*}
\begin{figure*}
 \begin{center}
  \begin{minipage}{1.0\textwidth}
  \begin{center}
  \includegraphics[width=0.9\textwidth,angle=0]{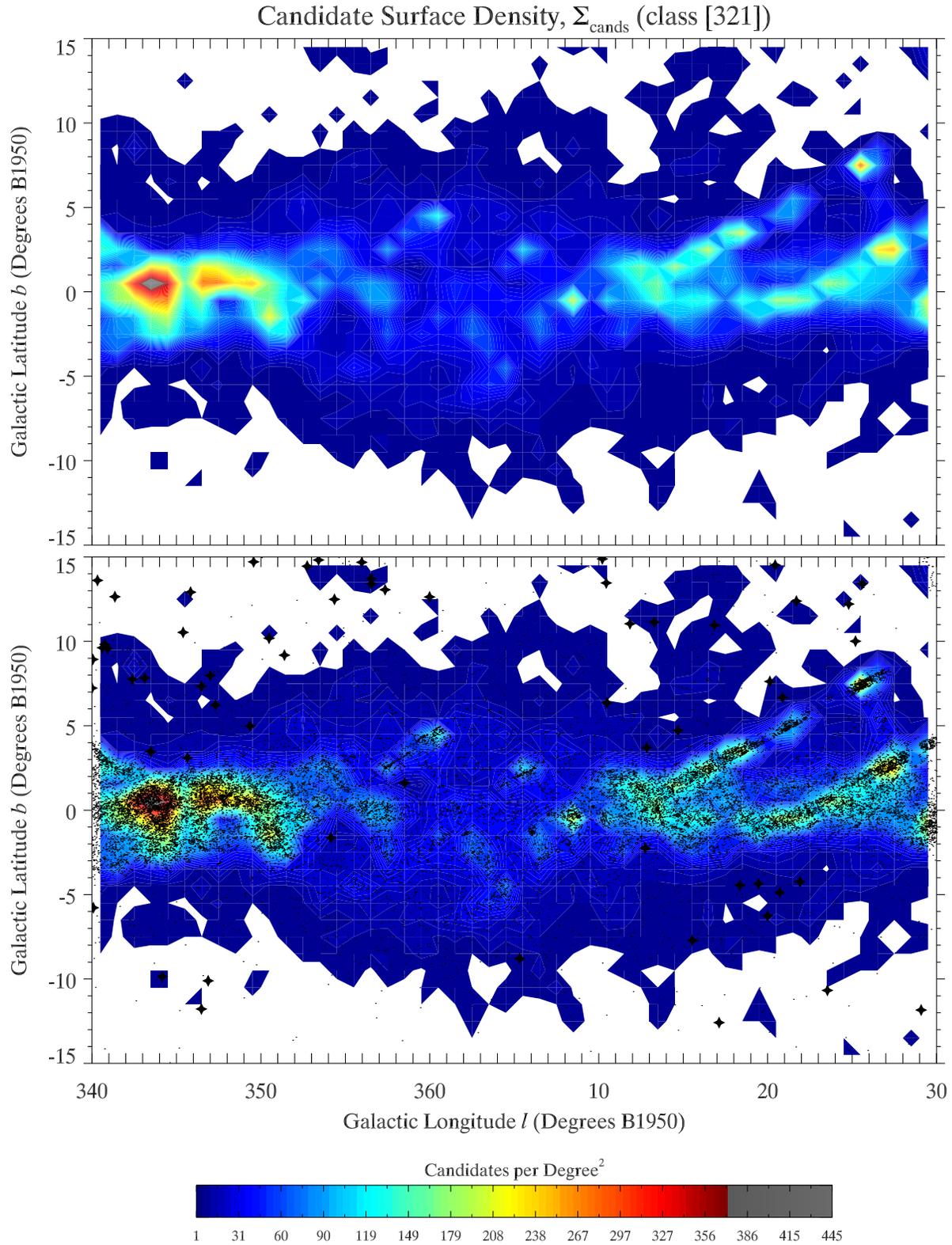}
  \caption{A colour-coded surface density contour map ($\Sigma_{\rm Cands}$) of all {\em class [321]} candidates obtained in the same way as for Fig.\,\protect\ref{surf_dens_t7to16}, but for the region $340^{\circ}\leqslant\Gall\leqslant360^{\circ}$ and $0^{\circ}\leqslant\Gall\leqslant30^{\circ}$ for $|b|\la 15^{\circ}$, and covering 1500\,deg$^2$. The region of high density (red) on the left coincides with the Lupus molecular clouds -- one of the nearest regions of low-mass star formation. Note the linear features tracing higher density, primarily between $10^{\circ}\leqslant\Gall\leqslant30^{\circ}$ (see text for details). The base level colour is taken as one candidate per $1^{\circ}\times1^{\circ}$ resolution element the same value as used to invoke the overcrowding criteria. The axis are plotted isotropically in both panels.}
  \label{surf_dens_t18to24}
  \end{center}
  \end{minipage}
 \end{center}
\end{figure*}

The problems of overcrowding, and our effectiveness in dealing with it, are illustrated in both panels of Fig.\,\ref{surf_dens_t7to16} and Fig.\,\ref{surf_dens_t18to24}, which show colour-coded contour maps of the candidate surface density, $\Sigma_{\rm Cands}$, for an 800\,deg$^2$ region along the Galactic plane within $|b|\la 5^{\circ}$, and a 1500\,deg$^2$ region within $|b|\la 15^{\circ}$ respectively. These surface density maps are based on all {\em class [321]} candidates (55,782) before the overcrowding criteria have been applied, illustrating the degree of structure in the distribution of these candidates, as well as indicating the shear number involved. The base level contour in Fig.\,\ref{surf_dens_t7to16} and Fig.\,\ref{surf_dens_t18to24} is set at $\Sigma_{\rm Cands}=1$, the same value that is used to force the method to switch to using the overcrowding selection process. The contouring shows that there are still significantly large areas with low values of $\Sigma_{\rm Cands}\leqslant1$ for longitudes between $260^{\circ}\leqslant\Gall\leqslant340^{\circ}$, with the lower panels of both Fig.\,\ref{surf_dens_t7to16} and Fig.\,\ref{surf_dens_t18to24} over-plotting the actual {\em class [321]} candidates. This confirms that there is a large degree of clumping on small scales of one square degree or less, presumably from dense regions of high reddening such as in small isolated dark clouds \citep[i.e., see][]{sml_dark_clouds95}. Also plotted on the lower panels of these two Figures are the final UCD catalogue members (as black filled star symbols) that trace a more homogeneous distribution of lower surface density.

It is interesting to note that the region of high surface density (red colour) in Fig.\,\ref{surf_dens_t7to16} centred around $\Gall\approx325^{\circ}$ and $\mbox{\em b}=0^{\circ}$, is a reddening hot spot known as the Norma dark cloud -- a giant molecular cloud and site of massive star formation. Another region of high reddening showing a high surface density is located at about $\Gall\approx343^{\circ}$ and $\mbox{\em b}=0^{\circ}$ (left side in Fig.\,\ref{surf_dens_t18to24}), and coincides with the Lupus molecular clouds -- one of nearest regions of low-mass star formation and part of the Sco-Cen OB association lying at a distance of 150\,pc \citep[e.g. see][and references therein]{lupus_MC_dist00}. Consequently, at these location there exists a large fraction of {\em class [321]} candidates with only near-IR detections (23.6 per cent).

On the right of Fig.\,\ref{surf_dens_t18to24} (primarily between $10^{\circ}\leqslant\Gall\leqslant30^{\circ}$) there appear to be curious linear features that look artificial in nature. On further investigation these features appear to run in declination (constant in $\alpha$) and trace the East and West overlap regions of the $4^{\circ}\times4^{\circ}$ degree SSS scanned photographic plates. However, plotting the SSA scanned source catalogue data for these regions reveals there to be an over density of detections in these overlap regions, caused by spurious detection records from a combination of two or more plates in these overcrowded and reddened regions of sky (N. Hambly, priv. comm.).

What appears to be happening is that the SSA query returns less optical counterparts in these overlap regions, most likely due to the constraint on the quality flags rejecting the spurious and confused detections. The linear features seen in Fig.\,\ref{surf_dens_t18to24} are then caused by an over density of selected candidates with only near-IR photometry along these overlap regions, which cannot be matched to optical data. These linear features, along with the other over-densities of near-IR only candidates, will serve to limit the efficiency of our method in detecting genuine UCDs in these difficult regions.

\begin{figure*}
 \begin{center}
  \begin{minipage}{1.0\textwidth}
  \begin{center}
  \includegraphics[width=0.92\textwidth,angle=0]{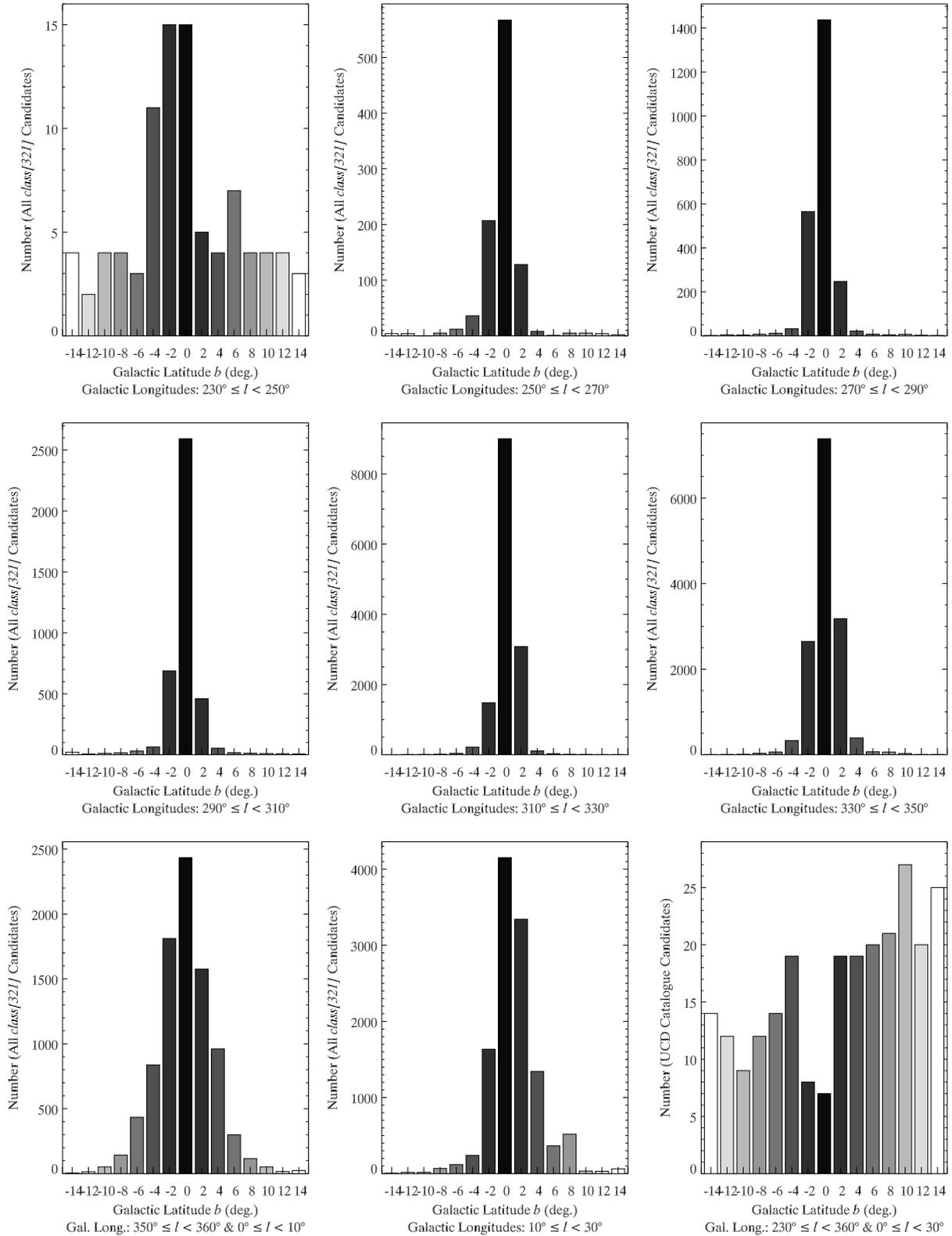}
  \caption{Panels 1 -- 8 increasing by column then row are histograms in consecutive ranges in Galactic longitude showing the number distribution of {\em class [321]} candidates as a function of Galactic latitude before treatment for overcrowding. Panel 9 bottom right: the distribution of all final UCD catalogue members for the whole area searched. The histograms show how the distribution of all candidates deviates strongly from a homogeneous distribution as a function of Galactic longitude and latitude, indicating many are contaminants. However, the UCD catalogue members are more evenly distributed with Galactic latitude, with the exception of the Galactic mid-plane ($|b|\la 2^{\circ}$).}
  \label{hist_321cands}
  \end{center}
  \end{minipage}
 \end{center}
\end{figure*}

To assess the degree of success our method has in dealing with overcrowding and reddening, we studied the spatial distribution of the {\em class [321]} candidates before application of the overcrowding criteria, with that of the final UCD catalogue members selected from the remaining candidates after application. We plotted the {\em class [321]} candidates as a function of Galactic latitude for seven consecutive ranges in Galactic longitude, which are shown as histograms in Fig.\,\ref{hist_321cands}. It is clear from these histograms that the number of candidates increases dramatically within $|b|\la 4^{\circ}$ eastwards of $\Gall\ga 300^{\circ}$, indicating that a large majority will be highly reddened contaminants due to the high stellar densities seen through the disk as the Galactic centre is approached.

In the lower right histogram of Fig.\,\ref{hist_321cands} we have also plotted the distribution of all our UCD catalogue members (\catnum: for $230^{\circ}\leqslant\Gall\leqslant30^{\circ}$). This shows that the UCD catalogue members have been selected from a much more homogeneous distribution of sky over the area searched, more in-line with what would be expected for a UCD population within a 100\,pc radius. Interestingly, this histogram suggests that the efficiency of our candidate selection process slightly favours northern Galactic latitudes, for reasons which are currently unclear.

\subsection{Visual Checking: Refining the Candidate Sample}\label{meth:vis}

Visually checking all the sky tile candidates obtained from the automated method represented a significant investment of time, requiring somewhere in the region of $\sim13,000$ separate images to be examined. Here we discuss our approach used to produce the final catalogue by removing additional contaminants, the majority due to the following:

\begin{enumerate}
\item associated with plate artefact's/defects.
\item associated with small isolated dark clouds, or in highly reddened regions and showing no proper-motion.
\item associated with regions of bright nebulosity (if no proper-motion is evident).
\item detections associated with bright stellar halos.
\item bright variable stars (e.g., giant LPVs) without SSA listed detections (i.e., through confusion).
\item photometrically confused/blended objects in highly crowded regions showing no sign of proper-motion.
\end{enumerate}

\subsubsection{Visual Checking Procedure}

For each candidate we examined $3\mbox{\arcmin}\times3\mbox{\arcmin}$ FITS images for each \Bj, \Rone, \Rtwo, and \In~band obtained from the SSS. The positions of the candidate 2MASS catalogue entries were overlaid on the images to identify the optical counterpart; highlight any astrometric calibration problems if present (i.e., calibration offsets not being confused with proper-motion); and determine by-eye if the candidate displays any proper-motion (if SSA database values were unavailable), or has a consistent proper-motion with an SSA measured value. In many cases 2MASS images were also checked and 'blinked' compared with the SSA images to aid in this process. 

For candidates where the optical counterparts were easily identifiable and isolated (i.e., not photometrically blended with nearby objects), the checking process consisted of verifying that the SSA magnitudes for each band had: stellar profiles, brightness differences consistent with very red optical colours of UCDs (i.e., to identify spurious photometric measurements), and that any discernible proper-motion was consistent with epoch differences. For these situations the SSA proper-motion measurements were used if available and deemed reliable if having small uncertainties. Unfortunately, well behaved situations like this were quite uncommon, especially so for regions covering the Galactic plane ($|\mbox{{\em b}}|\la5^{\circ}$). So, for the majority of cases many problems were encountered that required different approaches to be taken to accept, or reject, a potential catalogue candidate. Two particular recurrent problems were identified:

\begin{enumerate}
\item  SSA {\em R}- and/or \Ib~photometry for objects in the initial candidate list is missing, or is obviously incorrect compared to the brightness of the optical counterpart in the SSS image(s); this maybe due the object faintness, proximity to nearby objects (i.e., $\la 10$\arcsec), or an SSA miss-match due to high proper-motion and/or high stellar densities. 
\item  An object would have an optical detection in one or all of the \Bj, \Rone, \Rtwo, and \In~images (and may or may not have SSA magnitudes), but show no sign of proper-motion. This situation usually occurred when: (i) the variability of an LPV star was caught in each non-contemporaneous band such that the derived optical and optical-NIR colours pass each criterion, (ii) an object with only one faint {\em R}- or \Ib~detection that passes the colour criterion -- this was a common occurrence and such candidates are highly reddened distant objects, (iii) objects (bright or faint) that are photometrically blended with very nearby objects in crowded fields, and thus have incorrectly measured magnitudes.
\end{enumerate}

Solutions to these two recurrent problems were devised which were included as part of the method:
\begin{enumerate}
\item  In this situation the photometry available in the SSS image FITS detection tables was utilised. However, sometimes \In~detections had incorrect pairings with the corresponding \Rone~and/or \Rtwo~detections, especially in highly crowded regions or for higher proper-motion objects. Once optical counterparts were correctly identified, FITS table {\em R}$_F$-band(s) and \In~band magnitudes were then converted onto the Cousins system (as discussed in \textsection\,\ref{prep:conv}), with the resulting colour indices then subjected to the same criteria as applied in the initial selection process (details given in \textsection\,\ref{meth:opt_nir}). In these cases proper motions (where evident) were calculated from the astrometric differences between the 2MASS detection (from the PSC) and an SSS optical detection(s). The SSS optical detection with the longest temporal baseline and bright enough to give a reliable centroid was used. 
\item  The solution for the second case was to impose an astrometric criterion based on the argument that if any measured difference in the object positions between two (2MASS and SSA) epochs is less than the typical SSA astrometric uncertainty of $0.3\arcsec$, then the object is unlikely to have any significant real proper-motion. If an object under test had a motion greater than this lower limit, and the resulting \rpm~value passed the selection criteria (see \textsection\,\ref{opt_nir:rpm}), then the candidate was retained. One potential problem with this solution is that caused by small epoch differences of less than ten years. As mentioned, the average {\em R}$_F$ and \In~plate epoch difference to 2MASS is $\sim16$ years, but in some regions the epoch difference between the \Rtwo, and \In~plates can be significantly less than this (e.g., as low as two years in some cases). Potentially this could lead to some genuine UCD candidates being rejected. However, 2MASS / UKST/ESO epoch differences of less than ten years are not common, so for the majority of cases this was not an issue.
\end{enumerate}

Both of these approaches were effective in greatly reducing the numbers of contaminants in the initial candidate list, particually from longitudes of $260^{\circ}<\ell<340^{\circ}$ for $|\mbox{{\em b}}|\la5^{\circ}$, and all regions eastward of $\ell=340^{\circ}$.

There were a number of objects during this search for which no optical detections could be identified with the 2MASS position in any of the SSS image pass-bands. Reasons for this may include the following:
\begin{enumerate}
\item variability of contaminant LPV stars which have been coincidentally observed in each optical pass-band at the stars minimum has resulted in them being too faint to detect.
\item a fast moving Solar system object such as a minor planet or Near Earth Object (NEO).
\item a genuine high proper-motion nearby UCD, and/or one of late spectral type.
\end{enumerate}

To help address these possibilities the DENIS database and {\em I}-, {\em J}-, and {\em K}-band images were checked to see if any detections were visible at the 2MASS locations. If no DENIS database entry or detection was visible in the DENIS {\em J}-band (with a limiting magnitude of $\sim16.5\,\mbox{\rm mag}$ -- similar to 2MASS), then we assumed the candidate was a transient Solar system object or variable star, and it was rejected. However, if a detection was visible at (or near) the 2MASS position then the object was retained.

\subsubsection{Examples of Visually Based Rejections}

\begin{figure*}
  \begin{center}
   \begin{minipage}{1.0\textwidth}
    \begin{center}
     \subfigure[An LPV star: \Rone~image]{\includegraphics[scale=0.35,angle=0]{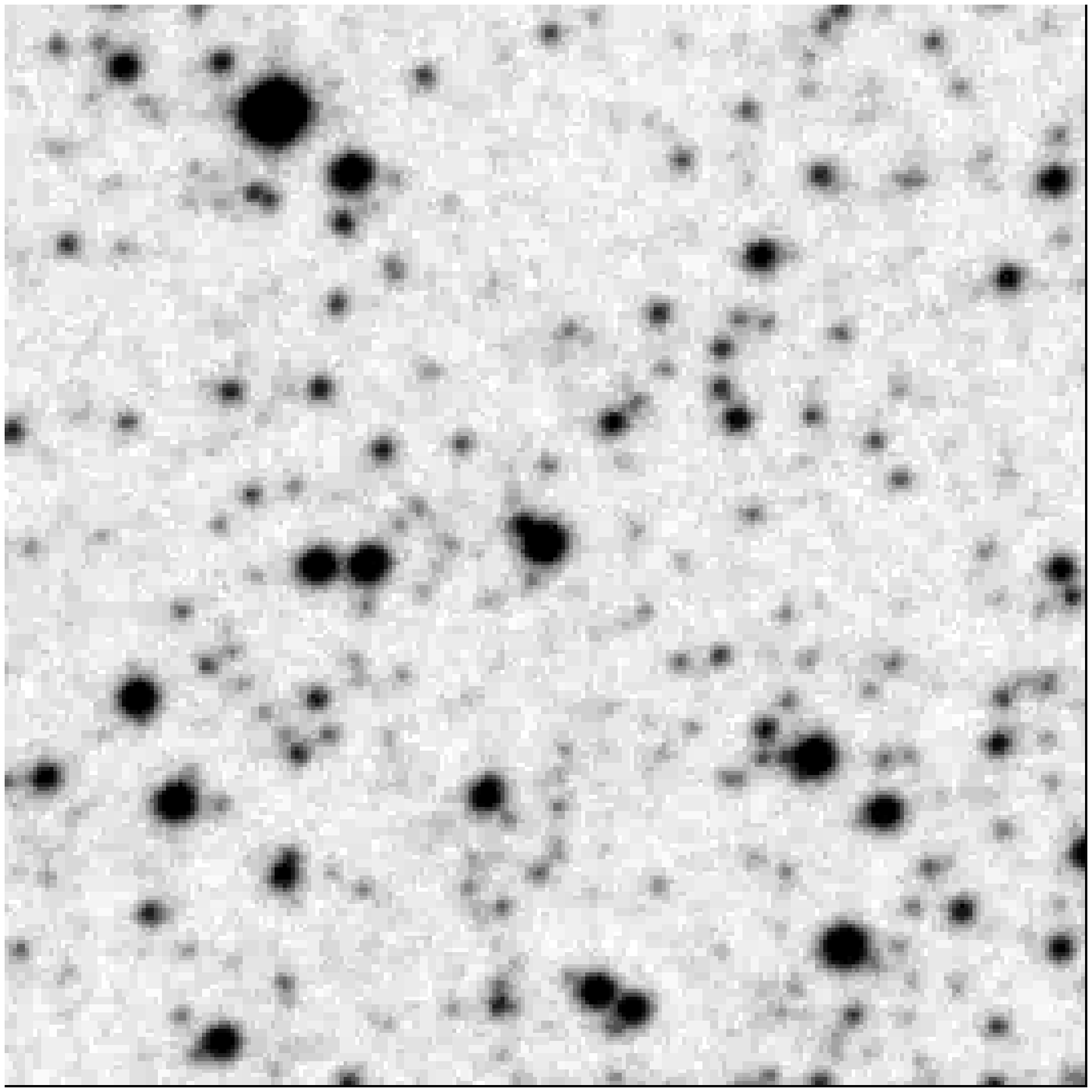}\label{lpv_r1}}
     \subfigure[An LPV star: \Rtwo~image]{\includegraphics[scale=0.35,angle=0]{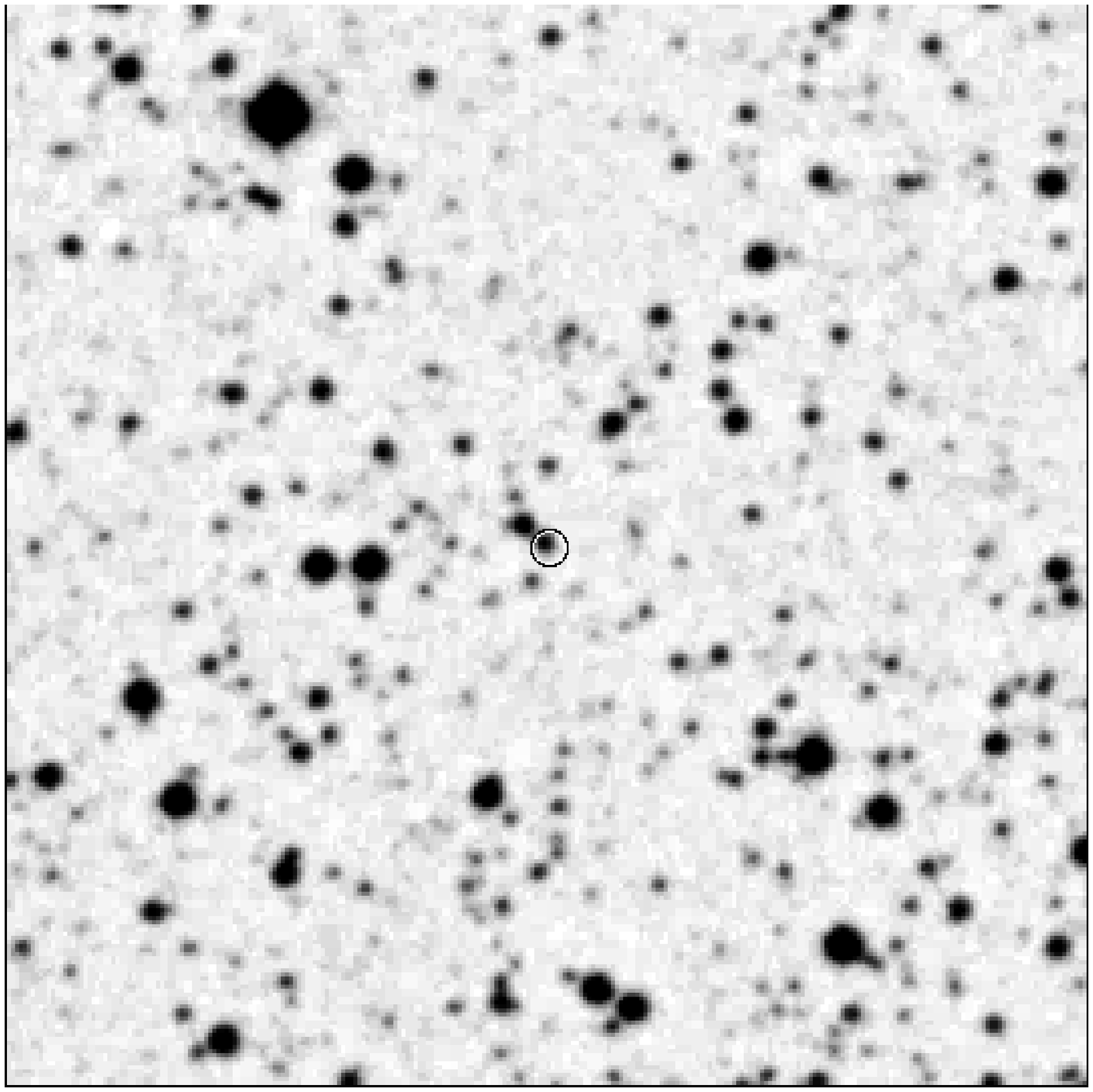}\label{lpv_r2}}\\
     \subfigure[Dark Molecular Cloud BHR74: \Bj~image]{\includegraphics[scale=0.35,angle=0]{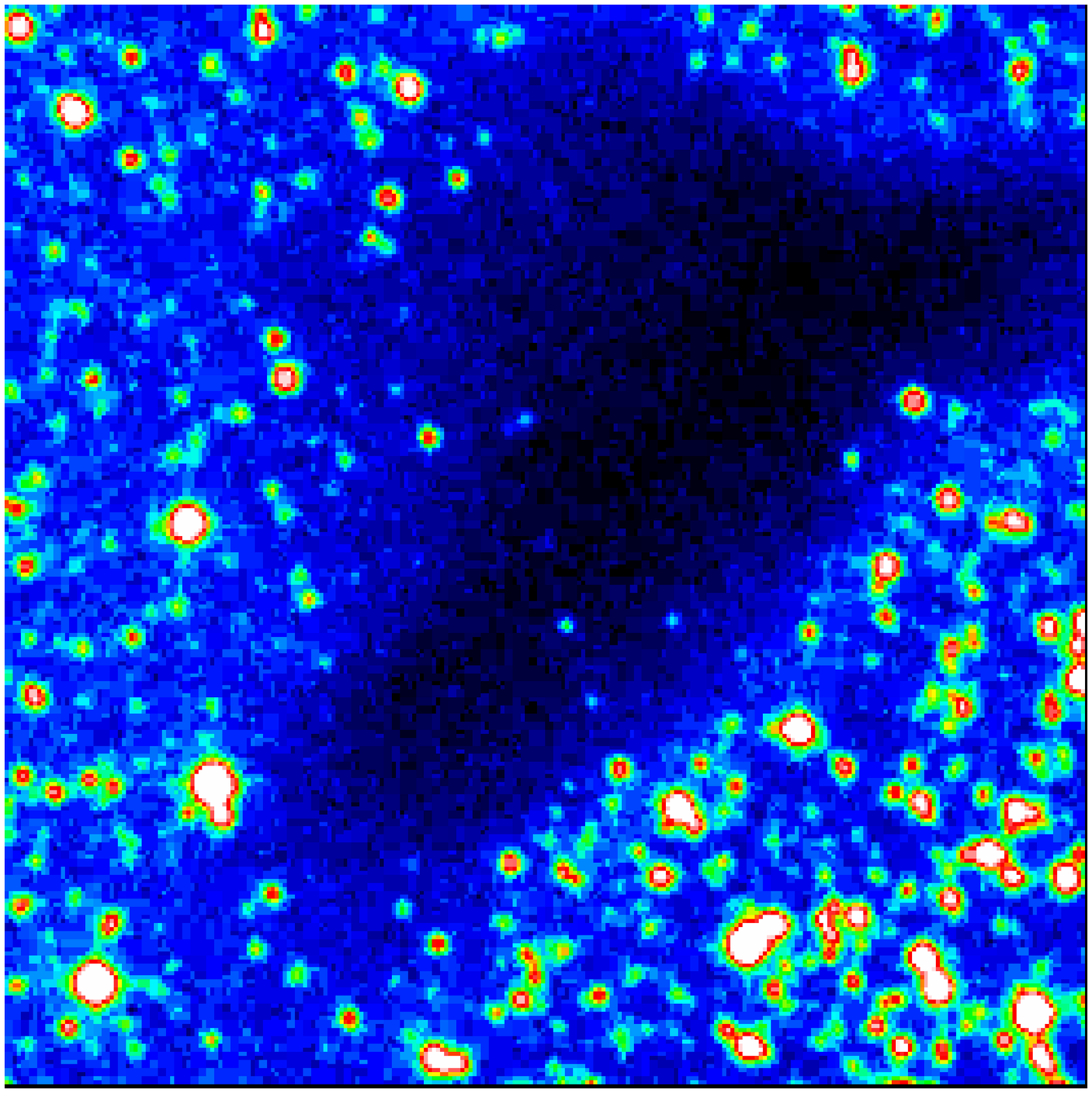}\label{darkneb_B}}
     \subfigure[Dark Molecular Cloud BHR74: \In~image]{\includegraphics[scale=0.35,angle=0]{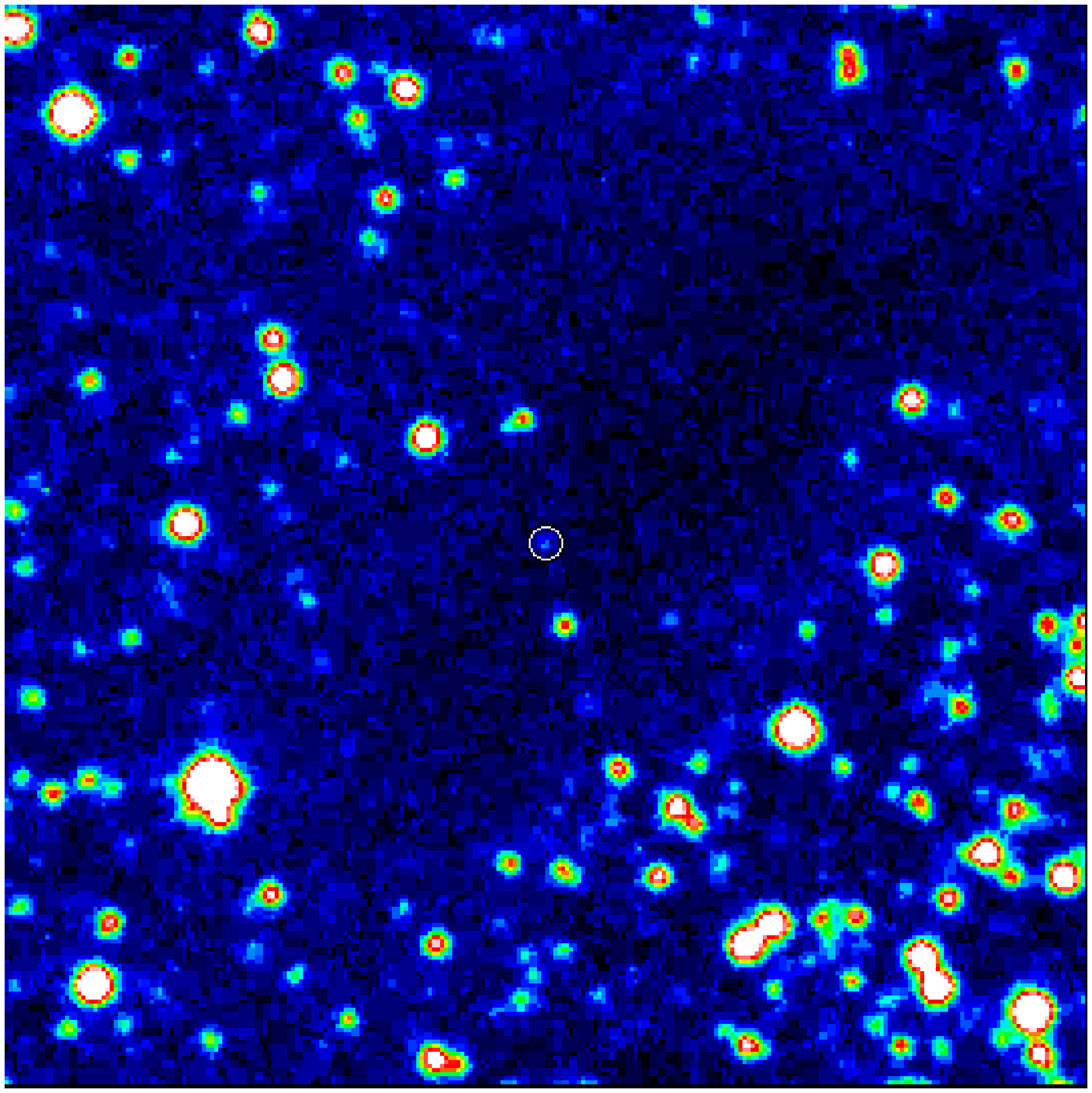}\label{darkneb_I}}
     \caption{Examples of some objects rejected during the visual checking stage in creating the Galactic plane catalogue. In panels \subref{lpv_r1} and \subref{lpv_r2} is an example of an LPV star which displays a large increase in brightness between the two {\em R}$_F$ plates (circled in panel \subref{lpv_r1}) that have an epoch difference of 5.75 years. This LPV object had no SSA optical magnitudes listed, and was rejected due to its variability. Panels \subref{darkneb_B} and \subref{darkneb_I} show an example of a small dark molecular cloud causing a distant luminous star to be highly reddened. This object only has an SSA \In~magnitude listed that passed the \IK~criterion. Panel \subref{darkneb_B} is a \Bj~image, while panel \subref{darkneb_I} is an \In~image showing that the rejected candidate (circled 2MASS position) has no proper-motion. This object was rejected due to its \rpm~value being below the criterion limit. All SSS images are $3\arcmin\times3\arcmin$~in size. North is up and East to the left.}
    \end{center}
   \end{minipage}
  \end{center}
  \label{exam_probs_fig1}
\end{figure*}
\begin{figure*}
  \begin{center}
   \begin{minipage}{1.0\textwidth}
    \begin{center}
     \subfigure[Bright stellar halo: \Rone~image]{\includegraphics[scale=0.35,angle=0]{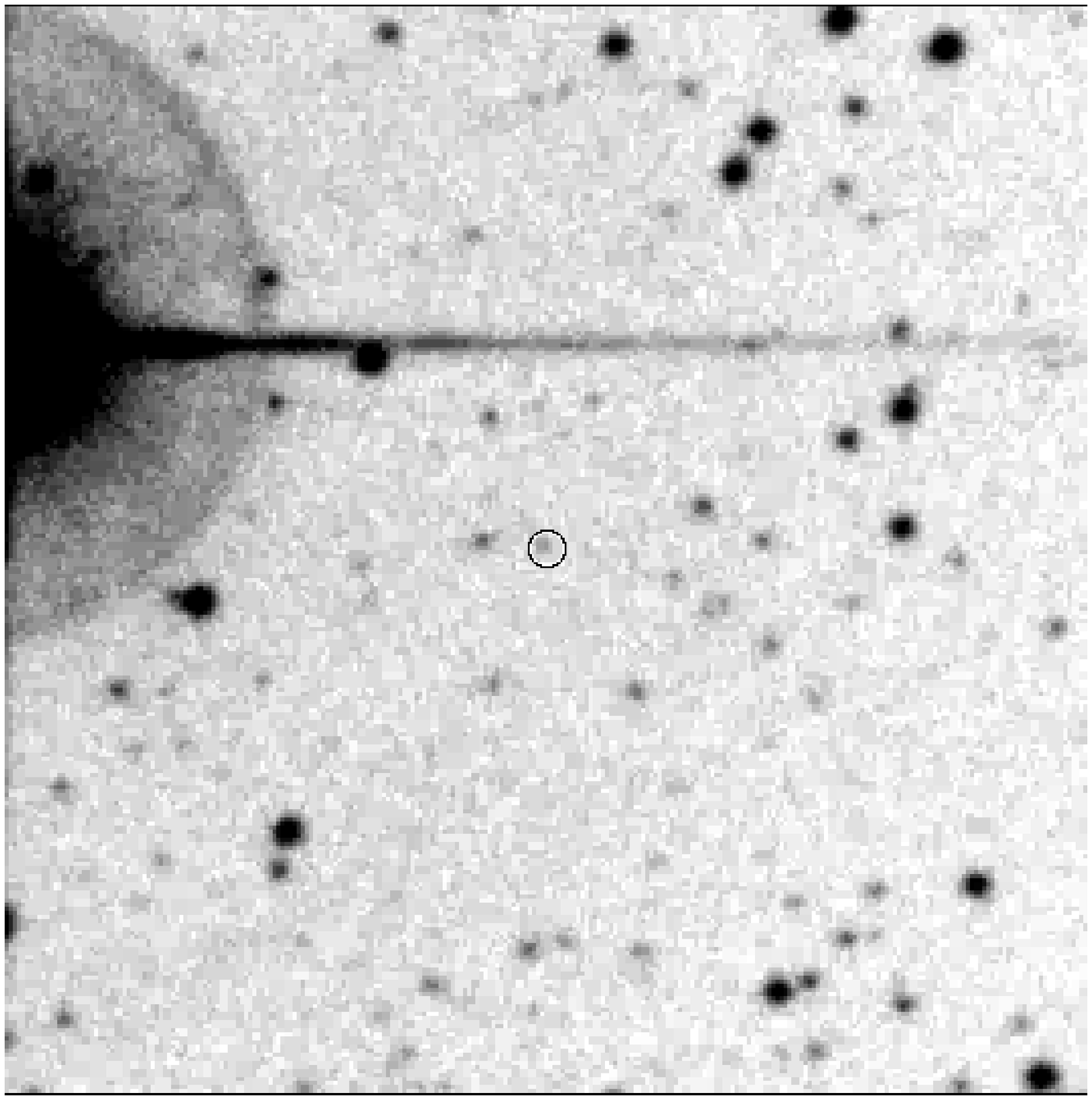}\label{ring_R1}}
     \subfigure[Bright stellar halo: \In~image]{\includegraphics[scale=0.35,angle=0]{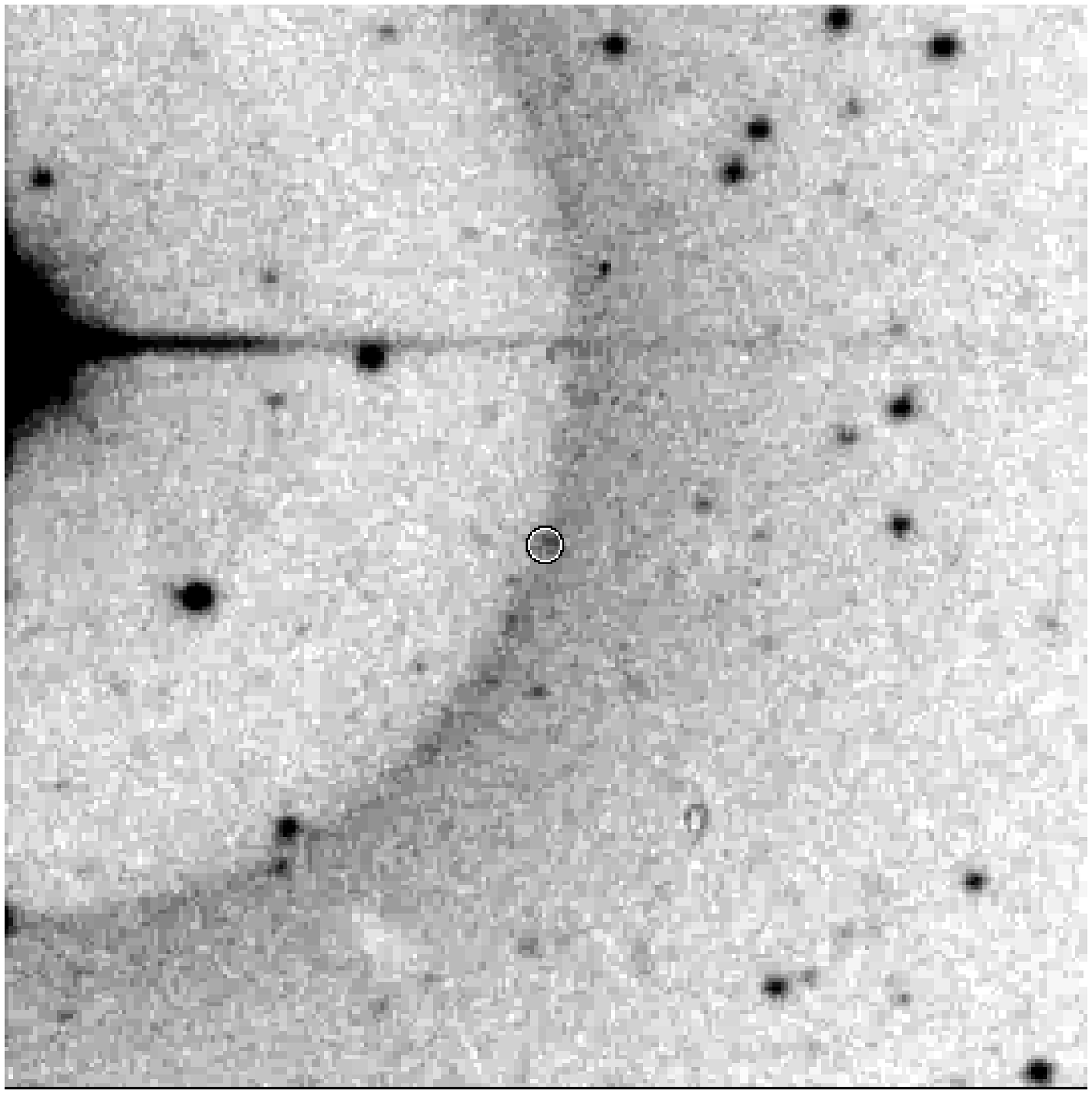}\label{ring_I}}\\
     \subfigure[Photometric blending: \Bj~image]{\includegraphics[scale=0.35,angle=0]{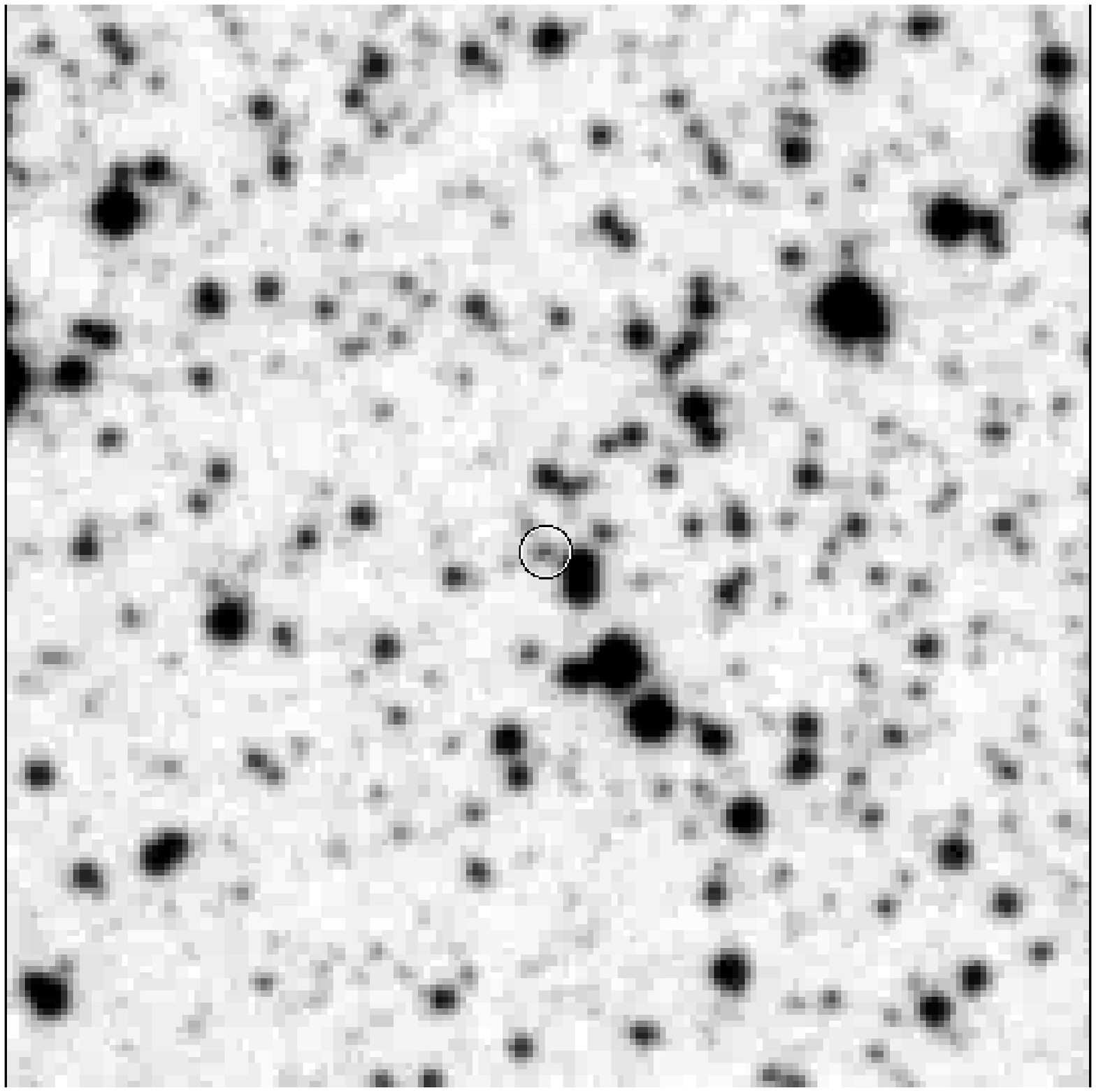}\label{blend_B}}
     \subfigure[Photometric blending: \In~image]{\includegraphics[scale=0.35,angle=0]{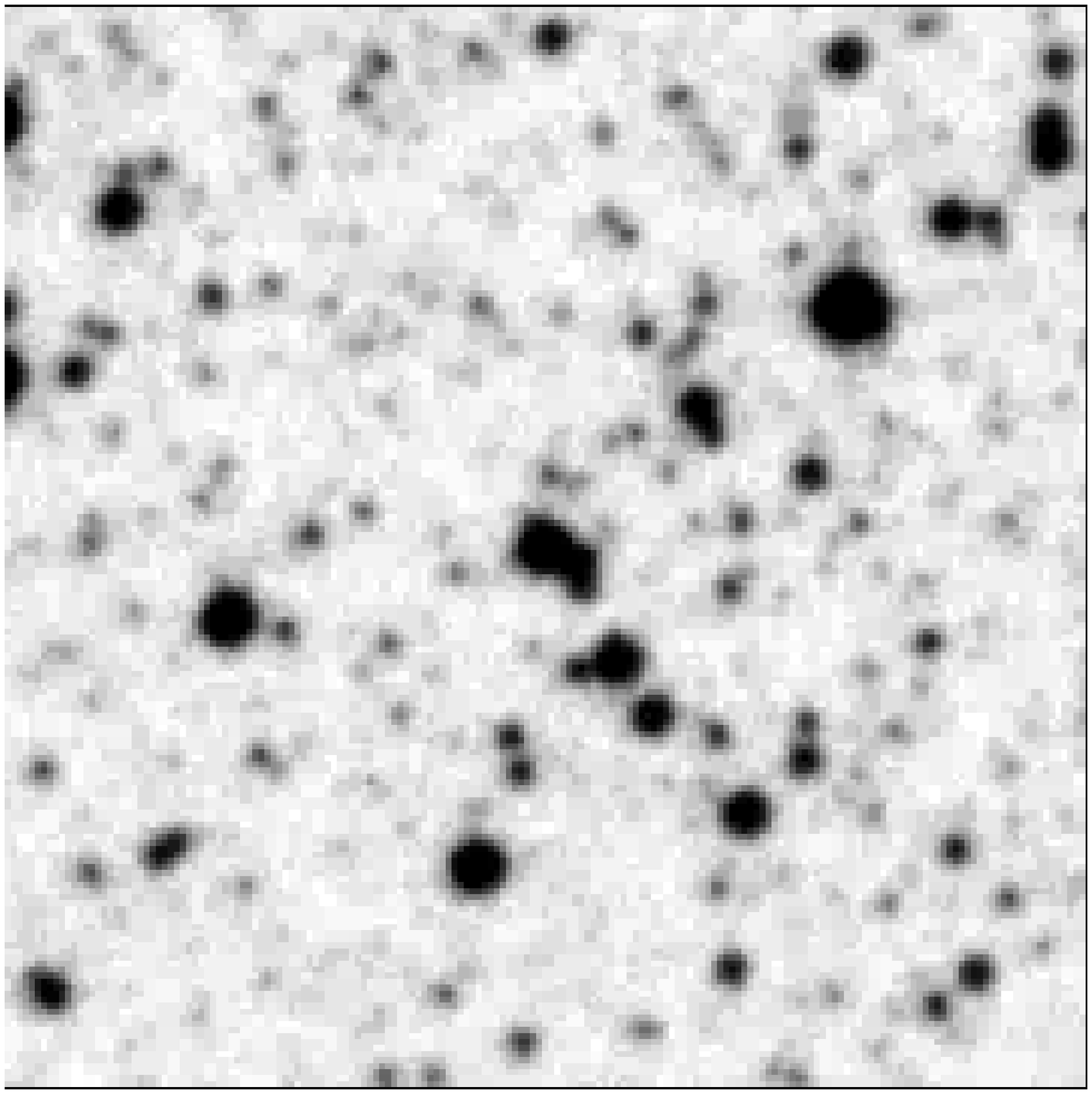}\label{blend_I}}
     \caption{This figure gives more examples of objects rejected during visual checking. The upper two panels \subref{ring_R1} and \subref{ring_I} show an object which lies in the halo of a bright star. This object has {\em R}$_F$ and \In~SSA magnitudes listed, but no \In~magnitude is used due to the error flag warning status. The object appears to show motion between the \Rone~and \In~images, but relative to the 2MASS position (identified by a circle in panel \subref{ring_I}) the motion is seen to be an artefact of the halo and is therefore rejected. The two lower panels \subref{blend_B} and \subref{blend_I} show a non-variable luminous star, but in the \In~image (panel \subref{blend_I}) and in the {\em R}$_F$ image (not shown), the brighter detection and proximity to its neighbours blends the photometry rendering it inaccurate. The \Bj~image in this example (panel \subref{blend_B}) was used to reject it as it shows no proper-motion. This object only had an SSA \Bj~magnitude listed which passed the \BK~criterion. the upper two images are $3\arcmin\times3\arcmin$~in size, while the lower two are $2\arcmin\times2\arcmin$. North is up and East to the left.}
    \end{center}
   \end{minipage}
  \end{center}
  \label{exam_probs_fig2}
\end{figure*}

Figures \ref{lpv_r1}-\ref{darkneb_I} and \ref{ring_R1}-\ref{blend_I} show examples of typical problem objects encountered during the visual checking stage. The problem of photometric blending is highlighted in Fig.\,\ref{lpv_r1}-\ref{lpv_r2} and Fig.\,\ref{blend_B}-\ref{blend_I} which show how the proximity of these objects to their close neighbours blends their photometry, rendering their magnitudes inaccurate or absent in the SSA. In the former two Figures (\ref{lpv_r1} and \ref{lpv_r2}) the object in question appears to be a LPV due to its large brightness difference between the SSS {\em R}$_F$-band images. This object had no SSA optical magnitudes listed, and was rejected due to its variability after inspection. For the latter case in Figures \ref{blend_B} and \ref{blend_I} this candidate appeared to be a non-variable giant which only had an SSA \Bj~magnitude listed, again probably due to photometric blending in both the {\em R}$_F$ and \In~detections. The \Bj~image in this example (panel \ref{blend_B}) was used to reject this candidate, as its measured proper-motion gave an reduced proper-motion value outside the \rpm~criterion.

Figures \ref{darkneb_B} and \ref{darkneb_I} show one of the better examples of a number of small dark molecular clouds that were encountered, causing distant luminous stars to be reddened into the near-IR colour selection planes. The circled object lies in the most dense part of this cloud and has only an SSA \In~magnitude listed which passed the \IK~criterion. This object was also rejected due to its \rpm~value falling below the criterion, using the approach of the `second solution' mentioned above. The last example pertains to a candidate which lies in the halo of a bright star, which can be seen prominently in the \In~image of panel \ref{ring_I}. This object has {\em R}$_F$ and \In~SSA magnitudes listed, but no \In~magnitude was used due to the error flag warning status. The object appears to show motion between the \Rone~and \In~images relative to the 2MASS position identified by a circle in panel \ref{ring_R1} and \ref{ring_I}. However, the motion is not consistent with the SSS and 2MASS epochs and is therefore an artefact of the halo -- this object was rejected.


\section{The Southern Galactic Plane Ultra-Cool Dwarf Candidate Catalogue (SGPUCD)} \label{ucd_cat}

Our candidate UCD catalogue contains \catnum~objects and is presented in Table\,\ref{cat_table}\footnote{please refer to on-line version for the full catalogue table}, which provides the primary astrometric, photometric, and proper-motion data for each catalogue member, as well as spectral type and distance estimates derived for both the spectroscopic and photometric samples. We indicate the distinction between the photometric and spectroscopic members in column 17 by the use of the following keys: `P' = Photometric analysis, `S' = Spectroscopic analysis, and also indicate if a catalogue member has been previously discovered or independently identified using; `K' = Known object. For some of these known objects we use the spectral type, distance, or proper-motion data provided in the literature, as indicated in the table footnotes (see Table\,\ref{known_table} and \textsection\,\ref{cat:known} for details of these objects). 

For each object listed in Table\,\ref{cat_table} we present near-IR 2MASS photometry and optical/near-IR colours using optical data from the SSA, where the {\em R}$_{\rm F}$- and \In-band photometry is transformed onto the standard Cousins system if both are available \citep{bessell86}. Column 11 gives the values of reduced proper-motion (\rpm) used in the candidate selection for where proper motions were available from the SSA.

During the visual checking stage of the candidate selection, proper motions were determined by visual examination of the 2MASS images and SSS {\em R}$_{\rm F}$- and \In-band optical images, and also DENIS \Ib~or {{\em J}-band} images. We indicate here in Table\,\ref{cat_table} if proper motion was identified for the catalogue members in column 12 (Visual PM) denoted by `Yes', and if no optical SSA or DENIS counterpart could be identified we denote this by `\textemdash'.

\begin{sidewaystable*}\relsize{-2}
\begin{minipage}{1.0\textwidth}
\begin{center}
  \caption[]{: A truncated version of the full Southern Galactic Plane UCD catalogue (SGPUCD) containing \catnum~members (see \textsection\,\protect\ref{online_cat} for details of the full on-line table version): astrometric, photometric, and proper-motion data are presented, as well as spectral types and distances derived for both the spectroscopic and photometric samples.}
 \label{cat_table}
\begin{tabular}{@{\extracolsep{\fill}}l*{15}{c}}
  \toprule\addlinespace
  \multirow{3}{1.0cm}{SGPUCD No.[\#]$^a$} & \multirow{3}{2cm}{2MASS\,J Designation$^b$} & \multicolumn{3}{c}{2MASS Photometry$^c$} & \multicolumn{2}{c}{SSA/2MASS Colours$^d$} & \multirow{3}*{\rpm$^e$} & \multirow{3}*{Visual$^f$ PM} & \multicolumn{3}{c}{Proper Motion (arcsec \peryr)$^e$} & \multirow{3}*{Sp-T} & \multirow{3}{1.0cm}{Member$^g$ Type} & \multirow{3}*{d\,(pc)} & \multirow{3}*{Refs}\\ \addlinespace\cmidrule(l{1.75em}r{1.75em}){3-5}\cmidrule(l{1.75em}r{1.75em}){6-7}\cmidrule(l{1.75em}r{1.75em}){10-12}
   & & $J$ & \JH & \JK & \IK & \RI & & & $\mu_{\rm tot}$ & $\mu_{\alpha\cos(\delta)}$ & $\mu_{\delta}$ & & & &\\
  \midrule
  1  &  06164933-1411434  &  15.146  &  0.671  &  1.109  &   4.255        &  \textemdash   &  \textemdash     &   Yes   &  \textemdash        &   \textemdash  &   \textemdash &    M7.0V$\pm$  0.5 & P      &    74.39$\pm$  4.59  & \\
  2  &  06300140-1840143  &  12.681  &  0.751  &  1.220  &  \textemdash   &  \textemdash   &  \textemdash     &   Yes   &  0.586              &    0.298       &   -0.505      &    M9.0V$\pm$  0.5 & S/K    & $17.50^{+1.2}_{-1.1}$&  (1)\\
  3  &  06362726-1226531  &  15.332  &  0.963  &  1.702  &  \textemdash   &  \textemdash   &  \textemdash     &   Yes   &  \textemdash        &   \textemdash  &   \textemdash &     L4.5$\pm$  4.0 & P      &    25.17$\pm$ 14.94  & \\
  4  &  06400355-1449104  &  14.143  &  0.666  &  1.095  &   4.187        &   2.275        &   14.70          &   Yes   &  0.054              &   -0.054       &    0.001      &    M7.5V$\pm$  0.5 & P      &    45.14$\pm$  2.23  & \\
  5  &  06431685-1843375  &  13.009  &  0.724  &  1.206  &   4.279        &   2.634        &   13.08          &   Yes   &  0.234              &    0.188       &   -0.139      &    M8.0V$\pm$  0.5 & S      & $23.50^{+2.2}_{-1.7}$& \\
  6  &  06450029-2333259  &  15.548  &  0.575  &  1.169  &   4.081        &  \textemdash   &  \textemdash     &   Yes   &  \textemdash        &   \textemdash  &   \textemdash &    M7.5V$\pm$  1.0 & P      &    85.56$\pm$ 14.27  & \\
  7  &  06465202-3244011  &  15.032  &  0.679  &  1.184  &   4.044        &   2.169        &   16.02          &   Yes   &  \textemdash        &   \textemdash  &   \textemdash &    M7.0V$\pm$  0.5 & P      &    72.79$\pm$  2.91  & \\
  8  &  06482289-2916280  &  14.315  &  0.627  &  1.082  &  \textemdash   &  \textemdash   &  \textemdash     &   Yes   &  \textemdash        &   \textemdash  &   \textemdash &    M7.0V$\pm$  0.5 & P      &    49.59$\pm$  2.93  & \\
  9  &  06495677-2104472  &  15.257  &  0.785  &  1.329  &   4.327        &  \textemdash   &  \textemdash     &   Yes   &  \textemdash        &   \textemdash  &   \textemdash &    M7.5V$\pm$  1.5 & P      &    70.79$\pm$ 14.74  & \\
 10  &  06512977-1446150  &  13.811  &  0.732  &  1.154  &  \textemdash   &  \textemdash   &  \textemdash     &   Yes   &  \textemdash        &   \textemdash  &   \textemdash &    M7.5V$\pm$  0.5 & S      & $37.10^{+4.1}_{-3.1}$& \\
 11  &  06525508-1614191  &  14.504  &  0.723  &  1.408  &   5.057        &   2.222        &   13.44          &   Yes   &  0.084              &   -0.067       &    0.051      &    M8.0V$\pm$  0.5 & P      &    48.52$\pm$  5.21  & \\
 12  &  06533530-2129406  &  15.121  &  0.853  &  1.397  &  \textemdash   &  \textemdash   &  \textemdash     &   Yes   &  \textemdash        &   \textemdash  &   \textemdash &    M9.0V$\pm$  2.0 & P      &    53.34$\pm$ 17.41  & \\
 13  &  06573154-1940057  &  15.238  &  0.742  &  1.234  &   4.329        &  \textemdash   &  \textemdash     &   Yes   &  \textemdash        &   \textemdash  &   \textemdash &    M8.0V$\pm$  1.5 & P      &    65.29$\pm$ 16.54  & \\
 14  &  07102532-0633215  &  14.627  &  0.725  &  1.179  &   4.158        &   2.460        &   12.98          &   Yes   &  \textemdash        &   \textemdash  &   \textemdash &    M7.5V$\pm$  1.5 & P      &    53.68$\pm$ 10.60  & \\
 15  &  07134978-0656011  &  15.431  &  0.748  &  1.403  &   5.105        &  \textemdash   &  \textemdash     &   Yes   &  \textemdash        &   \textemdash  &   \textemdash &     L3.0$\pm$  3.5 & P      &    32.35$\pm$ 17.55  & \\
 16  &  07164790-0630369  &  13.899  &  0.831  &  1.334  &  \textemdash   &  \textemdash   &  \textemdash     &   Yes   &  0.152              &   -0.052       &    0.143      &     L0.0$\pm$  0.5 & S/K    &  $27.2^{+1.7}_{-1.6}$&  (1)\\
 17  &  07230144-1616209  &  14.187  &  0.719  &  1.181  &   4.923        &   2.571        &   16.72          &   Yes   &  \textemdash        &   \textemdash  &   \textemdash &     L0.0$\pm$  1.0 & P      &    29.81$\pm$  5.14  & \\
 18  &  07233749-2300321  &  15.667  &  0.709  &  1.376  &  \textemdash   &  \textemdash   &  \textemdash     &   Yes   &  \textemdash        &   \textemdash  &   \textemdash &     L1.0$\pm$  3.0 & P      &    52.49$\pm$ 22.80  & \\
 19  &  07234556-1236532  &  14.294  &  0.704  &  1.128  &   4.333        &   2.513        &   12.14          &   Yes   &  0.060              &    0.001       &   -0.060      &    M8.0V$\pm$  0.5 & P      &    44.96$\pm$  3.65  & \\
 20  &  07252641-1025122  &  15.179  &  0.661  &  1.123  &  \textemdash   &  \textemdash   &  \textemdash     &   Yes   &  0.110              &    0.042       &   -0.102      &    M8.0V$\pm$  1.5 & P      &    63.27$\pm$ 16.20  & \\
  \bottomrule\addlinespace
 \end{tabular}
\end{center}
\end{minipage}\relsize{+2}
 {\em $^a$} The entry number of the SGPUCD members.\\
 {\em $^b$} The 2MASS designation as obtained from the Point Source Catalogue: 2MASS\,Jhhmmss[.]s$\pm$ddmmss.\\
 {\em $^c$} 2MASS photometry obtained from the Point Source Catalogue: j\_m, h\_m, and k\_m parameters.\\
 {\em $^d$} SSA optical colours: {\em R}$_{\rm F}$- and \In-band photometry is transformed onto the standard Cousins system where both are available \citep{bessell86}, otherwise SSA {\em R}$_{\rm F}$ and \In~magnitudes are used.\\
 {\em $^e$} The reduced proper-motion (\rpm) calculated using the 2MASS \Ks~magnitude, and the proper-motion data from the SSA in units of arcseconds \peryr.\\
 {\em $^f$} Possible proper motions determined by visual examination of the 2MASS images, SSS {\em R}$_{\rm F}$- and \In-band optical images, and/or DENIS \Ib~or {{\em J}-band} images.\\
 {\em $^g$} The mode by which catalogue membership was based, and how the subsequent spectral types and distances were calculated: `P' = Photometric analysis,`S' = Spectroscopic analysis. Also, we indicate if a catalogue member was previously discovered using; `K' = Known object.\\ 
 {\em $^{\dagger}$} Indicates that the proper-motion data are taken from \citet{phan_bao_gp07}.\\
 {\em $^{\ddagger}$} Indicates that the proper-motion data, spectral types, and distances have been taken from the these authors where available.\\
 References: (1) \citet{phan_bao_gp07}; (2) \citet{SSSPM0829_02}; (3) \citet{SSSPM0829_05}; (4) \citet{chamI_marti04}; (5) \citet{chamI_luhman05}; (6) \citet{chamII_Vuong01}; (7) \citet{2M1520m44_Ken06}; (8) \citet{2M1520m44_Burgy07}; (9) \citet{lupus3_obj03}; (10) \citet{chamI_luhman05}; (11) \citet{Luhman07_my_obj}; (12) \citet{my_paper}: (13) \citet{Xray_m9_t10_03}: (14) \citet{Kirkp2010_2m_pm}.\\
\end{sidewaystable*}

\subsection{Known Objects Identified in this Survey} \label{cat:known}
The UCD catalogue members identified as previously known objects in SIMBAD or {\sc http://dwarfarchives.org}, or that have been independently identified by \citet{phan_bao_gp07}, are listed in Table\,\ref{known_table} by their 2MASS name and other known identifiers. We primarily list the relevant data for each object sourced from those references given in the last column of Table\,\ref{known_table}, but also from this work where indicated.  

Objects identified by \citet{phan_bao_gp07} have been highlighted as shaded rows in Table\,\ref{known_table}, with the data presented for these objects originating entirely from those authors. Note that 2MASS\,J1126-5003 was first discovered in \citet{my_paper} from this work, but is included in Table\,\ref{known_table} as previously published object.

\begin{sidewaystable*}\relsize{-1}
\begin{minipage}{1.0\textwidth}
\begin{center}
 \caption[]{: UCDs identified in this Galactic plane catalogue that have been independently or previously discovered. Data for each object taken from the referenced authors, unless where indicated otherwise.}
 \label{known_table}
 \begin{tabular}{@{\extracolsep{\fill}}lcccccccccc}
  \toprule
2MASSJ & \multicolumn{4}{c}{} &  & \multicolumn{3}{c}{arcseconds\,\peryr} & \multicolumn{2}{c}{}\\\cmidrule(l{.75em}r{.75em}){7-9}
Designation$^a$ & Other name & $\mbox{\em I}-\mbox{\em J}$ & $M_{\rm J}$ & SpT & d (pc) & $\mu_{\rm tot}$ & $\mu_{\alpha\cos(\delta)}$ & $\mu_{\delta}$ & Notes & Refs.\\
\midrule
06300140-1840143 & DENIS-P\,0630-1840 & 3.17 & 11.28 & M8.5V$\pm$1.0 & 19.3$\pm$3.0 & 0.613 & 0.350  & -0.503 &\textemdash & 1\\
07164790-0630369 & DENIS-P\,0716-0630 & 3.55 & 12.19 & L1$\pm$1.0    & 22.0$\pm$3.7 & 0.122 & -0.016 & 0.121  &\textemdash & 1\\
07511645-2530432 & DENIS-P\,0751-2530 & 3.31 & 12.38 & L1.5$\pm$1.0  & 14.8$\pm$2.9 & 0.896 & -0.885 & 0.142 &\textemdash & 1\\
08123170-2444423 & DENIS-P\,0812-2444 & 3.38 & 12.38 & L1.5$\pm$1.0  & 20.1$\pm$4.3 & 0.191 & 0.096 & -0.165 &\textemdash & 1\\
08230313-4912012 & DENIS-P\,0823-4912 & 3.56 & 12.38 & L1.5$\pm$1.0  & 17.4$\pm$3.6 & 0.138 & -0.137 & 0.017 &\textemdash & 1\\
\multirow{2}{*}{08283419-1309198} & DENIS-P\,0828-1309 & \multirow{2}{*}{3.37$^b$} & \multirow{2}{*}{12.19$^b$} & \multirow{2}{*}{L2$^{c}\pm0.5$} & \multirow{2}{*}{11.6$^{c}\pm$1.4} & \multirow{2}{*}{0.593$^c$} & \multirow{2}{*}{-0.593$^c$} & \multirow{2}{*}{0.014$^c$} &\multirow{2}{*}{\textemdash} & \multirow{2}{*}{1,2,3}\\
 & SSSPM\,J0828-1309 & & & & & & & & & \\
10482788-5254180 & DENIS-P\,1048-5254 & 3.26 & 12.38 & L1.5$\pm$1 & 21.0$\pm$4.6 & 0.182 & -0.179 & 0.033 &\textemdash & 1\\
11085176-7632502 & \textemdash & $3.98^{f}$ &\textemdash & M7.25/M8 &\textemdash &\textemdash &\textemdash &\textemdash & Chamaeleon I Brown Dwarf & 4,5\\
11085497-7632410 &\textemdash & $3.59^{f}$ &\textemdash & M5.5 &\textemdash &\textemdash &\textemdash &\textemdash & Chamaeleon I Brown Dwarf & 4\\
11104006-7630547 &\textemdash & $3.62^{f}$ & 8.5$\pm0.2^{f,g}$ & M7.25$^h$(M$8.0^{f}$) & 160$\pm15^{i}$ &\textemdash &\textemdash &\textemdash & Chamaeleon I Brown Dwarf & 10,11\\ 
11263991-5003550 & DENIS-P\,1126-5003 & $3.85^b$ & 14.44$^{f}$ & L9.0$\pm1.0^{f}$ & $8.2^{+2.1}_{-1.5}$$^{f}$ & 1.650$^{f}$ & -1.580$^{f}$ & 0.450$^{f}$ & Blue L dwarf & 1,12\\
\multirow{2}{*}{11592743-5247188} & DENIS-P\,1159-5247 & \multirow{2}{*}{3.12} & \multirow{2}{*}{11.47} & \multirow{2}{*}{M9.0V$\pm$1} & \multirow{2}{*}{9.6$\pm$1.9} & \multirow{2}{*}{1.085} & \multirow{2}{*}{-1.077} & \multirow{2}{*}{-0.131} & \multirow{2}{*}{X-Ray flaring + strong H$\alpha$ emission} & \multirow{2}{*}{13}\\
& 1RXS\,J115928.5-524717 & & & & & & & & &\\
12531092-5709248 & DENIS-P\,1253-5709 & 3.29 & 12.01 & L0.5$\pm$1.0 & 19.4$\pm$4.0 & 1.634 & -1.575 & -0.435 &\textemdash & 1\\
\multirow{2}{*}{13030905-7755596} & DENIS-P 1303-7756 & \multirow{2}{*}{ $3.06^{f}$} & \multirow{2}{*}{\textemdash} & \multirow{2}{*}{\textemdash} & \multirow{2}{*}{\textemdash} & \multirow{2}{*}{\textemdash} & \multirow{2}{*}{\textemdash} & \multirow{2}{*}{\textemdash} & \multirow{2}{*}{Chamaeleon II PMS object} & \multirow{2}{*}{6}\\
 & C51 & & & & & & & & &\\
15200224-4422419$^{d}$ & DENIS-P\,1520-4422 & $3.45^b$ & $12.19^b$ & L$1.5^{e}$/L$4.5^{e}$ & $19^{e}\pm2$ & $0.733^{e}$ & $-0.634^{e}$ & $-0.367^{e}$ & Binary separation $1\farcs174\pm0\farcs016$ & 1,7,8\\
\multirow{2}{*}{16081603-3903042} & Par-Lup3-1/cc1 & \multirow{2}{*}{ $3.67^{f}$} & \multirow{2}{*}{\textemdash} & \multirow{2}{*}{M5V} & \multirow{2}{*}{\textemdash} & \multirow{2}{*}{\textemdash} & \multirow{2}{*}{\textemdash} &\multirow{2}{*}{\textemdash}& \multirow{2}{*}{M5V/M7.5V binary: Lupus 3 YSO} & \multirow{2}{*}{9}\\
 & CFB2003 & & & & & & & & &\\
17054744-5441513 & DENIS-P\,1705-5441 & 3.20 & 11.28 & M8.5V$\pm$1.0 & 27.2$\pm$5.7 & 0.078 & -0.072 & 0.030 & \textemdash & 1\\
17343053-1151388 & \textemdash & 3.940 & 11.5$_{-0.2}^{+0.1}$ & M9.0V$\pm$0.5 & 21.30$^{+1.5}_{-1.3}$ & 0.427 & 0.133 & -0.405 & 2MASS Near-IR proper motion survey & 14\\
17453466-1640538 & DENIS-P\,1745-1640 & 3.37 & 12.38 & L1.5$\pm$1.0  & 18.7$\pm$4.1 & 0.161 & 0.116 & -0.111 & \textemdash & 1\\ 
17562963-4518224 & DENIS-P\,1756-4518 & 3.14 & 11.47 & M9.0V$\pm$1.0 & 14.8$\pm$3.0 & 0.194 & 0.064 & -0.183 &\textemdash & 1\\
17502484-0016151$^{d}$ & \textemdash & 4.10 & 14.7$\pm0.23$ & L5.5$\pm0.5$ & $8.0^{+0.9}_{-0.8}$ & 0.491& -0.440 & 0.218 & \textemdash & 7\\
19090821-1937479 & DENIS-P\,1909-1937 & 3.58 & 12.19 & L1$\pm$1 & 26.9$\pm$6.0 & 0.158 & -0.064 & -0.145 &\textemdash & 1\\
  \bottomrule\addlinespace
 \end{tabular}
\end{center}
\end{minipage}\relsize{+1}
 {{\em $^a$} The 2MASS designation as obtained from the Point Source Catalogue: 2MASS Jhhmmss[.]s$\pm$ddmmss.\\
{\em $^b$} These data are taken from \citet{phan_bao_gp07}.\\
{\em $^c$} These data are taken from \citet{SSSPM0829_05}.\\
{\em $^d$} Discovered by \citet{2M1520m44_Ken06}.\\
{\em $^e$} These data are taken from \citet{2M1520m44_Burgy07}.\\
{\em $^f$} $\mbox{\em I}-\mbox{\em J}$ or other data are from this work.\\
{\em $^g$} Magnitude is on the CIT system.\\
{\em $^h$} From \citet{Luhman07_my_obj}.\\
{\em $^i$} Distance to Cham-I taken from \citet{cham_I_dist}.\\
References: (1) \citet{phan_bao_gp07}; (2) \citet{SSSPM0829_02}; (3) \citet{SSSPM0829_05}; (4) \citet{chamI_marti04}; (5) \citet{chamI_luhman05}; (6) \citet{chamII_Vuong01}; (7) \citet{2M1520m44_Ken06}; (8) \citet{2M1520m44_Burgy07}; (9) \citet{lupus3_obj03}; (10) \citet{chamI_luhman05}; (11) \citet{Luhman07_my_obj}; (12) \citet{my_paper}: (13) \citet{Xray_m9_t10_03}: (14) \citet{Kirkp2010_2m_pm}.}
\end{sidewaystable*}

\subsection{Catalogue Sky Coverage and Selection Numbers} \label{cat:areas}

\begin{figure*}
 \begin{center}
\begin{minipage}{1\textwidth}
\begin{center}
  \includegraphics[width=0.70\textwidth,angle=90]{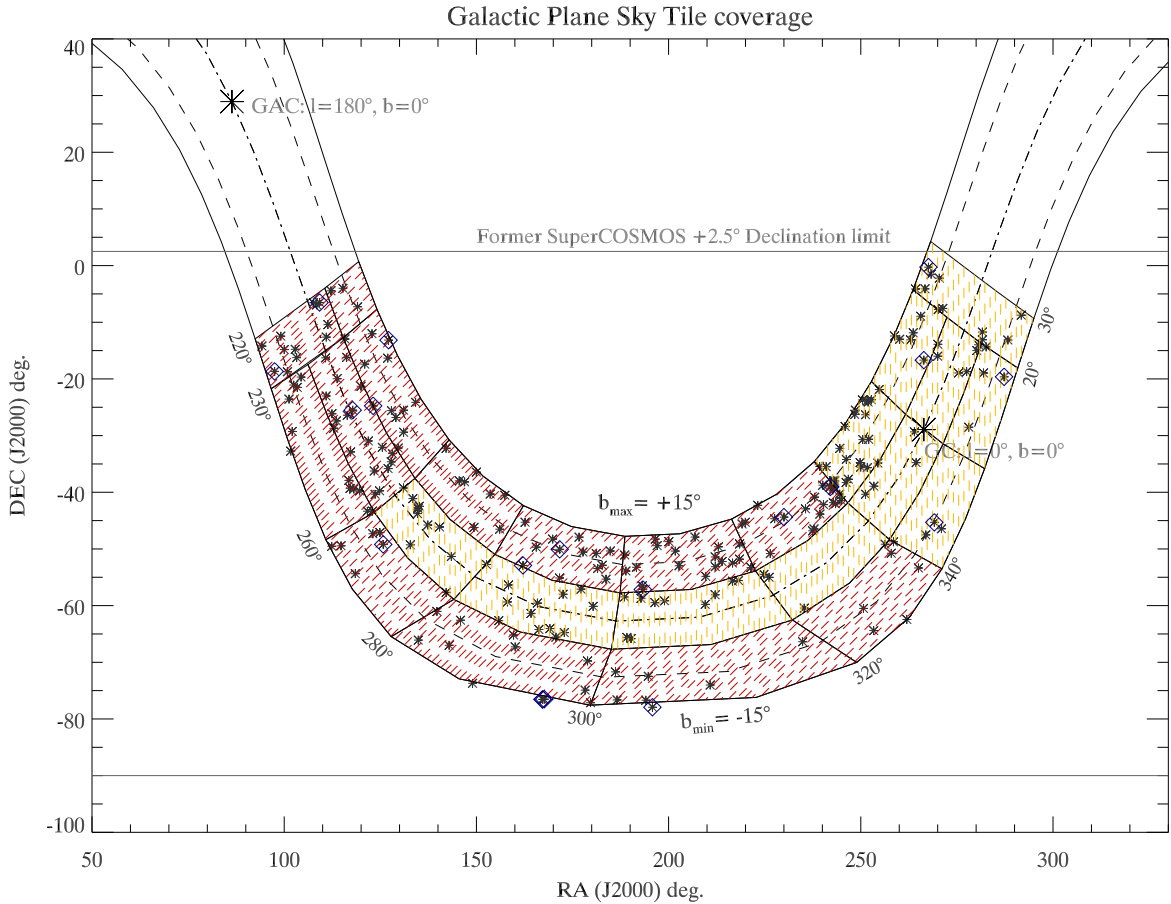}
  \caption{A map of the individual sky tile locations searched during the creation of UCD catalogue candidate showing the distribution of the catalogue members (See Table\,\protect\ref{area_table} for details). Each sky tile denoted by a red boxed section, with: orange crosshatching represents tiles using the overcrowding criteria; red crosshatched tiles were no overcrowding criteria was applied. The UCD catalogue members are plotted as black asterisks with the ones surrounded by blue squares indicating that they previously known or independently identified objects (see Table\,\protect\ref{known_table} for details). The dashed lines running parallel to the Galactic plane represent $\mbox{{\em b}}=\pm10^{\circ}$ lines in Galactic latitude. The positions of the Galactic Centre (GC), and the Galactic Anti-Centre (GAC) are marked, as well as the former SuperCOSMOS northen declination limit of $\delta < 2.5^{\circ}$ which applied to SSA data used in this work.}
  \label{gal_tile_map}
 \end{center}
\end{minipage}
 \end{center}
\end{figure*}

We applied our method to 24 discrete areas (sky tiles) at low Galactic latitudes and through the mid-plane, ranging in area from between 100\,deg$^2$ to 300\,deg$^2$, with most being 200\,deg$^2$. In Table\,\ref{area_table} we present the details of each sky tile along with a breakdown of the results for successive stages through the selection process. A map of the sky tile coverage over the range of Galactic longitude searched is shown in Fig.\,\ref{gal_tile_map} beginning at $\Gall=220^{\circ}$. The regions where the overcrowding criteria have been applied ($\Sigma_{\rm Cands}>1$) have been highlighted as shaded rows in Table\,\ref{area_table}, and are also represented by orange crosshatching in Fig.\,\ref{gal_tile_map}. 

Due to the selection criteria being different for these regions, we have broken down the contribution made to the total area searched in these tiles, into that from which candidates have been selected using the primary criteria (without the $\mbox{\Ks}\la12.5$ photometric `clipping'), and from that which invoked the photometric `clipping' criterion. The `clipped' area is split into three parts: 
\begin{enumerate}
\item  `clipped' selected area (column 8 in Table\,\ref{area_table}): The area from which candidates have been selected using the overcrowding photometric clipping criterion.
\item Non selected area (column 9 in Table\,\ref{area_table}): The area within the sky tile that contains no {\em class[321]} candidates, either from primary selection, or the photometric clipped selection ($\mbox{\Ks}\la12.5$).
\item Unused `clipped' area (column 10 in Table\,\ref{area_table}): The area within the sky tile that contains photometrically clipped ($\mbox{\Ks}\leqslant12.5$) candidates, that have not been included due to the candidate limit being reached for that tile (equal to the number of square degrees in total area).
\end{enumerate}
The sum of these three areas plus the primary selection area in column 6, is equal to the total available for a given tile (column 5). However, for those tiles where the overcrowding criteria was not applied ($\Sigma_{\rm Cands}<1$) the area searched is simply the total available within that tile (i.e., the primary selected area), i.e., an area not included in our analysis. 

For each tile in Table\,\ref{area_table} are listed five columns that contain the number of candidates after the following key stages in the selection process:
\begin{enumerate}
\item Column 11: Objects returned from the 2MASS query surviving the initial near-IR selection criteria.
\item Column 12: Candidates which passed all the optical photometric and reduced proper-motion selection criteria.
\item Column 13: The number of {\em class[321]} candidates selected after the classification scheme was applied ({\em class[0]} rejected) in the tile.
\item Column 14: For tiles where $\Sigma_{\rm Cands}>1$ this is the number of {\em class[321]} candidates remaining after the overcrowding criteria have been applied, with a maximum candidate number equal to the number of square degrees in that tile (i.e., the one per deg$^2$ criterion: see \textsection\,\ref{meth:overcrow}).
\item Column 15: The final number of candidates from the sky tile after the visual checking stage.
\end{enumerate}

The last column in Table\,\ref{area_table} lists values of the candidate surface density (based on the numbers in column 13 of Table\,\ref{area_table}), used in activating the overcrowding criteria when ${\Sigma_{\rm Cands}>1}$ (see \textsection\,\ref{meth:overcrow}). It is interesting to note how the value of $\Sigma_{\rm Cands}$ varies with Galactic $\ell$~and {\em b}, as it remains below a value of one as far east as $\ell=340^{\circ}$ for $|\mbox{{\em b}}|>5^{\circ}$. Tile number 16 has the greatest candidate surface density ($\Sigma_{\rm Cands}=68.18$) and lies some $30^{\circ}$ west of the Galactic centre. This region covers the large reddening hot spot seen in Fig.\,\ref{surf_dens_t7to16} known as the Norma dark molecular cloud. Consequently this region has a large fraction of candidates with only near-IR detections ($\sim24$ per cent). 

\begin{sidewaystable*}
\relsize{-1}
\begin{minipage}{\textwidth}
\begin{center}
 \caption[]{: Details of candidate numbers and area coverage for the individual sky tile regions. Note that the shaded rows indicate tiles that have the overcrowding criteria applied to them (i.e., $\Sigma_{\rm Cands}>1$).}
 \label{area_table}
 \begin{tabular}{l*{15}{c}}
  \toprule\addlinespace
  & \multicolumn{4}{c}{Deg. (Galactic)} & \multicolumn{5}{c}{Area (deg$^2$)} & \multicolumn{5}{c}{Candidate Numbers} & \\
 \cmidrule(l{.75em}r{.75em}){2-5}\cmidrule(l{.75em}r{.75em}){11-15}\cmidrule(l{.75em}r{.75em}){6-10}
 \multirow{2}{0.75cm}{Tile No.\#}& \multirow{2}{0.75cm}{$\Gall_{\rm min}$} & \multirow{2}{0.75cm}{$\Gall_{\rm max}$} & \multirow{2}{0.75cm}{{\em b}$_{\rm min}$} & \multirow{2}{0.75cm}{{\em b}$_{\rm max}$} &\multirow{2}{0.8cm}{Total$^a$ in tile} & \multirow{2}{1cm}{Primary$^b$ selected} & \multirow{2}{1cm}{`clipped'$^c$ selected} & \multirow{2}{1cm}{Non$^d$ selected} & \multirow{2}{1cm}{Unused$^e$ `clipped'} & \multirow{2}{1cm}{2MASS$^f$ sources} & \multirow{2}{1.25cm}{Optical$^g$ selected} & \multirow{2}{1.25cm}{{\em class[321]} sources$^h$} & \multirow{2}{1.25cm}{{\em class[321]} clipped$^i$} & \multirow{2}{1.25cm}{Final$^j$ number} & \multirow{2}{1cm}{${\Sigma_{\rm Cands}}^k$}\\\addlinespace\addlinespace
  \midrule
 1                      & 220 & 230 & $+5$  & $+15$ & 98.4  & 98.4  &\textemdash &\textemdash &\textemdash & 45,754  & 10     & 7      &\textemdash & 3  & 0.08\\
 2                      & 220 & 230 & $-15$ & $+5$  & 198.2 & 198.2 &\textemdash &\textemdash &\textemdash & 140,503 & 162    & 55     &\textemdash & 12 & 0.27\\
 3                      & 230 & 260 & $+5$  & $+15$ & 295.1 & 295.1 &\textemdash &\textemdash &\textemdash & 138,044 & 41     & 28     &\textemdash & 11 & 0.01\\
 4                      & 230 & 260 & $-5$  & $+5$  & 299.6 & 299.6 &\textemdash &\textemdash &\textemdash & 311,486 & 534    & 129    &\textemdash & 21 & 0.43\\
 5                      & 230 & 260 & $-15$ & $-5$  & 295.1 & 295.1 &\textemdash &\textemdash &\textemdash & 165,270 & 52     & 31     &\textemdash & 12 & 0.11\\
 6                      & 260 & 280 & $+5$  & $+15$ & 196.7 & 196.7 &\textemdash &\textemdash &\textemdash & 96,143  & 32     & 20     &\textemdash & 8  & 0.10\\
 \rowcolor[gray]{.75}7  & 260 & 280 & $-5$  & $+5$  & 199.7 & 78.9  & 6.0        & 99.8       & 15.0       & 375,162 & 6,726  & 1,788  & 200        & 9  & 8.96\\
 8                      & 260 & 280 & $-15$ & $-5$  & 196.7 & 196.7 &\textemdash &\textemdash &\textemdash & 129,077 & 50     & 22     &\textemdash & 7  & 0.11\\
 9                      & 280 & 300 & $+5$  & $+15$ & 196.7 & 196.7 &\textemdash &\textemdash &\textemdash & 114,718 & 46     & 33     &\textemdash & 14 & 0.17\\
 \rowcolor[gray]{.75}10 & 280 & 300 & $-5$  & $+5$  & 199.7 & 94.8  & 0.0        & 51.9       & 53.0       & 312,020 & 7,579  & 2,672  & 196        & 12 & 13.28\\
 11                     & 280 & 300 & $-15$ & $-5$  & 196.7 & 196.7 &\textemdash &\textemdash &\textemdash & 149,261 & 75     & 54     &\textemdash & 12 & 0.27\\
 12                     & 300 & 320 & $+5$  & $+15$ & 196.7 & 196.7 &\textemdash &\textemdash &\textemdash & 159,372 & 92     & 63     &\textemdash & 20 & 0.32\\
 \rowcolor[gray]{.75}13 & 300 & 320 & $-5$  & $+5$  & 199.7 & 73.8  & 5.0        & 28.9       & 92.0       & 386,707 & 20,791 & 8,124  & 199        & 10 & 40.67\\
 14                     & 300 & 320 & $-15$ & $-5$  & 196.7 & 196.7 &\textemdash &\textemdash &\textemdash & 157,900 & 131    & 98     &\textemdash & 7  & 0.50\\
 15                     & 320 & 340 & $+5$  & $+15$ & 196.7 & 196.7 &\textemdash &\textemdash &\textemdash & 180,523 & 202    & 112    &\textemdash & 18 & 0.57\\
 \rowcolor[gray]{.75}16 & 320 & 340 & $-5$  & $+5$  & 199.7 & 62.8  & 1.0        & 18.0       & 117.9      & 395,060 & 30,211 & 13,616 & 197        & 4  & 68.18\\
 17                     & 320 & 340 & $-15$ & $-5$  & 196.7 & 196.7 &\textemdash &\textemdash &\textemdash & 128,314 & 90     & 83     &\textemdash & 6  & 0.42\\
 \rowcolor[gray]{.75}18 & 340 & 360 & $+5$  & $+15$ & 196.7 & 87.8  & 0.0        & 91.1       & 17.9       & 222,342 & 1,023  & 376    & 194        & 25 & 1.95\\
 \rowcolor[gray]{.75}19 & 340 & 360 & $-5$  & $+5$  & 199.7 & 35.9  & 2.0        & 10.0       & 151.9      & 374,028 & 24,820 & 12,985 & 191        & 5  & 65.16\\
 \rowcolor[gray]{.75}20 & 340 & 360 & $-15$ & $-5$  & 196.7 & 80.0  & 2.0        & 103.8      & 10.9       & 109,697 & 561    & 312    & 195        & 4  & 1.59\\
 \rowcolor[gray]{.75}21 & 0   & 20  & $+5$  & $+15$ & 196.7 & 88.6  & 12.9       & 76.4       & 18.9       & 262,453 & 1,153  & 416    & 197        & 6  & 2.11\\
 \rowcolor[gray]{.75}22 & 0   & 20  & $-5$  & $+5$  & 199.7 & 32.9  & 1.0        & 4.0        & 161.8      & 335,012 & 15,231 & 8,635  & 193        & 5  & 43.23\\
\rowcolor[gray]{.75} 23 & 0   & 20  & $-15$ & $-5$  & 196.7 & 74.9  & 9.9        & 82.1       & 29.8       & 137,402 & 1,424  & 579    & 195        & 4  & 2.95\\
 \rowcolor[gray]{.75}24 & 20  & 30  & $-15$ & $+15$ & 296.6 & 102.8 & 3.0        & 115.9      & 74.9       & 483,616 & 14,781 & 6,264  & 289        & 11 & 21.13\\
  \midrule
 {\bf Totals:}&\textemdash &\textemdash &\textemdash &\textemdash & 5041.6 & 3573.2 & 42.8 & 681.9 & 744.0 & 5,309,864 & 125,817 & 56,502 & 2,246 & \catnum &\textemdash \\
  \bottomrule\addlinespace
 \end{tabular}
\end{center}
\end{minipage}\relsize{+1}
{({\em $^a$}) Total area on the sky of each tile region used for the primary near-IR 2MASS query (adjusted for spherical geometry).\\
({\em $^b$}) The area within the sky tile for which candidates have been selected using the primary near-IR and optical criteria, without any overcrowding photometric clipping criteria being applied (see \textsection\,\ref{meth:overcrow}).\\
({\em $^c$}) The area within the sky tile from which candidates have been selected using the additional overcrowding photometric clipping criteria, for where $\Sigma_{\rm Cands}>1$ (see \textsection\,\ref{meth:overcrow}).\\
 ({\em $^d$}) The area within the sky tile which no {\em class[321]} candidates were found from the primary selection when using the additional overcrowding and photometric clipping criteria.\\
({\em $^e$}) The area within the sky tile that was found to contain photometrically clipped ($\mbox{\Ks}\leqslant12.5$) candidates, but have not been included due to the candidate number limit already being reached for that tile (see \textsection\,\ref{meth:overcrow}).\\
({\em $^f$}) The number of near-IR selected sources returned from the 2MASS query that survived the near-IR selection criteria as well as the quality flag rejection and SNR filtering (see \textsection\,\ref{nir:sel}).\\
({\em $^g$}) The number of candidates which survived all the near-IR/optical and reduced proper-motion selection criteria, but before the candidate classification scheme and overcrowding criteria were applied.\\
({\em $^h$}) The total number of {\em class[321]} candidates in the tile before the overcrowding criteria were applied.\\
({\em $^i$}) The number of {\em class[321]} candidates remaining after the overcrowding and photometric clipping criteria $\mbox{\Ks}\leqslant12.5$) were applied, for the maximum number limit of that tile (i.e., typically one per deg$^2$: see \textsection\,\ref{meth:overcrow}).\\
({\em $^j$}) The final number of {\em class[321]} candidates from the sky tile after the visual checking stage.\\
({\em $^k$}) The surface density of all {\em class[321]} sources within the whole sky tile used in activating the overcrowding criteria when ${\Sigma_{\rm Cands}>1}$ (see \textsection\,\ref{meth:overcrow}).}
\end{sidewaystable*}

\section{Properties of the Catalogue members}

\subsection{Near-IR and Optical Spectroscopy}

We obtained near-IR spectroscopy for sixteen of our catalogue candidates, using both the SOFI spectrograph mounted on the ESO maintained 3.6-metre New Technology Telescope (NTT) situated at La Silla Chile, and the IRIS2 spectrograph mounted on the 3.9-metre Anglo Australian Telescope (AAT) at Siding Spring Australia. Thirteen spectra where obtained using SOFI on two separate observing runs\footnote{ESO programmes ID 076.C-0382 and ID 077C.0117.} on 2006 January 17--19, and on 2006 April 7--9. The spectrum of another candidate\footnote{ESO programme ID 279.C-5039A.} was obtained using SOFI from directors discretionary time on 2007 August 26. A further three spectra\footnote{service mode programmes IR086 and IR091.} were taken at the AAT on 2007 May 25--26.

The blue {\em JH} grism was used for the 2006 SOFI observations, with the August 2007 DDT spectrum also including the red {\em HK} grism. The wavelength coverage using just the {\em JH} grism is 0.95--1.65$\,\mu$m, while including the {\em HK} grism extended this continuous range to $\sim2.5\,\mu$m. A slit width of 0.6\asec was used for all observations giving a resolving power of $R\approx1000$ and a dispersion\footnote{All quoted dispersions are linearised wavelength values after reduction.} of 6.96\AA pixel$^{-1}$. For all these SOFI observations the seeing was good, and generally between 0.5--1.0 arcsec.

The spectra acquired under AAT service time were obtained using both the {\em J$_{\rm long}$} (SAPPHIRE\_240) and {\em H}$_{\rm s}$ (SAPPHIRE\_316) grisms with a slit width of 1.0 arcsec, giving a resolving power of $R\approx2400$. The wavelength coverage in the {\em J}-band is between 1.11--1.35$\,\mu$m with a dispersion of 2.362\AA pixel$^{-1}$, and for the {\em H}-band between 1.47--1.83$\,\mu$m with a dispersion of 3.517\AA pixel$^{-1}$. Standard calibration Xenon arc frames and spectroscopic flats were also taken, and the seeing was between 0.8--1.0\asec with photometric observing conditions.

Near-IR image reduction and spectral extraction were performed using standard packages within the {\sc iraf} environment after the AB and BA nodded pairs were subtracted, however, the telluric corrections and flux calibrations were made using the IDL based {\sc xtellcor\_general} routine provided by the NASA InfraRed Telescope Facility (IRTF), as part of the {\sc spextool} data reduction package \citep*[see][for details]{vacca}. Telluric standard stars of spectral types between B9V and A0V were observed before and after the target acquisitions, that are required for this procedure. Flux calibration was achieved also within this package by using the known Johnson-Morgan {\em B} and {\em V} magnitudes of the telluric standards which were obtained from SIMBAD, resulting in a F$(\lambda)$ flux scale for all our spectra.

We obtained a red optical spectrum for one other catalogue candidate using the IMACS/F2 spectrograph mounted on the 6.5-metre Baade/Magellan telescope situated at Las Campanas Chile, over the period 2009 September 30 to October 2. We used the following instrumental setup: the red grism (200 lines mm) with a 0.7 arcsecond wide longslit mask aligned to the parallactic angle, providing a wavelength coverage of 0.500--1$\,\mu$m with an average resolving power of R$\sim1000$, and a dispersion of about 2\,\AA pixel$^{-1}$. The OG590 longpass filter was used to eliminate second order light shortward of 0.5$\,\mu$m. The primary flux standards GD\,71 \citep{Hamuy92} and ESO68-8 were observed twice per night using the same instrumental setup. Flat-field quartz lamp exposures were used for pixel response calibration. The spectrum was extracted and calibrated using standard packages within the {\sc iraf} environment, with the dispersion solutions obtained from OH sky line spectra extracted using the same dispersion object trace; solutions were accurate to about 0.3--0.5\,\AA.

\subsection{Spectral Types and Distances}\label{Anal:spec_type}

All of the sixteen near-IR spectral we obtained show spectral features typical of late-type field dwarfs. To spectral type these objects we used a combination of spectral indices and template fitting of optically defined spectral standards. We used spectral indices defined in the {\em J}- and {\em H}-bands by \citet{reid,mcLean,burg_uni_class} for older field dwarfs (0.5--10\,Gyr). For the L spectral types we used the spectral indices of \citet{burg_uni_class} that are reliable from L0 to T8 as the primary indicator of spectral type, while the older \citet{reid,mcLean} indices were used for the M-type dwarfs which are also valid up to late-L. The spectral range of the H$_2$O-B index defined by \citet{mcLean} falls outside that of the AAT {\em H}-band coverage, so was therefore omitted for those objects affected.

A summary of the results from the spectral indices are given in Table\,\ref{spectro_table}, showing the actual index values along with their implied spectral types in rounded brackets; the average index derived spectral type; the best fitting template spectral type, and final adopted spectral types determined for each object. For two objects (the shaded rows in Table\,\ref{spectro_table}), the index values quoted in square brackets denote that these values are not reliable indicators of spectral type. For 2MASS\,J11263991-5003550 (hereafter 2M1126-5003) this is due to this object being a blue L dwarf, for which a detailed analysis of its spectral features was presented in \citet{my_paper}. For 2M1110-7630 -- a young low surface gravity Chamaeleon-I brown dwarf (see \textsection\,\ref{Anal:chamI_obj}) -- the spectral indices will only serve as a guide since they are defined for the older field population with higher surface gravities ($\log g\geqslant4.5$).

Finally, we consider an optical spectrum of 2MASS\,J18300760-1842361. This object was selected for spectroscopic follow-up due to its apparent brightness ($\mbox{\Ks}=8.8$ magnitude) and therefore its possible proximity if a genuine UCD; this was despite being flagged as variable in our dual \Rf~test. We note this object passed all the near-IR, \BK, \RK, and \IK~colour criteria, but would not have been selected by our method if it had not been a possible nearby interesting object. However, late-M dwarf template spectra failed to give a good match to this object, with strong TiO absorption bands seen blue-ward of $0.720\mu$m, and also at $\sim0.840\mu$m and $\sim0.890\mu$m. There is a noticeable absence of the K\,{\sc i} doublet at $\sim0.767\mu$m and $\sim0.769\mu$m, and absence of FeH and CrH between $\sim0.860\mu$m and $\sim0.880\mu$m. These combined observed spectral features are not normally evident in either the older field dwarf population, or in young late-M dwarfs \citep{kirkp_youg_obj}. A `by-eye' comparison to M-type giant template spectra gave a good fit for an M7\,III giant, which is shown in Fig.\,\ref{M7_giant}. Consequently, this object was removed from our catalogue and is not considered further in our analysis.

\begin{sidewaystable*}
\relsize{-1}
\begin{minipage}{1.0\textwidth}
\begin{center}
 \caption[]{: Observational and spectral typing details of the candidate ultra-cool dwarf spectroscopic sub-sample. Note that spectral index results enclosed by square brackets (the two shaded rows) are not reliable indicators of implied spectral type for the objects: in the case of 2M1126 this is due to the object being a blue L dwarf \protect\citep[see][for details]{my_paper,burgy2m1126}, and for 2M1110 this object is a young ($\sim$1--5\,Myr) brown dwarf with low surface gravity (see \textsection\,\protect\ref{Anal:chamI_obj}).}
 \label{spectro_table}
 \begin{tabular}{@{\extracolsep{\fill}}lllclllllccccc}
  \toprule
\multirow{3}{1.0cm}{2MASS Name$^a$} & \multirow{3}{0.75cm}{Obs.Date} & \multirow{3}{0.75cm}{Telescope} & \multirow{3}{0.5cm}{Band} & \multicolumn{5}{c}{Field dwarf indices} & \multirow{3}{0.75cm}{Index$^e$ SpT} & \multirow{3}{0.75cm}{Best$^f$ fitting} & \multirow{3}{0.5cm}{Final$^g$ SpT} & \multirow{3}{0.5cm}{\Mj$^h$} & \multirow{3}{0.5cm}{d(pc)$^j$}\\\cmidrule(l{1.em}r{1.em}){5-9}
\multicolumn{4}{c}{} & H$_2$OB$^b$ & H$_2$OA$^c$ & H$_2$OB$^c$ & H$_2$O-J$^d$ & H$_2$O-H$^d$ & \multicolumn{5}{c}{}\\
  \midrule
0630-1840 & 18 Jan.06 & NTT 3.6m & {\em JH} & 0.83(M9.3) & 0.73(M9.0) & 0.86(M9.9) & \textemdash & \textemdash & M9.4 & M9.0 & M9.0V$\pm$0.5 & 11.5$_{-0.2}^{+0.1}$ & 17.5$_{-1.1}^{+1.2}$\\[1ex]
0643-1843 & 18 Jan.06 & NTT 3.6m & {\em JH} & 0.90(M7.6) & 0.77(M7.9) & 0.95(M7.6) & \textemdash & \textemdash & M7.7 & M8.0 & M8.0V$\pm$0.5 & 11.2$_{-0.2}^{+0.2}$ & 23.5$_{-1.7}^{+2.2}$\\[1ex]
0651-1446 & 18 Jan.06 & NTT 3.6m & {\em JH} & 0.92(M7.0) & 0.79(M7.5) & 0.96(M7.4) & \textemdash & \textemdash & M7.3 & M8.0 & M7.5V$\pm$0.5 & 11.0$_{-0.2}^{+0.2}$ & 37.1$_{-3.1}^{+4.1}$\\[1ex]
0716-0630 & 18 Jan.06 & NTT 3.6m & {\em JH} & 0.81(L0.0) & 0.64(L1.4) & 0.81(L0.9) & \textemdash & \textemdash & L0.4 & L0.0 & L0.0$\pm$0.5 & 11.7$_{-0.1}^{+0.1}$ & 27.2$_{-1.6}^{+1.7}$\\[1ex]
0823-4912 & 26 May 07 & AAT 3.9m & {\em J}\&{\em H} & 0.68(L3.3) & 0.61(L2.2) & \textemdash & 0.77(L5.3) & 0.75(L4.8) & L3.9 & L3.0 & L4.0$\pm$1 & 13.1$_{-0.4}^{+0.5}$ & 12.4$_{-2.4}^{+2.6}$\\[1ex]
0827-1619 & 18 Jan.06 & NTT 3.6m & {\em JH} & 0.89(M7.8) & 0.82(M6.7) & 0.94(M7.9) & \textemdash & \textemdash & M7.5 & M8.0 & M8.0V$\pm$0.5 & 11.2$_{-0.2}^{+0.2}$ & 38.9$_{-2.8}^{+3.6}$\\[1ex]
0844-2924 & 18 Jan.06 & NTT 3.6m & {\em JH} & 0.71(L2.6) & 0.54(L4.0) & 0.67(L4.1) & 0.78(L4.8) & 0.74(L4.8) & L4.1 & L4 & L4.0$\pm$0.5 & 13.1$_{-0.2}^{+0.2}$ & 29.2$_{-2.9}^{+3.0}$\\[1ex]
0955-7342 & 26 May 07 & AAT 3.9m & {\em J}\&{\em H} & 0.83(M9.4) & 0.61(L2.0) & \textemdash & 0.86(L2.5) & \textemdash & L1.3 & L0.5& L0.0$\pm$1 & 11.7$_{-0.3}^{+0.3}$& 46.1$_{-5.4}^{+6.0}$\\[1ex]
1048-5254 & 08 Apr.06 & NTT 3.6m & {\em JH} & 0.75(L1.6) & 0.73(M9.0) & 0.79(L1.4) & 0.85(L2.9) & 0.82(L1.9) & L1.0    & L0-L1 & L0.5$\pm$0.5 & 11.9$_{-0.1}^{+0.1}$ & 27.0$_{-1.7}^{+1.7}$\\[1ex]
1050-4517 & 07 Apr.06 & NTT 3.6m & {\em JH} & 0.87(M8.4) & 0.75(M8.4) & 0.88(M9.2) & \textemdash & \textemdash & M8.7  & M9.0 & M9.0V$\pm$0.5 & 11.5$_{-0.2}^{+0.1}$ & 32.4$_{-1.9}^{+2.2}$\\[1ex]
\rowcolor[gray]{.75}1110-7630 & 25 May 07 & AAT 3.9m & {\em J}\&{\em H} & 0.90[M6.3] & 0.79[M7.6] & \textemdash & \textemdash & \textemdash & M7.0 & M8-M8.5 & M8.0$\pm$0.5 & $^{\dagger}7.9_{-0.2}^{+0.2}$ & $^{\ddagger}160_{-15}^{+15}$\\[1ex]
\rowcolor[gray]{.75}1126-5003 & 07 Apr.06 & NTT 3.6m & {\em JH} & 0.56[L6.6] & 0.356[L8.7] & 0.53[L7.3] & 0.72[<L8] & 0.65[L9.5] & L7-L9 & L7-T0 & L9$\pm$1 & $14.4_{+0.2}^{-0.2}$ & $8.2^{+2.1}_{-1.5}$\\[1ex]
1308-4925 & 08 Apr.06 & NTT 3.6m & {\em JH} & 0.90(M7.5) & 0.79(M7.4) & 0.91(M8.6) & \textemdash & \textemdash & M7.8 & M8-M9 & M8.5V$\pm$0.5 & 11.3$_{-0.2}^{+0.2}$ & 27.5$_{-1.8}^{+2.2}$\\[1ex]
1415-5212 & 09 Apr.06 & NTT 3.6m & {\em JH} & 0.83(M9.5) & 0.76(M8.3) & 0.79(L1.4) & \textemdash & \textemdash & M9.4 & M9.5  & M9.5V$\pm$0.5 & 11.6$_{-0.1}^{+0.1}$ & 41.3$_{-2.4}^{+2.6}$\\[1ex]
1426-5229 & 09 Apr.06 & NTT 3.6m & {\em JH} & 0.92(M7.0) & 0.85(M5.8) & 0.93(M8.2) & \textemdash & \textemdash & M7.0 & M8.0  & M8V$\pm$1.0 & 11.2$_{-0.4}^{+0.3}$ & 41.9$_{-5.6}^{+9.0}$\\[1ex]
1734-1151 & 26 Aug.07 & NTT 3.6m & {\em JHK} & 0.86(M8.5) & 0.71(M9.6) & 0.91(M8.8) & \textemdash & \textemdash & M9.0 & M9.0 & M9.0V$\pm$0.5 & 11.5$_{-0.2}^{+0.1}$ & 21.3$_{-1.3}^{+1.5}$\\[1ex]
1830-1842 & Sept--Oct 09 & Baade 6.5m & {\em RIZ} & \textemdash & \textemdash & \textemdash & \textemdash & \textemdash & \textemdash & M7.0 & M7.0\,III$\pm$1 & \textemdash & \textemdash\\[1ex]
  \bottomrule\addlinespace
 \end{tabular}
\end{center}
\end{minipage}\relsize{+1}
{({\em $^a$}) Abbreviated 2MASS designation obtained from the Point Source Catalogue: 2MASS Jhhmm$\pm$ddmm. ({\em $^b$}) This is the H$_2$O$^B$ index defined by \citet{reid}. ({\em $^c$}) These are the \citet{mcLean} H$_2$OA and H$_2$OB indices. ({\em $^d$}) These indices are defined by \citet{burg_uni_class}. ({\em $^e$}) This is the average of the separate spectral types derived from each given index. ({\em $^f$}) The optically classified spectral type of the best fitting template `standard' near-IR spectrum \citep[obtained from both][]{Cushing,mcLean} to that of the object spectrum. In all cases the template spectrum was first smoothed to the instrument resolution element of the object. ({\em $^g$}) Final spectral type assigned from both the spectral template fitting and the average of the H$_2$O spectral indices. ({\em $^h$}) Absolute {\em J}-band magnitude (2MASS) derived from the spectral type/\Mj~relation of \citet{Cruz03}, except for the two shaded rows. The uncertainty in \Mj~reflects only the uncertainty in the spectral type. ({\em $^j$}) Distance in parsecs derived from the \Mj~and 2MASS {\em J}-band magnitude, except for the two shaded rows. The uncertainty in the distance reflects only the uncertainty in \Mj. ($^{\dagger}$) \Mj~for this object is on the CIT photometric system, and derived from this work (see \textsection\,\ref{Anal:chamI_obj}). ($^{\ddagger}$) Distance to 2M1110 taken as the distance to the Chamaeleon-I dark cloud given by \citet{cham_I_dist}.}
\end{sidewaystable*}
\begin{figure}
 \begin{center}
  \includegraphics[width=0.48\textwidth,angle=0]{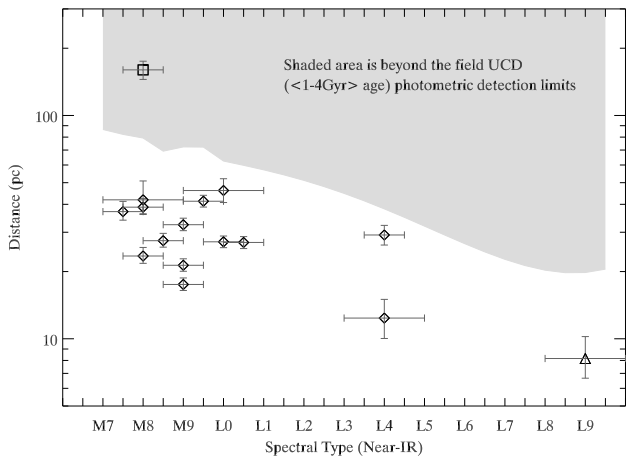}
  \caption{The distance vs. spectral-type distribution of the spectroscopic sub-sample using data given in Table\,\protect\ref{spectro_table}. Also shown in this plot are the approximate distance limits as a function of spectral type of this catalogue, derived from the 2MASS \Ks~photometric detection limits used in the method discussed in \textsection\,\protect\ref{nir:sel} and \textsection\,\protect\ref{meth:class}. See text in \textsection\,\protect\ref{Anal:spec_type} for details. Note the outlier in the shaded region with an apparent over-luminosity -- this is the young Chamaeleon-I M8 brown dwarf (see \textsection\,\protect\ref{Anal:chamI_obj}).}
  \label{spec_sample_plot}
 \end{center}
\end{figure}

\begin{sidewaysfigure*}
 \begin{center}
  \begin{minipage}{1.0\textwidth}
   \begin{center}
    \subfigure[][Spectra for six candidates obtained in 2006 January.]{\label{spec_jan06}\includegraphics[scale=0.6,angle=0]{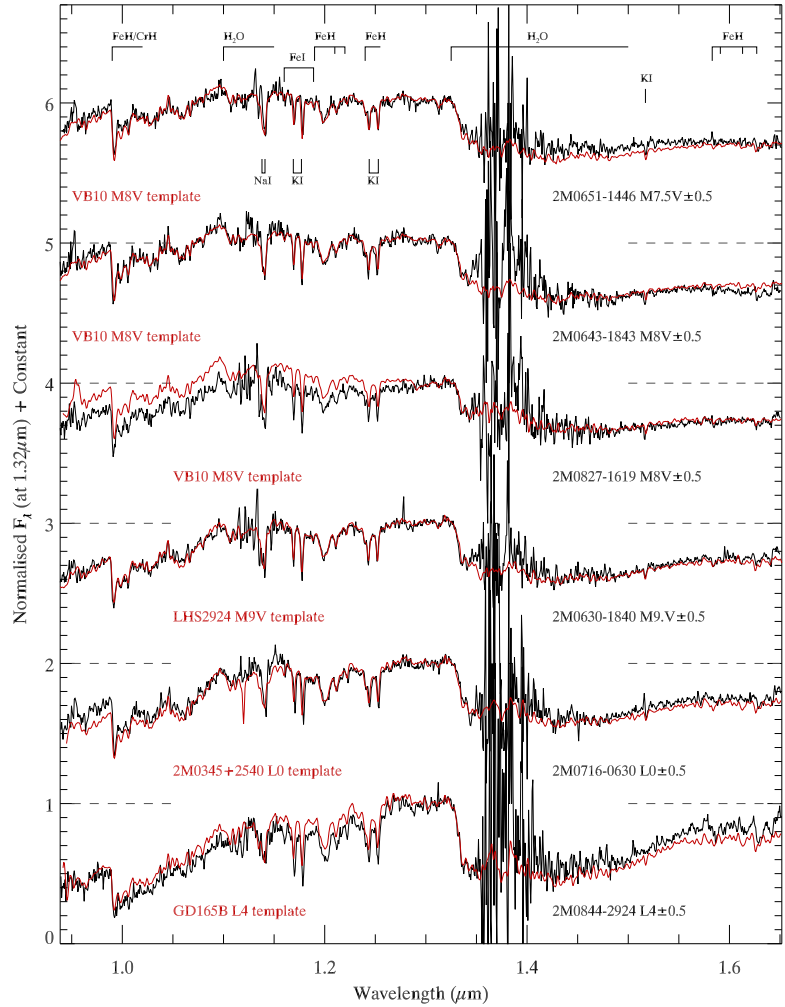}}\qquad
    \subfigure[][Spectra for six candidates obtained in 2006 April.]{\label{spec_apl06}\includegraphics[scale=0.6,angle=0]{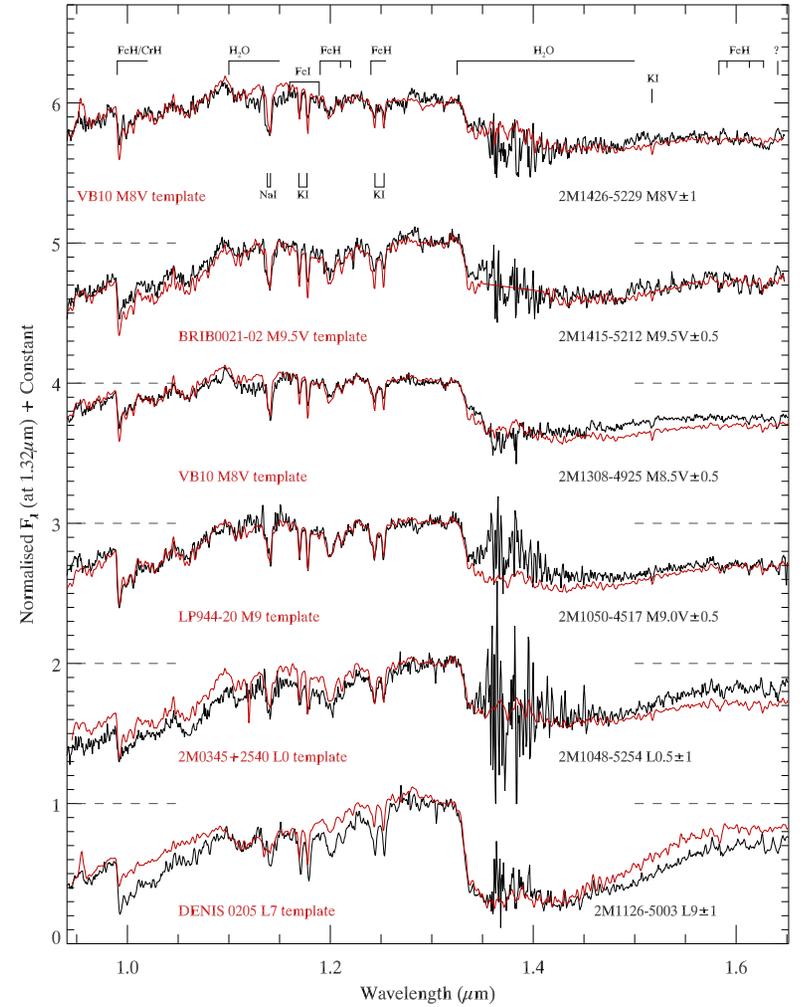}}
    \caption[]{: SOFI {\em JH}-band spectra for twelve candidates in descending order of spectral type. All spectra are corrected for telluric features, with the F$_\lambda$ flux scale normalised at $1.32\,\mu$m -- zero levels indicated by dashed lines. Template spectra of standard field dwarfs are shown overlain in red \protect\citep[templates obtained from][]{Cushing,mcLean}, which were found to give the best fit during the spectral typing of these candidates. In all cases the template spectra have been Gaussian smoothed to the same FWHM instrument resolution of the candidate spectra ($0.0018\,\mu$m). The main atomic and molecular absorption features are indicated at the top of the plot.}
   \label{slf1_8_spec_tmlt}
   \end{center}
  \end{minipage}
 \end{center}
\end{sidewaysfigure*}

\begin{figure*}
 \begin{center}
\begin{minipage}{1.0\textwidth}
\begin{center}
  \includegraphics[width=0.55\textwidth,angle=90]{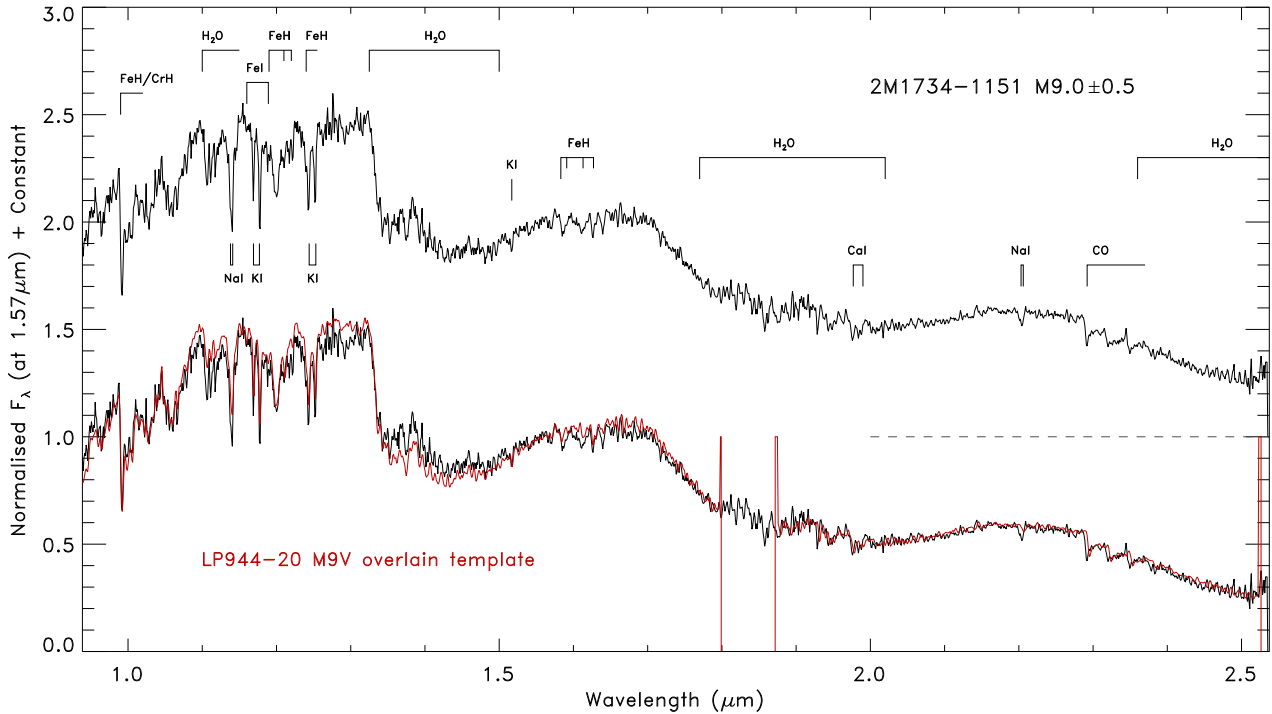}
  \caption{The SOFI {\em JHK}-band spectrum of 2M1734-1151 obtained in 2007 August, shown over-plotted with the M9V template spectrum in red that was found to give the best fit during spectral typing \protect\citep[obtained from the][IRTF spectral library]{Cushing}. The F$_\lambda$ flux scale has been normalised at $1.57\,\mu$m with the zero level indicated by a dashed line. The template spectrum has been Gaussian smoothed to the same FWHM instrument resolution of the candidate spectrum ($0.0025\,\mu$m). The main atomic and molecular absorption features across the {\em JHK} bands are indicated at the top of the plot.}
  \label{ddt_2m1734}
 \end{center}
\end{minipage}
 \end{center}
\end{figure*}

\begin{figure}
 \begin{center}
  \includegraphics[width=0.48\textwidth,angle=0]{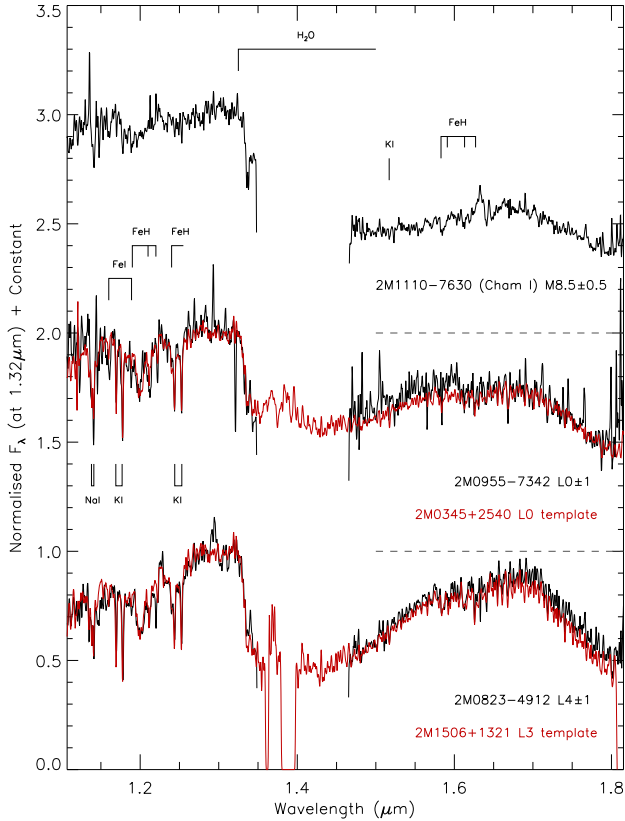}
  \caption{This figure presents the {\em J-} and {\em H}-band IRIS2 spectra obtained at the AAT in 2007 May for three candidates. Also shown over-plotted in red are template field L dwarf spectra for the lower two candidates that give the best fit during spectral typing. The top spectrum without an over-plotted template is of 2M1110-7630 -- a young Chamaeleon-I late-M type brown dwarf (discussed in \textsection\,\protect\ref{Anal:chamI_obj}). All spectra have been corrected for telluric features using A0V standards, with an F$_\lambda$ flux scale normalised at $1.32\,\mu$m. In all cases both the candidate and template spectra have been Gaussian smoothed to the FWHM instrument resolution of the IRIS2 {\em H-}band ($0.001\,\mu$m). The zero flux levels are indicated by dashed lines. The main atomic and molecular absorption features are indicated on the plot.}
  \label{all_aat_spec}
 \end{center}
\end{figure}

\begin{figure}
 \begin{center}
  \includegraphics[width=0.48\textwidth,angle=0]{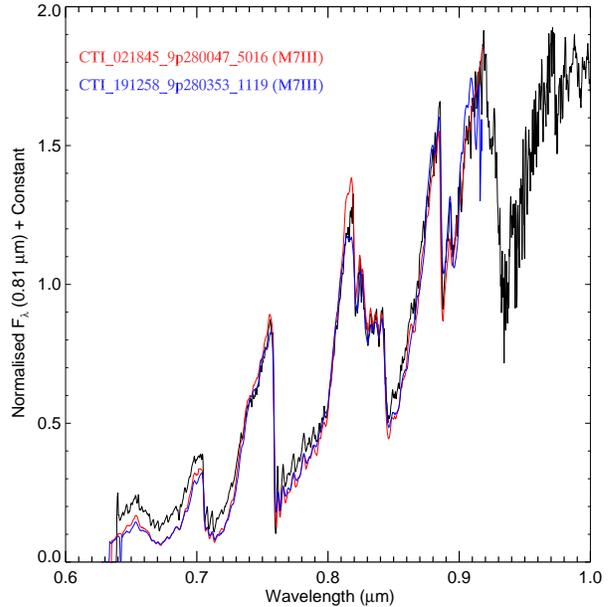}
  \caption{The red optical IMACS spectrum of 2MASS\,J18300760-1842361 (black line) over-plotted with two best fitting M7 giant template spectra.}
  \label{M7_giant}
 \end{center}
\end{figure}

Medium resolution near-IR template spectra of M- and L-type dwarfs which cover the whole of the {\em JHK} bands were obtained from both the IRTF spectral library \citep[SpeX instrument:][]{Cushing}, and from the NIRSPEC brown dwarf spectroscopic survey \citep{mcLean}. These templates were used to find the best fitting spectral types from a direct comparison with our objects. The template spectra were first smoothed to the same FWHM instrument resolution as the object spectrum, and their flux scales normalised at $1.32\,\mu$m for all spectra (except that of 2M1734-1151: normalised at $1.57\,\mu$m), before being over-plotted. The best fitting spectrum was the one that had the closest match in a `by-eye' comparison to the pseudo continuum flux levels across the spectrum, as well as the depths of the absorption features. A $\chi^2$ goodness-of-fit between the objects and templates were also performed as an aid in the fitting process. The final spectral types were determined by assigning more weighting towards the template fitting, but where discrepancies occurred between the two methods more reliance was placed on the average of the spectral indices instead. The spectra of all the confirmed UCDs and the best fitting spectral templates are shown in Fig.\,\ref{spec_jan06}, Fig.\,\ref{spec_apl06}, Fig.\,\ref{ddt_2m1734}, and in Fig.\,\ref{all_aat_spec} for the {\em JH}-band SOFI spectra, {\em JHK}-band SOFI spectrum, and AAT IRIS2 spectra, respectively.

We derive the spectro-photometric distances to our UCD spectroscopic sample (except for 2M1126-5003 and 2M1110-7630) by using the spectral type/\Mj~relation of \citet{Cruz03}, which are plotted in Fig.\,\ref{spec_sample_plot} as a function of spectral type (and given in Table\,\ref{spectro_table}). Also plotted in Fig.\,\ref{spec_sample_plot} for reference are the approximate UCD distance limits that our search is sensitive too (denoted by the shaded region), which follows from the 2MASS photometric detection limits across the UCD spectral type range (dependent on \JK~).

To determine the form of this region we obtained the \JK~colours of all L dwarfs listed in the {\sc http://dwarfarchives.org} database with well constrained spectral types ($\pm0.5$ sub-types) and with good photometry ($\mbox{SNR}\geqslant20$), which gave 54 in total. For the late-M dwarfs, a query of the SIMBAD database for known objects of M7V to M9.5V spectral sub-types having 2MASS photometry was made, which gave us mean \JK~values for $\sim5$ objects in each spectral sub-type. A linear ($\chi^{2}$ minimised) relation was fitted to the \JK/spectral-type data which was then used to find the {\em J} apparent magnitudes given the 2MASS limiting magnitude of $\mbox{\Ks}\simeq14.5$, for the range of spectral types required. The fit to the L dwarf \JK/spectral-type data gave; 
\begin{equation} \label{spt_jmk_rel}
\mbox{(\JK)}_{\rm 2mass}=(0.0711\cdot{\rm SpT})+0.4967 
\end{equation}
allowing the distance detection limits to be obtained by deriving the absolute {\em J} magnitudes for spectral types in the range of M6V to L9, using the \Mj/spectral-type relation of \citet{Cruz03}.

An interesting aspect of Fig.\,\ref{spec_sample_plot} is the data point plotted well into the shaded region that appears to be either over luminous or too distant to be detected. However, note that this is the young Chamealeon-I brown dwarf (2M1110-7630) some four magnitudes brighter in \Mj~compared to a typical higher mass M8V older field dwarf (see \textsection\,\ref{Anal:chamI_obj}).

\subsection{Photometric Spectral Types and Distances}

As the majority of the objects listed in our UCD candidate catalogue have not been spectroscopically confirmed, estimates of spectral types and distance of these candidates need to be determined by other means where possible.

This was achieved for most of the candidates by the use of the \Mj/(\ImJ)~and \Mj/spectral-type relations of \citet{dahn02}. Most of the candidates have SSA \In~detections (129), converted to \Ic, to give the needed $\mbox{\em I}-\mbox{\em J}$ colour of the relation. However, the SSA magnitudes typically have uncertainties of $\sigma\approx 0.3$\,mag, potentially leading to large scatter in the spectral types and distances. To alleviate this problem we obtained \Ib~magnitudes from the DENIS catalogue for 77 of the candidates without spectroscopy. The DENIS \Ib~uses a Gunn-{\em i} filter response with a limiting magnitude of $\simeq18$ \citep[0.82\microns~effective centre:][]{denis99}, and a filter response close enough to the standard \Ic~system that no transformation is necessary \citep[$\leqslant0.05$\,mag: e.g.,][]{phan_bao_gp07}. DENIS also includes uncertainties for each detection, which give a mean uncertainty of $\sigma=0.16$\,mag for our sample.

Using the \Mj/\ImJ~relation and the 2MASS {\em J}-band magnitudes, candidate distances and spectral types were determined from both the DENIS and SSA $\mbox{\em I}-\mbox{\em J}$ derived colours. Propagated uncertainties in distance and spectral type were calculated assuming an  SSA \In~uncertainty of $\sigma\approx 0.3$\,mag, or using the individual DENIS Gunn-{\em i} uncertainties. In all cases the {\em J}-band uncertainties obtained from 2MASS were also propagated through to the distance and spectral type uncertainties. All the photometric spectral types (and uncertainties) have been rounded to the nearest 0.5 sub-type.

\begin{figure}
 \begin{center}
  \includegraphics[width=0.48\textwidth,angle=0]{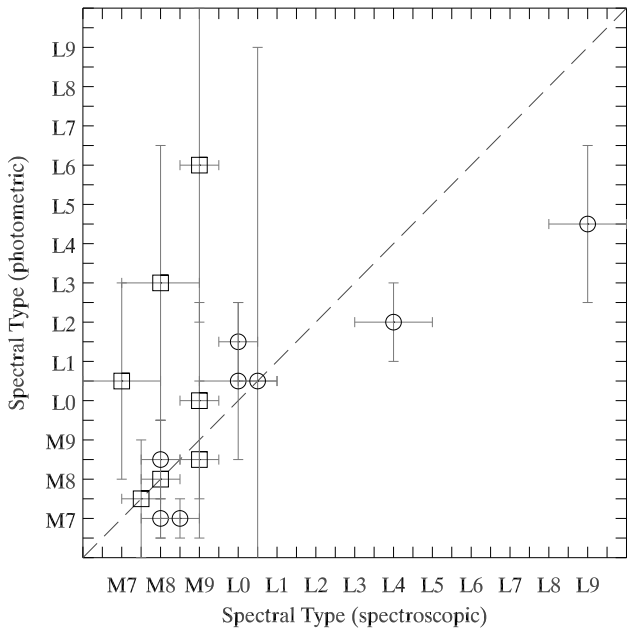}
  \caption{A SpT--SpT plot comparing the photometric and spectroscopic spectral types for objects in the spectroscopic sample. The circles and squares denote photometric spectral types obtained from the \Mj/\ImJ, and \Mj/spectral-type relation, of \protect\citet{dahn02} using both DENIS Gunn-$i$, and SSA \In~magnitudes respectively. Dashed line denotes the one-to-one correspondence between the two spectral types. Error bars are derived from the {\em I}- and {\em J}-band uncertainties as calculated for Table\,\protect\ref{spec_dist_table}. Spectral types and derived uncertainties are rounded to the nearest 0.5 sub-type.}
  \label{ssa_den_spec_diff}
 \end{center}
\end{figure}

In Fig.\,\ref{ssa_den_spec_diff} we compare the spectroscopic classifications with those of the photometric spectral types derived from $\mbox{\em I}-\mbox{\em J}$. In this figure most of the objects appear clustered around the dashed line indicating one-to-one correspondence, suggesting that both spectral types are generally in good agreement within the uncertainties. However, three of the late-M dwarfs having SSA \In~data show considerably overestimated photometric spectral types, most likely caused by photometric errors that scatter $\mbox{\em I}-\mbox{\em J}$ to a redder colour. When calculating the uncertainties for the photometric catalogue sample a typical uncertainty of $\sigma\approx 0.3$\,mag was assumed for all the SSA \In~magnitudes, while individual uncertainties for DENIS \Ib~and 2MASS {\em J}-band magnitudes were used.

The DENIS CCD \Ib~data shows much better correspondence with the spectroscopic classifications, with the apparent exception of one object -- 2M1126-5003. This object was found to be a blue L dwarf with a near-IR spectral type of L9 by \citet{my_paper}, and was originally considered to be an L--T transition object. However, analysis by \citet{burgy2m1126} found 2M1126-5003 to have an optical spectral type of L4.5, which places it in agreement with the photometric spectral type. This suggests that the \ImJ~colour may not be sensitive to the reduced condensate opacity of blue L dwarf atmospheres \citep{burgy2m1126}. We retain the spectral classification of L9 for 2M1126-5003 here for consistency with the way we have analysed all the catalogue members.

\begin{figure}
 \begin{center}
  \includegraphics[width=0.48\textwidth,angle=0]{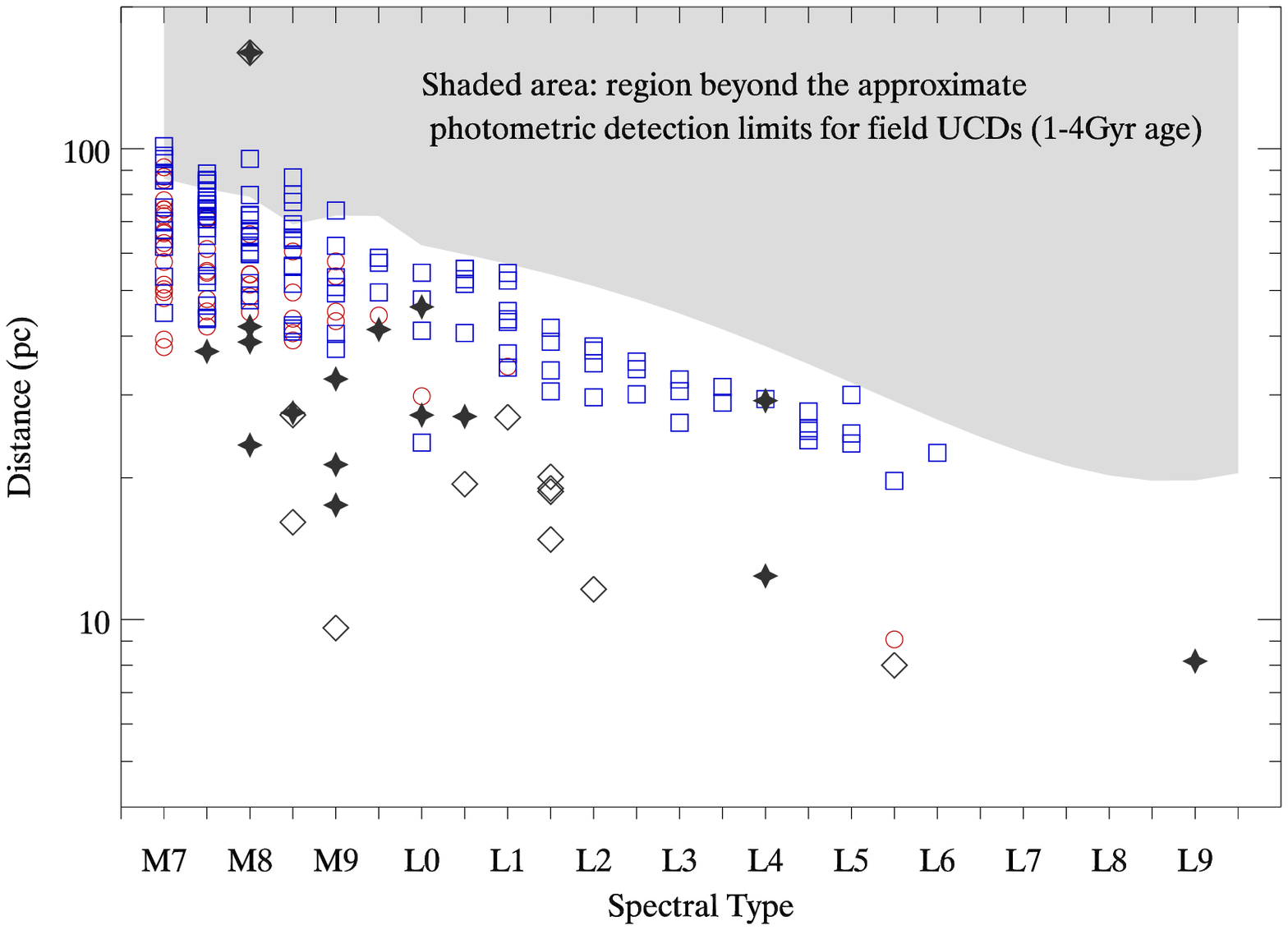}
  \caption{Photometric distances vs. spectral type distribution of the candidate UCD catalogue. The data are derived from the $\mbox{\em I}-\mbox{\em J}$ colour using the \Mj/\ImJ, and \Mj/spectral-type relations of \protect\citet{dahn02} for the range of M7V--L9. Candidates which use DENIS \Ib~magnitudes in the colour relation are denoted as red circles, and those using SSA \Ib~data as blue squares. Our spectroscopically confirmed UCDs and spectroscopically classified known objects are shown as filled grey stars and grey diamonds respectively. Approximate distance limits for the UCD search are also shown. All spectral types and uncertainties are rounded to the nearest 0.5 of a sub-type. See Table\,\protect\ref{spec_dist_table} for details of the photometric distances and spectral types including uncertainties.}
  \label{spt_dist_ssa_den}
 \end{center}
\end{figure}

The total number of UCD catalogue members for which we have reliable spectral types and distances derived from either spectroscopic or photometric $\mbox{\em I}-\mbox{\em J}$ data (from M7V to L9) is 194. The number with usable \Ib~magnitudes is 166 (not including the spectroscopic sample), split between 120 from the SSA, and 46 from DENIS. A further 28 candidates have $\mbox{\em I}-\mbox{\em J}$ data, but these fall outside the valid range of the \Mj/\ImJ, and \Mj/spectral-type relations of \citet{dahn02} and are not considered in the catalogue analysis. A plot of the resulting distance/spectral-type distribution is shown in Fig.\,\ref{spt_dist_ssa_den}, which differentiates the contribution between the spectroscopic, and photometric samples, by symbol and colour. For the photometric $\mbox{\em I}-\mbox{\em J}$ sample, those which use DENIS \Ib~detections are shown as red circles, while the blue squares denote objects using SSA \Ib~data. For the spectroscopic sample, those shown as filled grey stars are our near-IR spectroscopically confirmed UCDs, while the grey diamonds are other known objects we identified that have been spectroscopically classified. Table\,\ref{spec_dist_table} lists all the $\mbox{\em I}-\mbox{\em J}$ colours, spectral types, and distance estimates for the photometric sample where \Ib~band data was available, and also indicates those objects which have \Mj~and spectral type values that lie outside the valid range of the \citet{dahn02} relation.

\begin{sidewaystable*}
\relsize{-3}
\begin{minipage}{1.0\textwidth}
\begin{center}
 \caption{: A sample of spectral type and distance estimates taken from the full electronic table available on-line, of the southern Galactic plane photometric UCD catalogue candidate members (see \textsection\,\protect\ref{online_phot} for details of the on-line table version). The \Ib~magnitudes are from the SSA (\In~transformed to \Ic), or from DENIS (Gunn-$i$) which are indicated as such. Note that candidates with \ImJ~values that fall outside the valid colour range of the SpT/\ImJ~relation are denoted with spectral types of `<M7V' or `>L9' accordingly. Catalogue members with `Spec.' quoted for their spectral types and distances are part of our spectroscopic sample.}
 \label{spec_dist_table}
 \begin{tabular}{@{\extracolsep{\fill}}llll|llll|llll}
  \toprule
\multirow{2}{2.0cm}{2MASS Name$^a$} & \multirow{2}{1cm}{$\mbox{\em I}_{\rm c}-\mbox{\em J}$} & \multicolumn{2}{c}{Photometric} & \multirow{2}{2.0cm}{2MASS Name$^a$} & \multirow{2}{1cm}{$\mbox{\em I}_{\rm c}-\mbox{\em J}$} & \multicolumn{2}{c}{Photometric} & \multirow{2}{2.0cm}{2MASS Name$^a$} & \multirow{2}{1cm}{$\mbox{\em I}_{\rm c}-\mbox{\em J}$} & \multicolumn{2}{c}{Photometric} \\\cmidrule(l{1.5em}r{1.5em}){3-4}\cmidrule(l{1.5em}r{1.5em}){7-8}\cmidrule(l{1.5em}r{1.5em}){11-12}
  & & SpT$^b$ & Dist.$^c$ (pc) & & & SpT$^b$ & Dist.$^c$ (pc) & & & SpT$^b$ & Dist.$^c$ (pc)\\
  \midrule
0616-1411$\dagger$ & 2.814       & M7.0V $\pm$  0.5 &  74.39 $\pm$  4.59 & 0630-1840          & 3.147       & Spec.            & Spec.              & 0636-1226          & 3.810       &  L4.5 $\pm$  4.0 &  25.17 $\pm$ 14.94\\
0640-1449$\dagger$ & 2.896       & M7.5V $\pm$  0.5 &  45.14 $\pm$  2.23 & 0643-1843          & 3.070       & Spec.            & Spec.              & 0645-2333          & 2.910       & M7.5V $\pm$  1.0 &  85.56 $\pm$ 14.27\\
0646-3244$\dagger$ & 2.718       & M7.0V $\pm$  0.5 &  72.79 $\pm$  2.91 & 0648-2916$\dagger$ & 2.866       & M7.0V $\pm$  0.5 &  49.59 $\pm$  2.93 & 0649-2104          & 3.000       & M7.5V $\pm$  1.5 &  70.79 $\pm$ 14.74\\
0651-1446          & 2.990       & Spec.            & Spec.              & 0652-1614$\dagger$ & 3.043       & M8.0V $\pm$  0.5 &  48.52 $\pm$  5.21 & 0653-2129          & 3.250       & M9.0V $\pm$  2.0 &  53.34 $\pm$ 17.41\\
0657-1940          & 3.095       & M8.0V $\pm$  1.5 &  65.29 $\pm$ 16.54 & 0710-0633          & 2.980       & M7.5V $\pm$  1.5 &  53.68 $\pm$ 10.60 & 0713-0656          & 3.700       &  L3.0 $\pm$  3.5 &  32.35 $\pm$ 17.55\\
0716-0630          & \textemdash & Spec.            & Spec.              & 0723-1616$\dagger$ & 3.379       &  L0.0 $\pm$  1.0 &  29.81 $\pm$  5.14 & 0723-2300          & 3.465       &  L1.0 $\pm$  3.0 &  52.49 $\pm$ 22.80\\
0723-1236$\dagger$ & 3.015       & M8.0V $\pm$  0.5 &  44.96 $\pm$  3.65 & 0725-1025          & 3.100       & M8.0V $\pm$  1.5 &  63.27 $\pm$ 16.20 & 0728-0427$\dagger$ & 2.320       &  <M7V            & \textemdash       \\
0729-4931          & 3.060       & M8.0V $\pm$  1.5 &  72.13 $\pm$ 17.10 & 0729-2608$\dagger$ & 3.125       & M8.5V $\pm$  0.5 &  39.16 $\pm$  3.90 & 0731-2841          & 2.910       & M7.5V $\pm$  1.0 &  74.55 $\pm$ 12.30\\
0734-2724          & 3.430       &  L0.5 $\pm$  2.5 &  52.81 $\pm$ 21.93 & 0735-1957          & \textemdash & \textemdash      & \textemdash        & 0739-4926$\dagger$ & 3.000       & M7.5V $\pm$  0.5 &  54.55 $\pm$  5.04\\
0741-0359          & 3.140       & M8.5V $\pm$  1.5 &  40.89 $\pm$ 11.15 & 0743-1257          & 2.830       & M7.0V $\pm$  1.0 & 101.22 $\pm$ 13.05 & 0744-1611$\dagger$ & 2.818       & M7.0V $\pm$  0.5 &  74.27 $\pm$  4.69\\
0745-2624          & 3.210       & M9.0V $\pm$  2.0 &  62.21 $\pm$ 19.20 & 0747-3753          & 3.530       &  L1.5 $\pm$  3.0 &  38.90 $\pm$ 17.95 & 0748-3918$\dagger$ & 2.059       &  <M7V            & \textemdash       \\
0749-3252$\dagger$ & 3.338       & M9.5V $\pm$  1.5 &  44.23 $\pm$  9.83 & 0751-2530          & \textemdash & \textemdash      & \textemdash        & 0752-3925$\dagger$ & 3.216       & M9.0V $\pm$  1.0 &  43.03 $\pm$  7.19\\
0753-5421$\dagger$ & 2.790       & M7.0V $\pm$  0.5 &  61.55 $\pm$  2.74 & 0756-0715$\dagger$ & 3.159       & M8.5V $\pm$  1.0 &  43.55 $\pm$  5.40 & 0758-3945$\dagger$ & 3.493       &  L1.0 $\pm$  2.5 &  34.45 $\pm$ 12.19\\
0759-2117          & 3.000       & M7.5V $\pm$  1.5 &  43.75 $\pm$  9.06 & 0803-1723          & 2.810       & M7.0V $\pm$  1.0 &  75.05 $\pm$  8.88 & 0806-4320          & 3.710       &  L3.5 $\pm$  3.5 &  31.20 $\pm$ 17.08\\
0807-3056          & 3.690       &  L3.0 $\pm$  3.5 &  30.55 $\pm$ 16.55 & 0811-1201          & 3.000       & M7.5V $\pm$  1.5 &  77.52 $\pm$ 16.23 & 0811-4319          & 2.990       & M7.5V $\pm$  1.5 &  52.07 $\pm$ 10.55\\
0812-4719          & 3.320       & M9.5V $\pm$  2.5 &  58.65 $\pm$ 21.11 & 0812-2444          & \textemdash & \textemdash      & \textemdash        & 0813-3614          & 3.410       &  L0.5 $\pm$  2.5 &  55.56 $\pm$ 22.70\\
0814-4020$\dagger$ & 2.827       & M7.0V $\pm$  0.5 &  51.43 $\pm$  2.40 & 0819-3944$\dagger$ & 1.772       &  <M7V            & \textemdash        & 0819-4706          & 2.950       & M7.5V $\pm$  1.0 &  46.41 $\pm$  8.50\\
0822-3204          & 3.020       & M8.0V $\pm$  1.5 &  79.75 $\pm$ 17.65 & 0823-4912          & 4.480       & Spec.            & Spec.              & 0827-1619          & 2.750       & Spec.            & Spec.             \\
0828-3549$\dagger$ & 2.687       & M7.0V $\pm$  0.5 &  77.74 $\pm$  2.67 & 0828-1309          & \textemdash & \textemdash      & \textemdash        & 0831-3437          & 3.650       &  L2.5 $\pm$  3.5 &  30.08 $\pm$ 15.50\\
0831-2535$\dagger$ & 3.023       & M8.0V $\pm$  0.5 &  54.05 $\pm$  5.62 & 0832-3310$\dagger$ & 2.604       & M7.0V $\pm$  0.5 &  91.32 $\pm$  1.09 & 0836-2648          & 2.890       & M7.5V $\pm$  1.0 &  57.50 $\pm$  8.94\\
0838-3211$\dagger$ & 2.572       &  <M7V            & \textemdash        & 0844-3914$\dagger$ & 3.088       & M8.0V $\pm$  5.0 &  65.82 $\pm$ 54.26 & 0844-2530          & \textemdash & \textemdash      & \textemdash       \\
0844-2924          & 3.090       & Spec.            &  Spec.             & 0851-4916$\dagger$ & 3.079       & M8.0V $\pm$  1.0 &  54.17 $\pm$  7.18 & 0853-4106          & 3.920       &  L6.0 $\pm$  4.0 &  22.60 $\pm$ 14.74\\
0856-2408          & 2.980       & M7.5V $\pm$  1.5 &  65.29 $\pm$ 12.96 & 0858-4345$\dagger$ & 2.999       & M7.5V $\pm$  1.0 &  70.97 $\pm$ 11.97 & 0858-4240          & \textemdash & \textemdash      & \textemdash       \\
0859-6605          & 3.040       & M8.0V $\pm$  1.5 &  72.37 $\pm$ 16.47 & 0900-4227          & 3.421       &  L0.5 $\pm$  2.5 &  55.59 $\pm$ 22.73 & 0918-6101          & 3.440       &  L0.5 $\pm$  2.5 &  51.64 $\pm$ 21.61\\
0921-4607$\dagger$ & 2.522       &  <M7V            & \textemdash        & 0928-3214$\dagger$ & 3.115       & M8.5V $\pm$  1.0 &  60.48 $\pm$ 11.20 & 0928-5739          & 3.050       & M8.0V $\pm$  1.5 &  60.40 $\pm$ 13.95\\
0931-6659          & 2.810       & M7.0V $\pm$  1.0 &  87.57 $\pm$ 10.43 & 0933-5119          & 3.490       &  L1.0 $\pm$  3.0 &  45.24 $\pm$ 19.94 & 0942-3758          & 3.110       & M8.0V $\pm$  1.5 &  47.66 $\pm$ 12.32\\
0943-6234$\dagger$ & 2.747       & M7.0V $\pm$  0.5 &  85.92 $\pm$  5.36 & 0947-3810          & 3.260       & M9.0V $\pm$  2.0 &  37.56 $\pm$ 12.40 & 0955-7342          & 3.41        & Spec.            & Spec.             \\
0957-4612          & 3.090       & M8.0V $\pm$  1.5 &  66.87 $\pm$ 16.77 & 0959-3626          & \textemdash & \textemdash      & \textemdash        & 1010-5245          & 2.760       & M7.0V $\pm$  0.5 &  88.30 $\pm$  8.39\\
1014-4018          & 2.870       & M7.0V $\pm$  1.0 &  64.27 $\pm$  9.39 & 1023-6238          & 3.350       &  L0.0 $\pm$  2.5 &  41.07 $\pm$ 15.29 & 1024-4630          & 2.630       & M7.0V $\pm$  0.5 &  95.06 $\pm$  3.14\\
1029-4031          & 2.890       & M7.5V $\pm$  1.0 &  77.57 $\pm$ 12.13 & 1032-5920          & 3.590       &  L2.0 $\pm$  3.0 &  35.01 $\pm$ 17.19 & 1033-5621          & 3.030       & M8.0V $\pm$  1.5 &  95.14 $\pm$ 22.21\\
1038-6511          & 3.790       &  L4.5 $\pm$  3.5 &  25.48 $\pm$ 14.91 & 1040-6716          & 3.860       &  L5.0 $\pm$  4.0 &  23.65 $\pm$ 14.62 & 1048-5254          & 3.620       & Spec.            &  Spec.            \\
1050-4517          & 3.350       & Spec.            &  Spec.             & 1056-6122$\dagger$ & 1.281       &  <M7V            & \textemdash        & 1102-4940$\dagger$ & 2.509       &  <M7V            & \textemdash       \\
1103-5933          & \textemdash & \textemdash      & \textemdash        & 1104-6412          & \textemdash & \textemdash      & \textemdash        & 1108-7632          & \textemdash & \textemdash      & \textemdash       \\
1108-7632          & \textemdash & \textemdash      & \textemdash        & 1110-7630          & 3.650       & Spec.            & Spec.              & 1116-6403          & 2.980       & M7.5V $\pm$  1.5 &  85.87 $\pm$ 17.31\\
1119-4815$\dagger$ & 2.803       & M7.0V $\pm$  0.5 &  71.90 $\pm$  5.31 & 1122-6533          & 3.340       &  L0.0 $\pm$  2.5 &  23.75 $\pm$  8.70 & 1126-5507          & 3.620       &  L2.5 $\pm$  3.0 &  34.02 $\pm$ 17.08\\
1126-5003          & 3.700       & Spec.            & Spec.              & 1130-5759$\dagger$ & 2.935       & M7.5V $\pm$  0.5 &  41.91 $\pm$  2.29 & 1131-6446$\dagger$ & 3.122       & M8.5V $\pm$  0.5 &  40.57 $\pm$  4.07\\
1146-4754          & 3.180       & M8.5V $\pm$  2.0 &  51.68 $\pm$ 15.15 & 1148-5705          & \textemdash & \textemdash      & \textemdash        & 1150-5032$\dagger$ & 3.213       & M9.0V $\pm$  1.5 &  57.64 $\pm$ 12.62\\
1153-7454          & 2.950       & M7.5V $\pm$  1.0 &  73.63 $\pm$ 13.62 & 1155-6945$\dagger$ & 2.237       &  <M7V            & \textemdash        & 1158-7708          & 3.610       &  L2.0 $\pm$  3.0 &  37.28 $\pm$ 18.69\\
1159-5247          & \textemdash & \textemdash      & \textemdash        & 1201-6007          & 2.650       & M7.0V $\pm$  0.5 &  85.73 $\pm$  3.67 & 1204-5048          & 2.870       & M7.0V $\pm$  1.0 &  70.24 $\pm$ 10.26\\
1206-5326$\dagger$ & 2.755       & M7.0V $\pm$  0.5 &  63.04 $\pm$  2.55 & 1210-5252          & 2.920       & M7.5V $\pm$  1.0 &  74.28 $\pm$ 12.77 & 1214-5519          & \textemdash & \textemdash      & \textemdash       \\
1219-5021$\dagger$ & 2.757       & M7.0V $\pm$  0.5 &  39.32 $\pm$  1.01 & 1224-7141$\dagger$ & 3.161       & M8.5V $\pm$  1.0 &  49.50 $\pm$  9.05 & 1226-7638$\dagger$ & 3.033       & M8.0V $\pm$  0.5 &  51.45 $\pm$  5.48\\
 \bottomrule\addlinespace
\end{tabular}
\end{center}
\end{minipage}\relsize{+3}
({\em $^a$}) The abbreviated 2MASS designation obtained from the Point Source Catalogue: 2MASS Jhhmm$\pm$ddmm.\\
({\em $^b$}) Photometric spectral types obtained from the \Mj/\ImJ, and \Mj/SpT relations of \citet{dahn02} valid for UCD spectral types of M7V to $\sim$L9. The spectral types and uncertainties have been rounded to the nearest 0.5 of a sub-type. An uncertainty of {\em I}=0.3\,mag was assumed for all the distance and spectral type propagated errors for candidates with \In~SSA magnitudes, else for DENIS \Ib~data the individual object uncertainties were used.\\
({\em $^c$}) Photometric distance estimates are obtained from the distance modulus where the apparent {\em J}-band magnitude $mj$ is from 2MASS, and \Mj~is from the \Mj/\ImJ~relation \citep{dahn02}.\\
($\dagger$) This symbol indicates that the photometric spectral types and distances have been calculated from \ImJ~colours using DENIS \Ib~magnitudes (Gunn-$i$).
\end{sidewaystable*}

\begin{figure}
 \begin{center}
  \includegraphics[width=0.48\textwidth,angle=0]{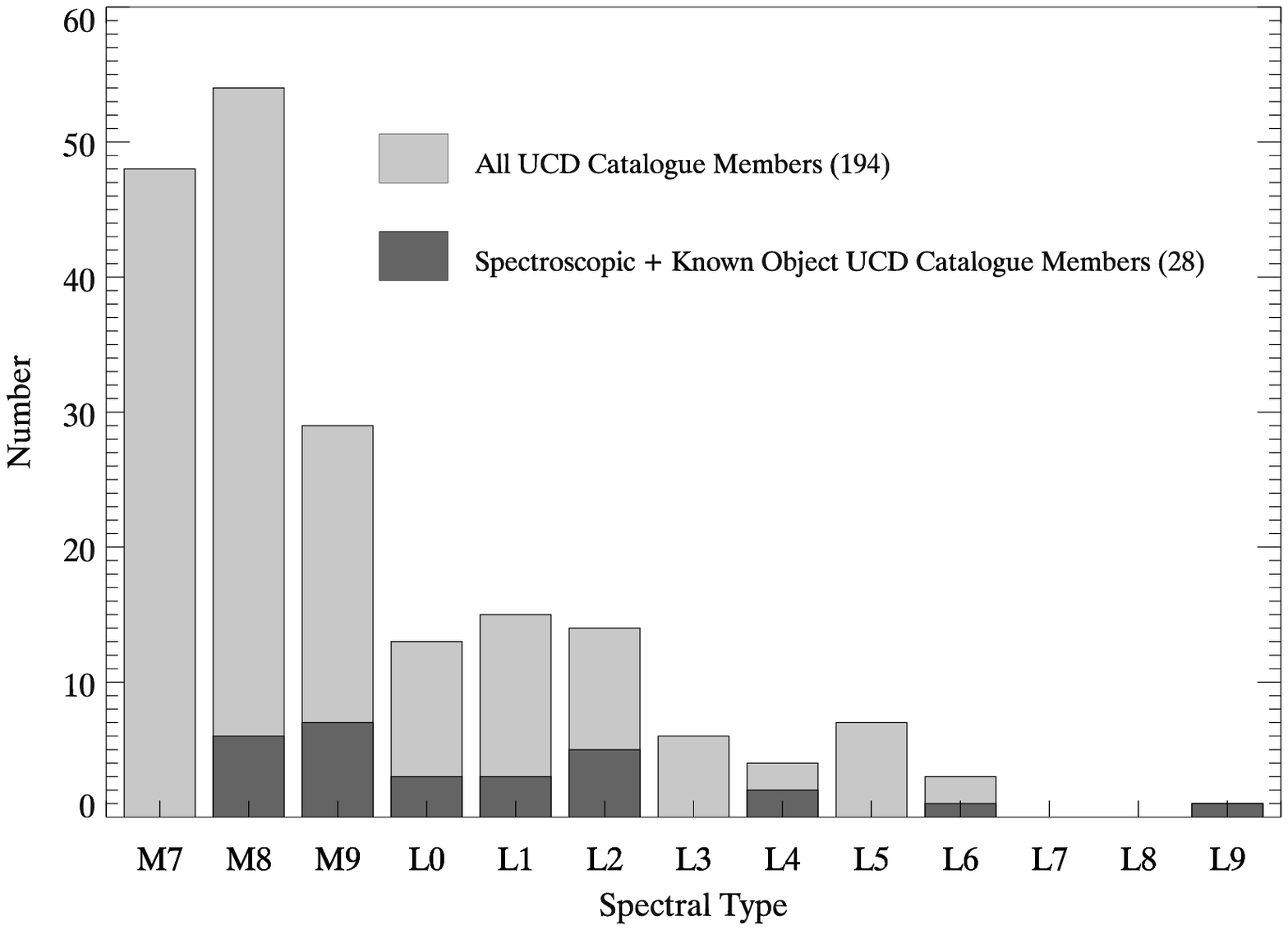}
  \caption{A stacked histogram showing the number distribution as a function spectral type of the photometric and spectroscopic UCD catalogue. Light shading represents all UCD catalogue members (photometric, spectroscopic sub-sample, and other known objects: 194). Dark shading represents both our spectroscopic sub-sample (16) and other known-object catalogue members (12).}
  \label{spt_dist_histo}
 \end{center}
\end{figure}

Fig.\,\ref{spt_dist_histo} shows a stacked histogram of the spectral-type distribution of our catalogue (without any correction for multiplicity or incompleteness in the sample), with light shading representing the whole catalogue, and dark shading the spectroscopically characterised sub-sample (our spectroscopic sub-sample plus other known objects). The distribution shows a general decrease in number from a peak at M8V down to late-L, but with small peaks centred at L1 and L5, with none occupying the L7 and L8 bins. The shape of our spectral-type distribution is broadly consistent with that derived in the 20\,pc spectral-type distribution obtained by \citet[][; see their fig.10]{cruz07}. However, our distribution does appear to have two small differences: an increase in number at L0, and no sharp increase seen at L7 compared to the distribution of \citet{cruz07}. These differences are most likely due to the sample of \citet{cruz07} being pseudo distance limited, so structure in the spectral type distribution will be more evident. Our sample is magnitude limited. In any case, the L0 and L7 frequencies in the \citet{cruz07} sample are small number statistics.

\subsection{{\em Hipparcos} Companions}
In order to identify any UCD catalogue members as possible wide companions to nearby known stars, which would provide good constraints (through association with the primary) for the UCD (e.g., distance, age, and metallicity), a cross-correlation with the \textit{Hipparcos} catalogue was made. The binary fraction of wide common proper-motion brown dwarf companions to stellar-mass primaries is estimated to be as high as $f_{\rm (s-bd)}=34^{+9}_{-6}$ per cent by \citet{piners06_benchmark}, for separations between $1,000\,\mbox{au}<a<5,000$\,au (these authors assume a mass function power-law to be $\alpha=1$), therefore such companions should be reasonably common.

We defined an angular radius around each \textit{Hipparcos} star that corresponds to a projected linear distance of 30,000\,au at the distance of each star. We chose a separation limit of 30,000\,au because star--star field binaries appear to exist with separations up to a few times $10^4$\,au \citep[see fig.9 of][]{Burgy_bin_sep_5000au}. Interestingly though, a young Castor moving group binary system has been found with a very wide separation of $\sim1$\,pc \citep[{$\alpha$}\,Librae + KU\,Librae:][]{Caballero_widest_bin}. However, in the field population low mass binary systems (with a total system mass $<1\mbox{\,M}_{\odot}$) appear to have smaller separations of no greater than $\sim7000$\,au \citep[e.g.,][]{Zhang_8wide_bins2010,Radigan_wide_field_bin}.

\begin{figure}
 \begin{center}
  \includegraphics[width=0.5\textwidth,angle=0]{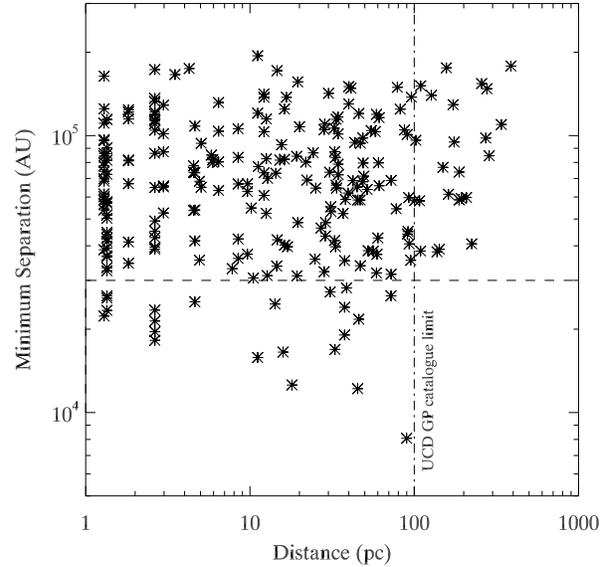}
  \caption{A $\log$--$\log$ plot of the closest physical projected separation of pairings (in units of au) found between each member of our Galactic plane UCD catalogue and stars contained in the \textit{Hipparcos} catalogue plotted against distance of the \textit{Hipparcos} stars. Vertical lines of plotted points indicate that multiple catalogue members were matched to the same \textit{Hipparcos} star (the nearest ones with large search areas). The 30,000\,au search limit of projected separation used to find possible companions is shown as a dashed horizontal line. The approximate catalogue distance limit for late-M dwarfs is shown as a dashed-dot line.}
  \label{hipp_test}
 \end{center}
\end{figure}

The cross-correlation produced a total of 30 matches with 19 \textit{Hipparcos} stars, and in Fig.\,\ref{hipp_test} we plot the minimum physical projected separation of pairings (in units of au) found between catalogue members and the matched \textit{Hipparcos} stars, against distance. For a few of the closest \textit{Hipparcos} stars, multiple catalogue members were matched (eight candidates were matched more than once) due to the larger sky area corresponding to 30,000\,au radii (i.e., $\sim3^{\circ}$ at 2.7\,pc), as denoted by the plotted points at close distances.

We rejected 23 candidate matched companions as their total proper motions were not consistent with those of their paired \textit{Hipparcos} stars. A further three candidate companions were rejected as their \ImJ~derived photometric distances were much further than the primary \textit{Hipparcos} star. This leaves four other possible \textit{Hipparcos} candidate companions:
\begin{description}
 \item []{\bf HIP\,48072/2M0947-3810} Although SSS \Rb~and \Ib~images show the proper motion of HIP\,48072 and 2M0947-3810 to be similar in RA, a slight southern motion is seen in DEC compared to the SSA value of $\mu_{\rm dec}=20\pm93\mbox{\,mas\,\peryr}$. However, given the large uncertainty, the SSA measurement is ignored. Due to the proximity of 2M0947-3810 to a nearby bright stellar halo its $\mbox{\em I}-\mbox{\em J}$ colour is not reliable, and we therefore derive its distance from the SpT/\JK~relation (see Eq.\,\ref{spt_jmk_rel} on page \pageref{spt_jmk_rel}) and \Mj/spectral-type relation of \citet{Cruz03} used in conjunction with the 2MASS {\em J}-band magnitude. The distance is consistent with that of HIP\,48072 to within the measurement uncertainties.
 \item []{\bf HIP\,48661/2M0955-7342} The SSA proper motion of 2M0955-7342 is $\mu_{\rm ra}=64\pm77\mbox{\,mas\,\peryr}$ and $\mu_{\rm dec}=217\pm83\mbox{\,mas\,\peryr}$, which have fairly large uncertainties. The distance estimate of 46\,pc for the UCD is too low based on the $\mbox{\em I}-\mbox{\em J}$ colour, however, if it is an unresolved binary itself its distance could be under-estimated. Assuming 2M0955-7342 is an equal mass binary will increase its luminosity by a factor of two, and adjusting its apparent magnitude by 0.75 mag increases its distance to 75\,pc, consistent with the primary \textit{Hipparcos} star distance. 2M0955-7342 is a member of our spectroscopic sample with a spectral type of L0$\pm$1, and may therefore be an interesting object.
 \item []{\bf HIP\,64765/2M1315-5908} The proper motions of HIP\,64765 and 2M1315-5908 are consistent from inspection of the SSS \In~image. The \ImJ~colour cannot be used to derive a distance estimate as the SSA \In~magnitude is not reliable. Using the SpT/\JK~relation again we obtain a spectral type of $\sim$M9V for 2M1315-5908 giving a photometric distance estimate of $\sim74$\,pc, which is somewhat less than the candidate primary \textit{Hipparcos} stars distance of 92\,pc. As in the previous example unresolved (equal mass) binary may explain this difference, so this pair is retained as a wide binary candidate.
 \item []{\bf HIP\,81181/2M1634-2531} 2M1634-2531 appears from visual inspection of SSS images to be a closely separated optical binary composed of both a redder and bluer component, which may possibly make the spectral type of M9V derived from the \ImJ~colour uncertain. For a spectral type M9V a  photometric distance estimate  of 40\,pc is obtained which is very similar to that of HIP\,81181. Due to the potential for this to be an interesting object, it was decided to retain it as a candidate binary, despite the uncertainty surrounding it.
\end{description}

A summary of the basic data for these four candidate binary systems are presented in Table\,\ref{Hipp_matches}. If binarity is confirmed from more accurate follow-up measurements, then age and metallicity constraints can be inferred for the UCD binary candidate companions from their \textit{Hipparcos} primaries.

Assuming these four candidate companions are confirmed as real binary members, we can estimate the implied companion fraction to main-sequence stars. For this, we applied the same approach as used by \citet{piners06_benchmark}, to first count the number of stars up to the search distance limits for each candidate, and then determine the companion fraction by summing the reciprocal for each of these numbers. For the distance limit, we chose the limit that corresponds to the \Mj~range over which our search is complete (see Table\,\ref{spc_dens_tbl}) for each candidate companion. We obtain a very low wide-companion fraction estimate of about $0.02$ per cent, considerably lower than the estimates of $2.7^{+7}_{-5}$ per cent by \citet{piners06_benchmark}, and $1.4\pm1.1$ per cent by \citep{gizis01_M_L_bin_frac}, for the L dwarf population. Based on the number of \textit{Hipparcos} stars within our search area of sky, assuming an average distance to the binary candidates of 50\,pc, and using a normalised Poisson probability distribution function, we should expect to find about 33 such UCD companions in our catalogue \citep[also assuming a late-M binary fraction of 40 per cent:][]{fischer92_Mdw_bin_frac}.

Our low wide-companion fraction estimate suggests that our search method is inefficient at finding companions around bright stars. This is most likely due to a combination of the different forms of incompleteness in our search method that we identify and discuss, resulting from the high stellar surface densities encountered at low Galactic latitude. A more in-depth analysis to identify the exact causes of our low wide-companion fraction estimate, is not justified by the current number of spectroscopic confirmed catalogue members, but this can be addressed at a later stage.

\begin{table*}
\begin{minipage}{0.8\textwidth}
\begin{center}
 \caption[]{Details of four wide binary candidate systems from a cross-correlation with the \textit{Hipparcos} catalogue and this Galactic plane UCD catalogue. Note that data for the \textit{Hipparcos} star is given on the left of the table, while details of the possible UCD companions are on the right.}
 \label{Hipp_matches}
 \begin{tabular}{@{\extracolsep{\fill}}ccccc|ccccc}
  \toprule
\multicolumn{5}{c}{Hipparcos Source Data} & \multicolumn{5}{c}{GP UCD Candidate Data}\\\cmidrule(l{.5cm}r{.5cm}){1-5}\cmidrule(l{.5cm}r{.5cm}){6-10}
\multirow{2}{0.75cm}{HIP No.} & \multirow{2}{0.75cm}{$\mu_{(\alpha)}$} & \multirow{2}{0.75cm}{$\mu_{(\delta)}$} & \multirow{1}{0.75cm}{Dist.} & \multirow{2}{*}{SpT} & 2MASSJ & \multirow{1}{0.75cm}{Sep.} & \multirow{2}{0.75cm}{SpT} & \multirow{2}{0.75cm}{$\mu_{(tot)}$$^b$} & \multirow{1}{0.75cm}{Dist.}\\
        &  &  & \multirow{1}{0.75cm}{(pc)} &  &Name$^a$&\multirow{1}{0.75cm}{($^{\prime\prime}$)}&  &  & \multirow{1}{0.75cm}{(pc)}\\
  \midrule
                    48072 & $-115.80$ &  $126.76$ & 38.74 & G2V         & 0947-3810 & 727.1 & $\sim$M9V$^f$  & $\sim$161 & $\sim$37$^f$         \\
                    48661 & $-170.40$ & $-10.08$  & 72.15 & G6V         & 0955-7342 & 365.6 & L0$\pm$1       & $\sim$168 & 46$\pm$6$^c$ (75)$^e$\\
                    64765 & $-33.19$  & $-7.31$   & 91.66 & F0IV/V      & 1315-5908 & 294.9 & $\sim$M9V$^f$? & $\sim$57  & 74$^f$               \\
                    81181 & $-127.69$ & $-267.63$ & 46.02 & \textemdash & 1634-2531 & 472.1 & $\sim$M9V      & $\sim$200 & 40$\pm$20$^d$        \\
 \bottomrule 
 \end{tabular}
\end{center}
{($^a$) Abbreviated 2MASS PSC designation: 2MASS Jhhmm$\pm$ddmm.\\
($^b$) The approximate proper motion measured between the 2MASS position and one of the SSS images (brightest and/or widest epoch difference), in units of mas \peryr.\\
($^c$) Distance derived from the \Mj/SpT relation of \citet{Cruz03} for this object, which is part of the spectroscopic sub-sample (see \textsection\,\ref{Anal:spec_type}).\\
($^d$) Distance based on the \ImJ~colour from the \Mj/\ImJ~relation of \citep{dahn02}.\\
($^e$) This corrected distance in parenthesis is the distance assuming the object is an unresolved binary of mass ratio $\approx1$, and thus components of equal luminosity (see text for details).\\
($^f$) Distance based on the 2MASS \JK~colour: the spectral type was obtained from the SpT/\JK~relation defined in Eq.\,\ref{spt_jmk_rel} of \textsection\,\ref{Anal:spec_type}, with the distance obtained from the \Mj/SpT relation of \citet{Cruz03} using the 2MASS {\em J}-band apparent magnitude.}
\end{minipage}
\end{table*}

\subsection{2M1110-7630: A Chamaeleon-I Brown Dwarf}\label{Anal:chamI_obj}
During the search of the $11^{\rm th}$ sky tile (see Table\,\ref{area_table}), which includes part of the Chamaeleon-I (hereafter Cham-I) dark cloud complex, two objects were identified that were found to be previously known Cham-I brown dwarfs/YSOs (see Table\,\ref{known_table} for object details) in SIMBAD. Another object not identified by SIMBAD (2MASS\,J11104006-7630547: hereafter 2M1110-7630) was located near to these known Cham-I brown dwarfs with a similar red colour of $\mbox{\ImJ}=3.62$.

We obtained a {\em J} and {\em H} band spectrum for 2M1110-7630 from which it is immediately apparent that it differs from `normal' older field late-M dwarfs, due to the lack of deep absorption features in the {\em J}-band such as FeH and the normally prominent K{\sc\,i} doublets. During subsequent analysis of this object it was realised that 2M1110-7630 had been previously identified as a Cham-I brown dwarf by \citet{Luhman07_my_obj}, but we include it in our analysis here as we obtain improved constraints for its physical parameters.

\begin{figure}
 \begin{center}
  \includegraphics[width=0.48\textwidth,angle=0]{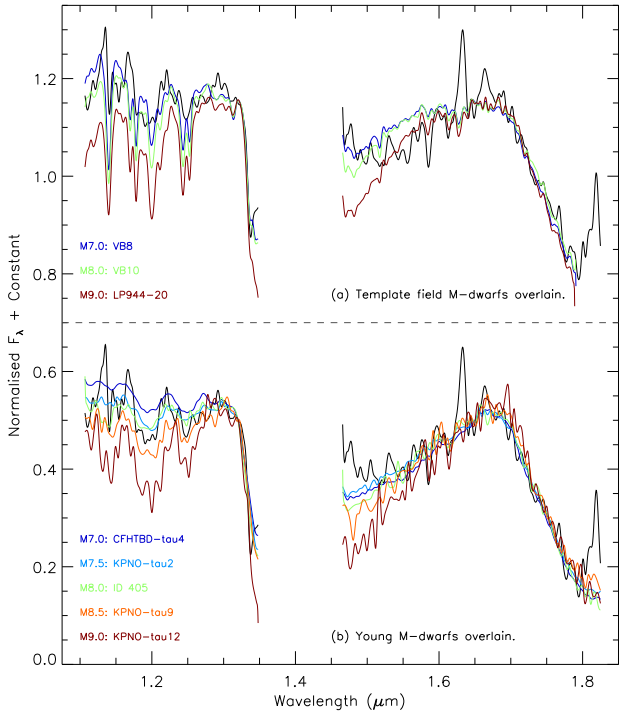}
  \caption{The {\em J-} and {\em H}-band de-reddened spectrum of 2M1110-7630. In the top plot (a) are shown older field M-dwarf spectra overlaid, while in the lower plot (b) late-M 1--3\,Myr Taurus and IC 348 brown dwarf spectra are overlaid, for a range of spectral sub-types. We confirm 2M1110-7630 as a young low gravity late-M type dwarf from this comparison -- see text for details. The spectrum of 2M1110-7630 and M-type field dwarf spectra have been smoothed to a $0.005\,\mu$m FWHM resolution element, similar to the young low gravity spectra. All the {\em J}- and {\em H}-band spectra have been normalised at $1.32\,\mu$m and $1.656\,\mu$m respectively. The large spike at $\sim 1.64\,\mu$m is a reduction artefact and not a real feature.}
  \label{chamI_spec}
 \end{center}
\end{figure}

In Fig.\,\ref{chamI_spec} we present the {\em J}- and {\em H}-band spectra of 2M1110-7630 which has been de-reddened using a total extinction of $A_J=0.59$ derived for 2M1110-7630 by \citet{Luhman07_my_obj}, and converted to a visual extinction ($A_V=2.1$) using the transformation of \citet{Rieke_Lebofsky85} and a ratio of total to selective extinction of $R=3.1$. The prominent K{\sc\,i} doublets at $\sim1.17\,\mu$m and $\sim1.25\,\mu$m seen in the field dwarfs are almost absent in 2M1110-7630, while the deep FeH absorption band at $\sim1.2\,\mu$m is missing. Indeed, \citet{Gorlova_BDgrav03} have shown that the {\em J}-band K{\sc\,i} doublets and FeH band absorption have a strong dependence on surface gravity, while the pseudo-equivalent width of the K{\sc\,i} $1.25\,\mu$m doublet can be used to measure the surface gravity in a systematic way. Other indicators of youth seen in the spectra are the presence of the VO (Vanadium Oxide) absorption band at $\sim 1.18\,\mu$m \citep{kirkp_youg_obj}, as well as the peaked triangular shape of the {\em H}-band \citep[e.g.,][]{lucas_h_peak}.

In Fig.\,\ref{chamI_spec} we compare the spectra of 2M1110-7630 against representative field (a), and young cluster (b), brown dwarf spectra for a range of late-M sub-types. All the {\em J} and {\em H} band spectra have been normalised at $1.32\,\mu$m and $1.656\,\mu$m respectively, with the field dwarf \citep[obtained from][]{Cushing} and 2M1110-7630 spectra smoothed to the same FWHM resolution element of $0.005\,\mu$m, consistent with the young brown dwarfs. These young objects are 1--3\,Myr Taurus and IC 348 brown dwarfs obtained from lucas (priv.comm.), which have spectral types derived from optical de-reddened spectra.

The presence of Na{\sc\,i} absorption is known to be a good indicator of late-type dwarf status, with young ($\sim1$--10\,Myr) cluster brown dwarfs showing Na{\sc\,i} absorption intermediate between those of older field dwarfs and giants of the same effective temperature. The Na{\sc\,i} absorption at $\sim1.14\,\mu$m is not as strong as in the field dwarfs, and therefore suggests that 2M1110-7630 is a low surface gravity young dwarf. It can be seen that the {\em J}-band spectral morphology is traced much better by the young M-type spectra in Fig.\,\ref{chamI_spec}b, especially for spectral types M8.0--M8.5, with the best fit to 2M1110-7630 being $\sim$M8.0 when taking the VO and Na{\sc\,i} absorption into account. Looking at the {\em H}-band, the blue wing of the H$_2$O absorption of 2M1110-7630 is too low compared to the field objects, but is a better match with the triangular shape of the young objects. From this comparison 2M1110-7630 appears to be a young late-M type dwarf with a spectral type of M8.0$\pm0.5$.

2M1110-7630 was found to be sub-stellar in nature with a mass of 0.04--0.05\,\Msun~by \citet{luhman_spt_temp}, and here we derive our own mass and {\em T}$_{\rm eff}$ estimates using the DUSTY model evolutionary tracks and isochrones for masses between 0.007-0.07\,\Msun~available from \citet{dusty00_Ch_model} and \citet{dusty02_Ba_model}. The spectral type of M8.0$\pm0.5$ gives an effective temperature of $\simeq 2710\,K$ for 2M1110-7630, and within the range 2600\,K to 2800\,K for the spectral type uncertainty using the SpT-{\em T}$_{\rm eff}$ relation of \citet{luhman_spt_temp}. Taking the known distance to the Cham-I dark cloud as $160\pm15$\,pc \citep{cham_I_dist}, the de-reddened absolute \Ks~magnitude can be obtained for 2M1110-7630 from the 2MASS apparent \Ks~magnitude. Once {\em M}$_{K_{\rm s}}$ is converted onto the CIT photometric system \citep[using the transformation of][to match the model data]{2m_colour_trans}, these two independently derived parameters can then be plotted onto the model isochrones.

\begin{figure}
 \begin{center}
  \includegraphics[width=0.5\textwidth,angle=0]{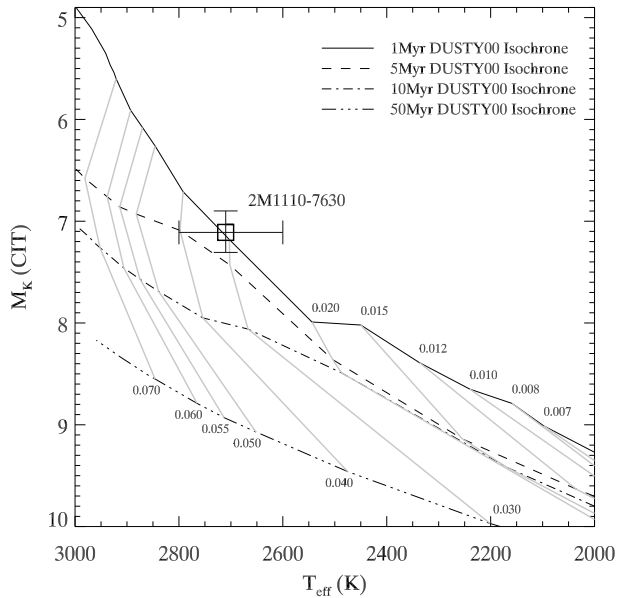}
  \caption{ {\em T}$_{\rm eff}$ and {\em M}${\rm _K}$ data for 2M1110-7630 are plotted as the large square and compared with theoretical DUSTY model evolutionary tracks and isochrones from \protect\citet{dusty00_Ch_model,dusty02_Ba_model} to estimate mass and age. Isochrones for 1,5,10 and 50\,Myrs are plotted (see key), while the lighter weight solid lines indicate lines of constant mass in the range of 0.007 to 0.070\,M$_{\odot}$. The uncertainty in {\em T}$_{\rm eff}$ is derived from the uncertainty in the spectral type, obtained from the SpT-{\em T}$_{\rm eff}$ relation of \protect\citet{luhman_spt_temp}. The uncertainty in {\em M}${\rm _K}$ is derived from the uncertainty in the distance to the Chamaeleon-I molecular cloud, taken as $160\pm15$\,pc \protect\citep{cham_I_dist}.}
  \label{chamI_model}
 \end{center}
\end{figure}

In Fig.\,\ref{chamI_model} the results for 2M1110-7630 are plotted on the isochrones for the ages of 1, 5, 10, and 50\,Myrs, and joined by the evolutionary tracks for the mass range mentioned above. This confirms that 2M1110-7630 is indeed sub-stellar with our mass estimate between 0.02--0.04\,\Msun, and with an age at the lower end of the range between $1-5$\,Myr, thus improving on the previous estimate of \citet{luhman_spt_temp}. The results of the 2M1110-7630 analysis are summarised in Table\,\ref{chamI_data}, along with those derived by \citet{luhman_spt_temp} for comparison.
\begin{table*}
\begin{minipage}{0.8\textwidth}
 \begin{center}
 \caption{A summary of the observable and model parameters derived and obtained for the late-M type Chamaeleon-I object 2M1110-7630, from this work and that of \citet{Luhman07_my_obj}.}
 \label{chamI_data}
 \begin{tabular}{@{\extracolsep{\fill}}l|ccccccc}
  \toprule
 \multirow{2}{2.0cm}{\bf Source} & \multirow{2}{1.75cm}{SpT (band)$^a$} & \multirow{2}{0.75cm}{2MASS \Ks~mag} & \multirow{2}{0.75cm}{{\em M}$_{K}$$^b$} & \multirow{2}{1.25cm}{$\log(L/L_\odot)$} & {\em T}$_{\rm eff}$ & \multirow{2}{0.75cm}{$\log g$} & \multirow{2}{0.75cm}{M/M$_\odot$}\\
 & & & & & (K) & &\\
  \midrule\addlinespace
 {\bf This work} & M8.0$\pm0.5$(NIR) & 13.34 & 7.11$^{+0.21}_{-0.20}$ & -2.13$^c$ & $2710^{+90}_{-110}$ & 3.72$^c$ & 0.03$\pm0.01$$^c$\\[1ex]
 \multirow{2}{2.0cm}{\bf Luhman (2007)} & \multirow{2}*{M7.25$\pm0.25$(Opt)} & \multirow{2}*{13.34} & \multirow{2}*{\textemdash} & \multirow{2}*{-2.09} & \multirow{2}*{2838} & \multirow{2}*{\textemdash} & \multirow{2}*{0.04/0.05}\\[3ex]
 \bottomrule\addlinespace
 \end{tabular}
\end{center}
{($^a$) Denotes whether 2M1110-7630 was spectral typed optically (Opt) or from the near-IR (NIR).\\
($^b$) {\em M}$_{K}$ is on the CIT photometric system and corrected for reddening (see text for details).\\
($^c$) These data taken from the best fit to the DUSTY models of \citet{dusty00_Ch_model,dusty02_Ba_model}. Luminosity is bolometric.}
\end{minipage}
\end{table*}

\subsection{A Nearby $\leqslant 20$\,pc sample and Red {\em $\mbox{\em I}-\mbox{\em J}$} Colours}

Our initial photometric candidate sample contained a number of objects with very red $\mbox{\em I}-\mbox{\em J}$ colours implying late-L spectral types, or with colours outside the valid range expected from the \Mj/(\ImJ) and \Mj/spectral-type relations of \citet{dahn02}. These colours were mostly derived using the \In, or converted \Ic~magnitudes, obtained from the SSS image FITS extension tables or the SSA. Twenty one such objects were identified, many with \JK~colours that appeared much bluer than expected for late-L spectral sub-types. These objects all had implied distances within 20\,pc.

However, an explanation for these red colours lies in the correction procedure for systematic plate errors of stellar colour distributions which are applied to the SSA \citep[see][]{ssa_p2b} photometry (obtained from the {\em sCorMag} query flag used in this work). As \Bj~is used as the reference passband for this correction, offsets are applied to the {\em R} and {\em I} band magnitudes. Although effective in correcting systematic plate errors, it also has the side effect of calibrating out any differential reddening across each plate. The implication here is that in the Galactic plane many regions will exhibit such reddening, resulting in systematic offsets to the {\em R} and {\em I} band magnitudes, and leading to the red colours we find for some objects.

We obtained DENIS \Ib~magnitudes for nine of these objects, and for all the remaining ones we obtained \Ib~magnitudes from the SSS that are not corrected for systematic plate errors. The results from the new $\mbox{\em I}-\mbox{\em J}$ colours are significantly different: twelve objects with previously implied spectral types of late-L or >L9 are now classified as late-M, with three of out of the original fourteen remaining as L dwarfs, five objects have new $\mbox{\em I}-\mbox{\em J}$ colours too blue for the \Mj/(\ImJ)~relation of \citet[][implying <M7V]{dahn02}. As $\mbox{\em I}-\mbox{\em J}$ is not a selection criterion we retain catalogue members with implied spectral types <M7V and >L9 (26 and 2 respectively).

With the new \Ib~photometry none of these previously red $\mbox{\em I}-\mbox{\em J}$ objects now have photometric distances of $\leqslant 20$\,pc. However, twelve others do and four of them with well constrained near-IR spectral types and spectro-photometric distances from our spectroscopic sample (see \textsection\,\ref{Anal:spec_type}). Details of these nearby UCDs are given in Table\,\ref{tbl_20pc}, and require parallax observations to potentially add them to the Solar neighbourhood 20\,pc census \citep[e.g.,][]{reid_cool_N10_08,cruz07}. Six of these objects were identified by \citet{phan_bao_gp07}, and one by \citet[][(2MASS\,J17343053-1151388)]{Kirkp2010_2m_pm}, and for any of these not part of our spectroscopic sample we have used the spectral type and distance estimates from these authors. One object is a known L5.5$\pm0.5$ field dwarf \citep{2M1520m44_Ken06} and for this we use the spectral type and distance estimates from the literature. Two other objects were found to be known Young Stellar Objects (YSOs): 2MASSJ\,11085176-7632502 part of Chamaeleon-I \citep[M7.25][]{Luhman07_my_obj}, and 2MASSJ\,13030905-7755596 part of Chamaeleon-II \citep[L1][]{chamII_Vuong01}. These two objects were therefore removed from this list as they belong to star forming complexes much further than 20\,pc distant. Finally, three previously unknown objects with distances potentially $\leqslant 20$\,pc are reported here for the first time: 2MASS\,J18000116-1559235, 2MASS\,J18300760-1842361, and 2MASS\,J14083421-5307378.

\begin{table}
 \begin{center}
 \caption[]{Members of the Galactic plane UCD catalogue with photometric distances $\leqslant 20$\,pc. Shaded rows indicate these objects are part of our spectroscopic sample (see \textsection\,\protect\ref{Anal:spec_type})}
 \label{tbl_20pc}
 \begin{tabular}{lccc}
  \toprule
2MASSJ   & (2MASS) &         &           \\
Name$^a$ & {\em J} & d\,(pc)$^b$ & SpT$^c$\\
  \midrule
\rowcolor[gray]{.75}0630-1840$\ddagger$    & 12.681 & $17.5_{-1.1}^{+1.2}$      & M9.0V$\pm0.5$\\         
                    0751-2530$\ddagger$    & 13.161 & $19.1\pm3.3$$\star$       & L1.5$\pm0.5$$\star$\\
\rowcolor[gray]{.75}0823-4912$\ddagger$    & 13.547 & $12.4_{-2.4}^{+2.6}$      & L4$\pm1$\\
                    0828-1309$\ddagger$    & 12.803 & $12.7\pm2.5$              & L1.0$\pm1.0$$\star$\\
\rowcolor[gray]{.75}1126-5003              & 13.997 & $8.2_{-1.5}^{+2.1}\dagger$& L9$\pm1\dagger$\\
                    1408-5307              & 15.155 & $19.7\pm12.4$             & L5.5$\pm4.0$\\
                    1520-4422$\ddagger$    & 13.228 & $16.2\pm3.2$$\star$       & L1.0$\pm1.0$$\star$\\
\rowcolor[gray]{.75}1734-1151$\star\star$     & 13.110 & $21.3^{+1.5}_{-1.3}$   & M9.0V$\pm0.5$\\
                    1750-0016$\star\star\star$& 13.294 & $8.0^{+0.9}_{-0.8}$    & L5.5$\pm0.5$\\
                    1756-4518$\ddagger$    & 12.386 & $14.8\pm3.0$$\star$       & M9.0V$\pm1.0$$\star$\\
                    1800-1559              & 13.431 & $9.1\pm2.1$               & L5.5$\pm1.5$\\
                    1830-1842              & 10.427 & $7.6\pm0.2$               & M7.5V$\pm0.5$\\
 \bottomrule
 \end{tabular}
 \end{center}
{($^a$) The abbreviated 2MASS designation obtained from the Point Source Catalogue: 2MASS Jhhmm$\pm$ddmm.\\
($^b$) Photometric spectral types obtained from the \Mj/(\ImJ), and \Mj/spectral-type relation of \citet{dahn02}, rounded to the nearest 0.5 of a sub-type.\\
($^c$) Photometric distance estimates obtained from the distance modulus where $mj$ is the 2MASS {\em J}-band magnitude and \Mj~is from the \Mj/\ImJ~relation \citep{dahn02}.\\
($\dagger$) Near-IR spectral type and distance estimates from \citet{my_paper}. Both Near-IR and optical \citep{burgy2m1126} spectral types give distance estimates under 20\,pc).\\
($\ddagger$) Indicates objects independently discovered by \citet{phan_bao_gp07}.\\
($\star$) Spectral types and distances from \citet{phan_bao_gp07}.\\
($\star\star$) Discovered by \citet{Kirkp2010_2m_pm}.\\
($\star\star\star$) Previously known object: spectral type and distance from \citet{2M1520m44_Ken06}.}
\end{table}

\section{Catalogue Completeness and Analysis}\label{completeness}

In this section we discuss, and establish the level of, the main causes of incompleteness in our catalogue, and derive space densities corrected for this incompleteness.

Our search for UCDs will exhibit incompleteness due to colour, reduced proper-motion, magnitude, and spatial selection criteria. Also, a search made at low Galactic latitude (especially in the southern hemisphere) will suffer additional completeness complications due to mismatches from the cross-correlation between the 2MASS and SSA databases, in regions of very high stellar densities and reddening.

\subsection{Photometric Incompleteness}

\subsubsection{Near-IR Colours}\label{nir_compl}

Only three out of the 186 reference $\mbox{SNR}\geqslant20$ L dwarfs fall outside the (\JH)/(\HK)~selection box limits. However, photometric errors for fainter objects, and the effects of unusual gravity and/or metallicity on these near-IR colours may scatter some UCDs outside our selection box.

\begin{figure}
 \begin{center}
  \includegraphics[width=0.48\textwidth,angle=0]{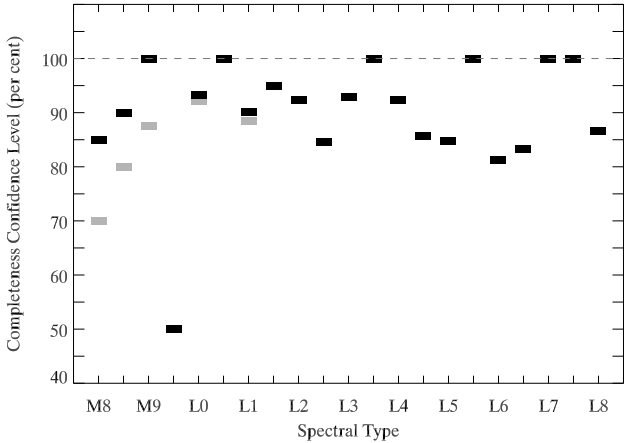}
  \caption{The near-IR colour selection completeness confidence levels as function of spectral subtype ($\mbox{M8}\leqslant \mbox{SpT}\leqslant \mbox{L8}$). Derived from 469 UCDs queried from {\sc http://dwarfarchives.org} with {\em JH}\Ks~photometry obtained from the 2MASS all-sky PSC. Dark solid bars represent the completeness level in the (\JH)/(\HK)~selection, while the light solid bars is completeness in the (\JH)/(\HK)~and \JK~selection. See text for details.}
  \label{nir_selec_complt}
 \end{center}
\end{figure}

To assess the likely degree of completeness using the (\JH)/(\HK)~selection box limits, we queried {\sc http://dwarfarchives.org} for all known dwarfs with well constrained optical spectral types of $\mbox{M8}\leqslant \mbox{SpT}\leqslant \mbox{L8}$. We obtained {\em JH}\Ks~photometry from the 2MASS all-sky PSC using the same quality flag selection as described in \textsection\,\ref{nir:sel}, and subjected the sample to the (\JH)/(\HK)~and \JK~selection criteria used in our search. We find that our overall near-IR colour selected completeness to be at the 91 per cent level, from a total number of 42 M-type dwarfs and 427 L dwarfs across these spectral subtype range. In Fig.\,\ref{nir_selec_complt} we plot the colour completeness confidence level for each 0.5 spectral subtype range as the dark solid bars for the (\JH)/(\HK)~selection, which shows scatter between $\sim 80$ and 100 per cent.

Our lower limit of $(\mbox{\JK})=1.075$ was carefully chosen primarily to allow dwarfs of spectral types M8V or later to pass through the criterion, but will also pass a percentage of M7V dwarfs. \citet{Cruz03} defined their \JK~colour lower limit as $(\mbox{\JK})=1.0$, and found an incompleteness level of $\sim10$ per cent for spectral types of M8V or later. For our $(\mbox{\JK})=1.075$ limit we find that only seven UCDs out of the 469 total fail our \JK~selection (1.5 per cent), bringing the overall completeness for both the (\JH)/(\HK)~and \JK~selection to a level of 89 per cent. However, combining the completeness for both the \JK~and \JH/\HK~selection across the spectral subtype range shows there to be a more systematic drop off in completeness for the earlier spectral types, and down to a level of 70 per cent for M8V (denoted as light solid bars Fig.\,\ref{nir_selec_complt}).

Interestingly, a query of the SIMBAD database for known M8 dwarfs reveals thirteen with 2MASS {\em J-} and \Ks-band photometry, none with a colour of $(\mbox{\JK})<1.09$, and a mean value of $(\mbox{\JK})=\langle1.18\rangle$.

Until more of our photometric sample have spectroscopic confirmation we can not be certain as to the true level of completeness for the earliest spectral type UCD in our candidate catalogue, therefore, no correction is made to the spectral type number distribution of Fig.\,\ref{spt_dist_histo} for near-IR photometric completeness.

\subsubsection{Optical and Optical-NIR Colours}\label{opti_nir_compl}

In Fig.\,\ref{bminusk} the \BK~colours indicate an overlap in the late-M and L dwarfs, with one L dwarf and a few late-M dwarfs falling outside the adopted limit of $(\mbox{\BK})\geqslant9.5$ (for $(\mbox{\JK})\geqslant1.075$). We note that of the twelve UCDs discovered by \citet{phan_bao_gp07} and were missed by our method (see \textsection\,\ref{missed_Bmag} below), four have \BK~colours outside our criterion. However, only one of these objects was rejected as a result of this colour criterion alone. As most UCDs will not have a \Bj~detection (i.e., too red and/or faint), significant incompleteness introduced by this colour is not expected in the catalogue.

The lower end of the \RK~and \IK~colour selection range criteria were conservatively adjusted to accommodate M8V and later dwarfs, but also to exclude as many contaminants as possible. As mentioned in \textsection\,\ref{opt_nir:rk_ik} the \IK~colour appears to be sensitive to small changes in \JK~with scatter evident in both these colours, with two of our 65 reference L dwarfs falling outside of the \IK~selection box in Fig.\,\ref{iminusk}. We found the \RK~colour to be better behaved in this respect as none of our 65 reference L dwarfs (with both \Rc~and \Ic~photometry) were outside the lower limits. We therefore note it is possible that some lower metallicity late-M dwarfs may be rejected due to this colour limit.

We do not expect the \RI~colour will be a source of incompleteness in the photometric selection, as again none of the 65 reference L dwarfs fell outside the \RI~limits, and they form a tight sequence within a range of one magnitude.

The \Rc~and \Ic~colours, and optical-NIR (2MASS) colours, of UCD spectral types in our selection range were explored in detail by \citet{Liebert06}. From a comparison with these authors work, any incompleteness from our optical-NIR colour selections is not expected to amount to more than a few percent, with the most problematic colour possibly being \IK~due to intrinsic scatter, especially concerning the late-M dwarfs.

\subsubsection{A Comparison With Known Objects}\label{missed_Bmag}

Although fifteen of the UCDs identified by \citet{phan_bao_gp07} were also identified in this work, twelve were missed by our method. Further investigation revealed the following reasons why these objects were not selected (listed by their DENIS identifier -- DENIS-P\,J[name]);
\begin{description}
\item []{{\bf 0615493-010041 (L2.5)}: Not within our selection area.}
\item []{{\bf 0644143-284141 (M9.5)}: The \BK, \RK, \IK, and \RI~colours all below the selection limits (SSA optical data).}
\item []{{\bf 0652197-253450 (L0.0)}: Below our 2MASS 'Prox' flag limit of 6\,arcsec (at 5.4 arcsec).}
\item []{{\bf 0805110-315811 (M8.0)}: \JK~value less than our lower limit.}
\item []{{\bf 1157480-484442 (L1.0)}: Does not fulfil the 2MASS cc\_flg flag selection requirement.}
\item []{{\bf 1159274-524718 (M9.0)}: With the SSA \In~magnitude (which is saturated) it was outside the \IK~colour selection limit. However, using DENIS \Ib~it passes this limit and was therefore included later in our catalogue.}
\item []{{\bf 1232178-685600 (M8.0)}: The \BK, \RK, and \RI~colours all below the selection limits (SSA optical data), and flagged as variable.}
\item []{{\bf 1347590-761005 (L0.0)}: Does not fulfil the 2MASS cc\_flg flag selection requirement.}
\item []{{\bf 1454078-660447 (L3.5)}: The \BK and \RI~colours are below the selection limits (SSA optical data).}
\item []{{\bf 1519016-741613 (M9.0)}: Outside the 2MASS \JH and \HK~colour selection criteria.}
\item []{{\bf 1733423-165449 (L1.0)}: Not returned in the 2MASS query covering this region.}
\item []{{\bf 1756561-480509 (L0.0)}: The \BK ~colour is below the selection limit (SSA optical data).}
\end{description}
It can be seen that these objects were not selected as they did not fulfil one of the astrometric, photometric, or quality selection requirements imposed by our method. Therefore, although they may represent an overall source of incompleteness in our UCD sample, we confirm that it is primarily well characterised photometric, and areal selection incompleteness which we discuss in this section (e.g., \textsection\,\ref{nir_compl}, \textsection\,\ref{opti_nir_compl}, and \textsection\,\ref{complete:2mass}).

Based on the eleven objects listed above (ignoring DENIS-P\,J0615493-010041) our selection appears to be $\sim41\mbox{per cent}$ incomplete compared to the UCD sample identified by \citet{phan_bao_gp07}. However, we also note that out of the seventeen UCDs in our spectroscopic sample, eleven of them (with spectral types $\geqslant\mbox{M7.5V}$) were missed by the method of \citet{phan_bao_gp07}, indicating an incompleteness of $\sim65\mbox{per cent}$ compared to that obtained by our selection method. The UCD selection method outlined in this paper, and that used by \citet{phan_bao_gp07} are different, and both are working in a very difficult region of the sky, so one cannot expect to recover the exact same object samples from both.

\subsection{Proper-Motion and Survey Cross-Matching} \label{ssa_xmatch}

The 5\asec cross-match radius between the 2MASS and SSA surveys implies that the efficiency of this method in detecting SSA optical counterparts is (on average) sensitive to them lying beyond a certain distance (assuming an average \Vtan) due to proper motion. This also implies a corresponding limiting UCD absolute magnitude at this distance. Given the average epoch difference between the ESO/SERC (1982) and 2MASS (1998) plates being $\sim16$ years, and using the 5\asec cross-match radius, gives a proper-motion limit of $\mu_{\rm tot}\simeq0.313$\asec\peryr), then the inferred distance limit can be found from Eq.\ref{eq:mdwfs}:
\begin{equation} \label{eq:mdwfs}
d\simeq\frac{\mu_{o}\cdot V_{\rm tan}}{\mu_{\rm tot}}\simeq21{\rm pc}
\end{equation}
Where the normalised proper motion at a distance of 1\,pc for 1\kms~is $\mu_{\rm o}=0.211\mbox{\asec\peryr}$, using a typical tangential velocity for disk stars of $\mbox{\em V}_{\rm tan}=31$\kms~\citep{dahn02}. The corresponding UCD absolute \Ks-band magnitude using the 2MASS photometric limit of $\mbox{\Ks}\approx14.5\,\mbox{\rm mag}$ at this distance is about {\em M}$_{K_{s}}\simeq12.9$, equivalent to a spectral type of about L8 \citep{b07}.

Thus, for regions with greater epoch differences an UCD with a lower proper motion, or one possessing a greater distance, would be required to allow an SSA optical detection. Therefore, a level of kinematic incompleteness may be expected to exist in our catalogue due to the 5\asec~cross-match radius imposed, and there remains the possibility that higher proper motion UCDs with fainter apparent magnitudes could be missed from our search criteria. However, as fainter UCD examples are most likely to have only measured \In-band magnitudes, the expected kinematic incompleteness will be lower as many \In~plates will have epoch differences considerably less than the average of $\sim16$yrs.

At present, any kinematic incompleteness in our catalogue is not easily quantifiable especially without a larger spectroscopically confirmed sample, but we expect most UCDs spanning the full spectral range (M8V to the L--T transition) passing the selection criteria to have an optical counterpart within 5\asec of the primary 2MASS location.

\subsection{Reduced Proper-Motion}

 To asses whether any incompleteness might be introduced from the \rpm~cut alone, due to both the tangential velocity distribution of UCDs and apparent \Ks~magnitude, a number of Monte-Carlo simulations were carried out to test for the probability of catalogue inclusion.

To achieve this, a population of test candidates with random distances up to a maximum of 100\,pc was created (based on a $n\propto d^{3}$ distribution), each assigned with a random spectral type in the range from M8V to L8. Next, a check was made to see if the random distance of the test candidate was closer than the maximum possible for its spectral type, given the limiting 2MASS magnitude of \Ks=14.5 of the catalogue. This was done by obtaining \Mj~from the \Mj/spectral-type relation of \citet{Cruz03}, and the apparent limiting {\em J} magnitude from its \JK~value obtained from the \JK/Spectral-type relation from Eq.\,\ref{spt_jmk_rel} (on page \pageref{spt_jmk_rel}). If the random distance was closer than the limiting test distance, then the test candidate was retained for the simulation.

Each test candidate was then assigned a \Vtan~value based on a normal distribution with the $\sigma$ and mean of the normal distribution taken from \citet{vrba}; 18.7\kms~and 30.0\kms~respectively. Next the proper motion was obtained from \[\mu=\frac{V_{\rm tan}}{d\cdot V_{\rm o}}\] where the conversion to \kms~and arcseconds yields the conversion factor $V_{\rm o}=4.75$, and by finding the apparent \Ks~magnitude from the distance modulus using \(\mbox{\Mk}=\mbox{\Mj}-\mbox{(\JK)}\). The final step was to find if the test candidate passed the \rpm~criterion using Eq.\,\ref{H}.

The results of this simulation for an arbitrary large $10^5$ tests show that 99.5 per cent of the test candidates passed the \rpm~criterion, indicating that any incompleteness introduced from this \rpm~cut is negligible and can be ignored.

\subsection{Areal Completeness}
\subsubsection{2MASS PSC Selection}\label{complete:2mass}

Our initial query of the 2MASS PSC for near-IR candidates used a proximity flag parameter set to $\geqslant 6$\asec to avoid confusion with nearby neighbours (the effective resolution of 2MASS). The area lost in our search due to this proximity radius can be significant for regions where the stellar densities are particularly high, such as the Galactic plane. The 2MASS executive summary documentation\footnote{see http://www.ipac.caltech.edu/2mass/releases/allsky/doc/sec2\_2.html for further details} shows that approximately 75 per cent of all detections in the PSC are within $|\mbox{{\em b}}|\la 15^{\circ}$, which is approximately 25 per cent of the sky, giving a high average source count density of $\sim9$ per arcmin$^2$ averaged over this low Galactic latitude region.  

Using a method based on that devised by \citet{deacon05_prox} we investigated the effect of areal incompleteness introduced by the 2MASS proximity flag for all regions we searched. We performed a set of simulations to test how many times a number of randomly placed point sources fell outside the proximity flag radius in an area of one arcmin$^2$, and if the point source fell inside the 6\asec radius the test failed. The simulations were carried out using a number density of point sources in the range of 1--100 per arcmin$^2$. The fraction which passed gave the probability of detecting a point source as a function of number density. The results indicate probability detection fractions in the range of 0.8--0.2 can be expected, corresponding to number densities from $\sim 5$--55 per arcmin$^2$ respectively.

\begin{figure}
 \begin{center}
  \includegraphics[width=0.48\textwidth,angle=0]{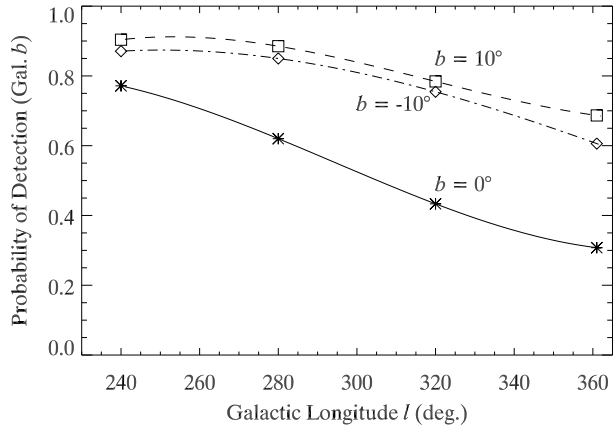}
  \caption{The probability of detecting 2MASS point sources along the Galactic plane for $240^{\circ}\leqslant \ell \leqslant360^{\circ}$ using a 2MASS proximity flag of 6 arcsec. The number of 2MASS PSC entries per arcmin$^2$ was determined for twelve regions along the Galactic plane at four locations in longitude  and for three Galactic latitudes ($\mbox{\em b}=10^{\circ}$, $0^{\circ}$ and $-10^{\circ}$), then multiplied by the probability of detection. $3^{\rm rd}$ order polynomials were fit along each Galactic latitude allowing estimates of the areal completeness to be made at the centres of each sky tile.}
  \label{prox6_probab_GP}
 \end{center}
\end{figure}

To determine the probability fractions to apply to each sky tile, we obtained the average 2MASS PSC source count number density for twelve test regions spread along four locations along the Galactic plane in longitude ($240^{\circ}\leqslant \ell \leqslant360^{\circ}$), and three locations in latitude ($\mbox{\em b}=10^{\circ}$, $0^{\circ}$ and $-10^{\circ}$). To calculate the corresponding probability fractions for each sky tile centre, $3^{\rm rd}$ order polynomials were fitted through the probability fractions of the test areas, along each Galactic latitude. Fig.\,\ref{prox6_probab_GP} shows these polynomials, which allow an interpolation to be made for each sky tile centre. For Galactic longitudes of $0^{\circ}\leqslant \ell \leqslant30^{\circ}$, the polynomial fits were assumed to be symmetrical about the Galactic centre.

To correct for areal incompleteness due to the 2MASS proximity flag the probability fractions for each sky tile centre were multiplied by the total area of each tile. The final corrected individual sky tile areas and total areal areas searched are given in Table\,\ref{areal_comp_tbl}. The total area searched per tile is the sum of each contribution from the primary, and overcrowded selection criteria, as well as from the area where no selections were made (from columns seven, eight, and nine in Table\,\ref{area_table}). The total corrected area quoted at the end of column six is used in the subsequent space density analysis.
\begin{table*}
\begin{minipage}{0.9\textwidth}
 \begin{center}
 \caption{A summary of the final sky tile areas searched in this catalogue before and after correction for areal incompleteness.}
 \label{areal_comp_tbl}
 \begin{tabular}{lccccc}
  \toprule
  \multirow{5}{1.1cm}{Tile$^a$ No.\#} & (Deg$^2$) & & \multicolumn{3}{c}{(Deg$^2$)}\\\addlinespace\cmidrule(l{.75em}r{.75em}){2-2}\cmidrule(l{.75em}r{.75em}){4-6}
  & \multirow{3}{2cm}{Total area (incomplete)$^b$} & \multirow{3}{2.75cm}{Tile detection probability fraction$^c$} & \multirow{3}{2.5cm}{Primary selection. Areal complete$^d$} & \multirow{3}{2.5cm}{Overcrowding selection (Areal complete)$^e$} & \multirow{3}{2.5cm}{Total area (Areal complete)$^f$}\\
  & & & & &\\
  & & & & &\\
  \midrule
 1 &  98.4 & 0.872 &  85.8 &  \textemdash &  85.8\\
 2 & 198.2 & 0.827 & 163.9 &  \textemdash & 163.9\\
 3 & 295.1 & 0.909 & 268.2 &  \textemdash & 268.2\\
 4 & 299.6 & 0.758 & 227.1 &  \textemdash & 227.1\\
 5 & 295.1 & 0.873 & 257.7 &  \textemdash & 257.7\\
 6 & 196.7 & 0.901 & 177.2 &  \textemdash & 177.2\\
 7 & 184.7 & 0.665 & 118.9 &   4.0        & 122.8\\
 8 & 196.7 & 0.863 & 169.8 &  \textemdash & 169.8\\
 9 & 196.7 & 0.864 & 170.0 &  \textemdash & 170.0\\
10 & 146.7 & 0.573 &  84.0 &   0.0        &  84.0\\
11 & 196.7 & 0.832 & 163.7 &  \textemdash & 163.7\\
12 & 196.7 & 0.812 & 159.8 &  \textemdash & 159.8\\
13 & 107.7 & 0.478 &  49.1 &   2.4        &  51.5\\
14 & 196.7 & 0.784 & 154.2 &  \textemdash & 154.2\\
15 & 196.7 & 0.756 & 148.8 &  \textemdash & 148.8\\
16 &  81.8 & 0.393 &  31.7 &   0.4        &  32.1\\
17 & 196.7 & 0.722 & 142.0 &  \textemdash & 142.0\\
18 & 178.9 & 0.707 & 126.5 &   0.0        & 126.5\\
19 &  47.9 & 0.329 &  15.1 &   0.7        &  15.8\\
20 & 185.8 & 0.649 & 119.3 &   1.3        & 120.6\\
21 & 177.9 & 0.707 & 116.7 &   9.1        & 125.8\\
22 &  37.9 & 0.329 &  12.1 &   0.3        &  12.5\\
23 & 166.9 & 0.649 & 101.9 &   6.4        & 108.4\\
24 & 221.7 & 0.607 & 132.8 &   1.8        & 134.6\\
 \midrule
Totals: & 4297.9 & \textemdash & 3196.3 & 26.4 & {\bf{3222.8}}\\
 \bottomrule
   \end{tabular}
  \end{center}
{({\em $^a$}) See Table\,\ref{area_table} for details of the sky tile regions.\\
({\em $^b$}) The total area searched for candidates without any areal incompleteness correction applied (the addition of columns 7, 8, and 9 in Table\,\ref{area_table}).\\
({\em $^c$}) This is the probability of detection fraction (the areal completeness) as a function of source density (sources per arcmin$^2$) derived for the centres of each tile (see Fig.\,\ref{prox6_probab_GP}, and text for details).\\
({\em $^d$}) The area searched for candidates from using only the primary selection criteria (without the overcrowding criteria), corrected for areal incompleteness.\\
({\em $^e$}) The area searched for candidates from using only the overcrowding criteria, corrected for areal incompleteness.\\
({\em $^f$}) The total final area searched for candidates from column 2 corrected for areal incompleteness. The total of this column is used in subsequent space density analysis.}
\end{minipage}
\end{table*}

\subsubsection{2MASS and SSA object mismatches}\label{mismatches}

Object mismatches between the 2MASS source and the true SSA optical counterpart could potentially lead to incompleteness in the catalogue sample for the crowded regions of the Galactic plane. UCDs with high proper motions that place them outside the 5\asec SSA search radius could be confused with an optical detection not associated with the 2MASS source. Additionally, mismatches may also occur for objects whose true optical counterparts are too faint to be detected in the SSA. Thus, an incorrect pairing between these two survey data-sets may cause a genuine UCD candidate to be rejected, by apparent optical/near-IR colour(s) that place them outside the selection criteria.

\begin{figure}
 \begin{center}
  \includegraphics[width=0.48\textwidth,angle=0]{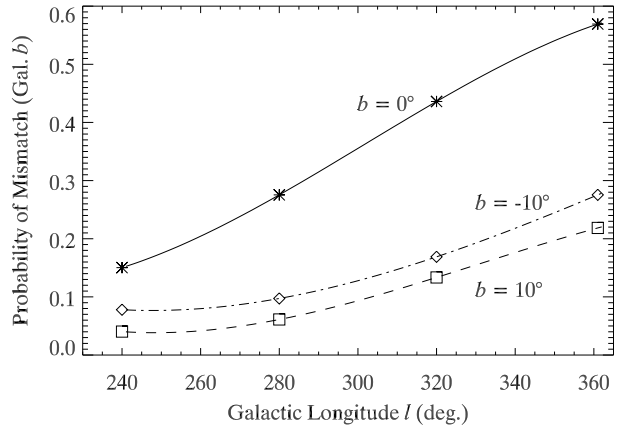}
  \caption{Results of a simulation of the probability of a mismatch between a 2MASS source and a spurious optical detection in the SSA, as a function of stellar density, and using a 5\asec search radius. The simulation was carried out for stellar surface densities in the range of 1 to 100 point sources per arcmin$^2$, and assumes the true optical counterpart is either too faint to be detected or has a high proper motion, placing it outside the search radius.}
  \label{mismatch_probab}
 \end{center}
\end{figure}

In Fig.\,\ref{mismatch_probab} we show the results of a simulation to calculate the probability of a mismatch between the 2MASS source and an SSA optical detection not associated with the source, using a 5\asec search radius, as a function of stellar surface density. The simulation was carried out using the same procedure outlined in detail in \textsection\,\ref{complete:2mass} where we find the probability of a mismatch as a function of point sources per arcmin$^2$ and Galactic latitude. The mismatch probability in this case is taken as 1-(probability of detection).

 These results indicate that incompleteness in our catalogue sample due to the 5\asec cross-match radius will be most significant ($>20$ per cent level) in the Galactic mid-plane within $\ell\pm 100^{\circ}$ of the Galactic centre, and also within $\ell\pm 30^{\circ}$ of the Galactic centre for higher Galactic latitudes of $|\mbox{{\em b}}| \la 10^{\circ}$.

\subsection{Volume Completeness}

\subsubsection{Cumulative Distribution} \label{compl_cuml_dist}

To gauge whether our catalogue sample is uniformly distributed with increasing space volume, we investigate the cumulative number distribution as a function of the volume (\(\log N\mbox{\,vs.\,}\log(r^3)\)) over four \Mj~ranges. The point where the distribution systematically deviates from the line of uniformity in each \Mj~range, may be used to indicate where incompleteness begins. Non-uniformity should ideally occur near the distance limits of each \Mj~range. 

The four \Mj~ranges were chosen to sample similar photometric distance limits within each range ($10.5\leqslant\mbox{\Mj}<11.5$, $11.5\leqslant\mbox{\Mj}<12.5$, $12.5\leqslant\mbox{\Mj}<14.0$, and $14.0\leqslant\mbox{\Mj}\leqslant14.9$), and also to provide relevance with the literature (e.g., \citet{cruz07}). Cumulative numbers were found for each successive distance shell over a distance range up to, and exceeding, the photometric limit expected for each \Mj~range.

\begin{figure*}
\begin{minipage}{0.9\textwidth}
 \begin{center}
  \includegraphics[width=0.8\textwidth,angle=0]{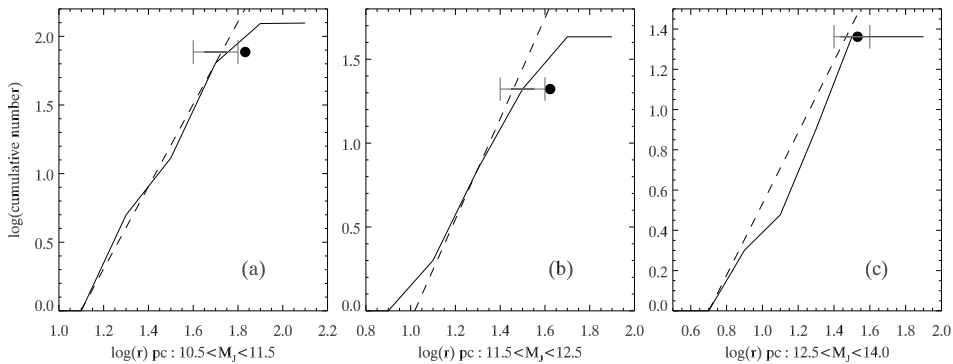}
  \caption{The cumulative number distribution for 191 members of our UCD catalogue as a function of space volume for three ranges in \Mj. In panels (a--c) increasing distance is binned into shells of 0.2\,dex. A uniform number distribution is represented by the dashed line (\(\log N\mbox{\,vs.\,}\log(r^3)\)). The solid circles indicate our adopted distance completeness limits obtained in conjunction with the volume statistic analysis (see \textsection\,\protect\ref{complete:vol_comp}). The bars denote the distance bin width nearest the point of deviation from uniformity.}
  \label{comp_dist_fig}
  \end{center}
 \end{minipage}
\end{figure*}

The three panels of Fig.\,\ref{comp_dist_fig}(a--c) show the distribution results as solid lines in each \Mj~range, along with the gradient expected for a uniform distribution (dashed line). All panels of Fig.\,\ref{comp_dist_fig} show a degree of uniformity that flattens off at progressively further distances, and with a decreasing total number of objects within each \Mj~range (125, 43, and 23 respectively). The faintest \Mj~range only contains one object within 10\,pc, and shows no uniformity, therefore we do not plot this \Mj~range. For all three \Mj~ranges shown in panels (a--c) we divide the distance into shells of bin-width equal to 0.2\,dex.

The solid circles in each panel of Fig.\,\ref{comp_dist_fig} are points representing our adopted distance completeness limits of our UCD catalogue, and were obtained in an iterative process between this cumulative number distribution investigation and the volume statistic analysis (see \textsection\,\ref{complete:vol_comp} below). This was achieved by taking into account the form of the volume statistic and space density distributions, and cross-identifying the bin where these cumulative distributions depart from uniformity. The number of objects represented by the plotted points in panels (a--c) are the cumulative number within the distance completeness limits indicated in Fig.\,\ref{vol_comp_sp_den}.

\subsubsection{Volume Statistic and Space Density}\label{complete:vol_comp}

\begin{figure*}
\begin{minipage}{0.9\textwidth}
 \begin{center}
  \includegraphics[width=0.6\textwidth,angle=90]{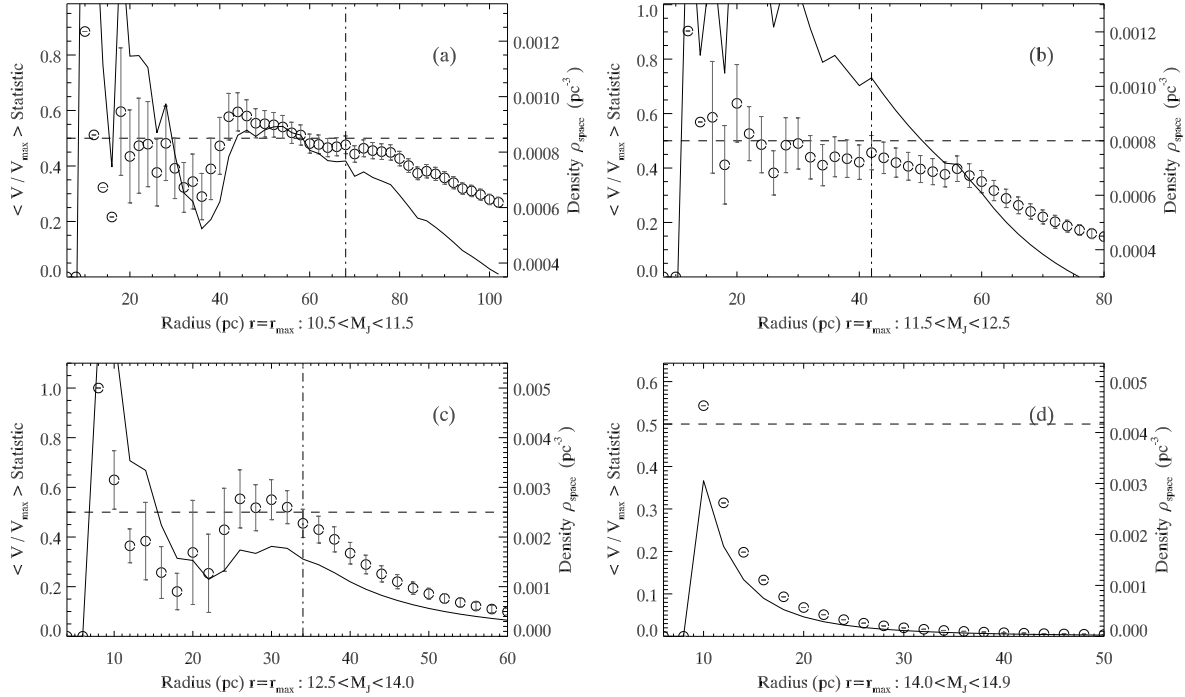}
  \caption{The volume statistic ($\langle V/V_{\rm max}\rangle$) and space densities ($\rho_{\rm (spc)}$) obtained using 191 objects from our Galactic plane UCD catalogue and over four ranges in \Mj. The $\langle V/V_{\rm max}\rangle$ values are plotted as open circles for the cumulative number of objects within a distance of ($r\leqslant r_{\rm max}$), and plotted at $r_{\rm max}$. Error bars give the $1\sigma$ standard error of the mean. Horizontal dashed lines represent a value of 0.5 for a uniform distribution (scale on left-hand vertical axises). The space densities for each \Mj~range are plotted as solid lines with values indicated by the right-hand vertical axises. The adopted completeness limits are shown as the vertical dash-dot lines. For the \Mj~range in panel (d) the sample is incomplete.}
  \label{vol_comp_sp_den}
 \end{center}
\end{minipage}
\end{figure*}

To help assess the degree of completeness throughout our catalogue, we also followed an approach used by \citet{reid95_vol_cmplt} \citep[originally devised by][]{schmidt75_vol_stat} referred to as the volume statistic ($\langle V/V_{\rm max}\rangle$), in conjunction with the cumulative number distribution. This differential statistic will produce a mean value of $\langle V/V_{\rm max}\rangle=0.5$ from the ratios of the individual objects, if the sample is uniformly distributed. Here, $V_{\rm max}$ takes the form of successive steps (shells) of increasing distance ($r_{\rm max}$), within which the cumulative number of objects is found. We apply this method using the same \Mj~ranges and maximum distance limits in \textsection\,\ref{compl_cuml_dist}, and the results are presented in Fig.\,\ref{vol_comp_sp_den}.

The average volume statistics are shown as open circles derived from the distances to individual objects that are less than the test distance ($r\leqslant r_{\rm max}$), and are plotted against $r_{\rm max}$ in each case in increments of 2\,pc. The line conforming to uniformity in the sample is indicated by a dashed horizontal line in each panel (a--d). We also over-plot the cumulative mean space densities ($\rho_{\rm (spc)}$) which uses the total (completeness corrected) sky area searched in the creation of our UCD catalogue (\catarea~deg$^2$: see Table\,\ref{areal_comp_tbl}). 

We derive distance limits over which we consider our UCD catalogue to be complete where $\langle V/V_{\rm max}\rangle$ begins to systematically drop below 0.5 (within the $1\sigma$ standard error of the mean), and where the space density is also beginning to systematically drop. Our adopted distance limits are indicated by vertical dashed-dot lines in panels (a--c). For the fainest \Mj~range in panel (d) no completeness limit can be reliably obtained as only one object exists here. However, to obtain a lower limit to the total space density we use a distance of 10\,pc, which is the distance bin containing the only object in this faintest\Mj~range, to calculate a space density.

Finally, we derive space densities from a total number of objects within our complete space volume (125), and these density values that we denoted as $\rho_{\rm (comp)}$ are listed in Table\,\ref{spc_dens_tbl}, along with the number of objects ($N_{\rm comp}$) found within each \Mj~range and corresponding complete distance limit. Also listed for comparison in Table\,\ref{spc_dens_tbl} are the space densities derived by \citet{cruz07} (in square brackets -- last column) from their 20\,pc 2M2U sample of M7--L8 UCDs derived from higher Galactic latitudes ($|\mbox{{\em b}}|>10^\circ$). 

We obtain a total space density of $(6.41\pm3.01)\times10^{-3}\,{\rm pc}^{-3}$ that is in agreement with the value of $(8.7\pm0.8)\times10^{-3}\,{\rm pc}^{-3}$ of \citet{cruz07}. However, some of our individual space densities over each \Mj~range differ slightly to those of the same authors. The most significant difference is the lower space density for the late-M dwarfs. We also note that our total space density has a large contribution from the faintest \Mj~range, which is highly uncertain. Omitting the faintest \Mj~range from the total space density gives $\rho_{\rm (comp)}=(3.4\pm0.6)\times10^{-3}\,{\rm pc}^{-3}$, a result more in line with that obtained by \citet{Gizis00}.

We now compare our results with the space density derived by \citet{phan_bao_gp07} using the DENIS data set over $4800\mbox{~deg}^2$ at low Galactic latitude. For the comparison we re-calculated our space density for the same \Mj~range used by these authors ($11.1\leqslant\mbox{\Mj}<13.1$: M8--L3.3), and we obtain a value of $\rho=(1.61\pm0.20)\times10^{-3}\,{\rm pc}^{-3}$, which is in very good agreement with the their result of $\rho=(1.64\pm0.46)\times10^{-3}\,{\rm pc}^{-3}$. As this \Mj~range covers our first two highest luminosity ranges in Table\,\ref{spc_dens_tbl}, we assumed a distance of 50\,pc as representing a complete sample in our calculation, but choosing higher and lower complete distances of 46\,pc and 56\,pc give respective space densities of $\rho=(1.85\pm0.24)\times10^{-3}\,{\rm pc}^{-3}$ and $\rho=(1.37\pm0.15)\times10^{-3}\,{\rm pc}^{-3}$, both still in good agreement with \citet{phan_bao_gp07}.

Although we have defined space volumes over which our UCD southern Galactic plane UCD catalogue is uniformly distributed, the reasons for our lower late-M dwarf ($10.5\leqslant\mbox{\Mj}<11.5$) space density in comparison with \citet{cruz07}, most likely stems from the following probable causes:

\begin{enumerate}
\item The volume limited sample from \citet{cruz07} was optimised for objects with spectral types $\geqslant \mbox{M7V}$ not M8V as in our selection requirement, and these authors find almost equal numbers of M7 and M8 dwarfs. Thus, the highest luminosity (spectral type) bin these authors consider may well have an increased space density estimate due to the M7V spectral type contribution.
\item The combination of individual causes potentially leading to incompleteness in our UCD catalogue as discussed in \textsection\,\ref{completeness}.
\end{enumerate}

\begin{table*}
 \begin{minipage}{0.85\textwidth}
  \begin{center}
  \caption[]{Space densities for the Galactic plane UCD catalogue over four \Mj~magnitude ranges.}
 \label{spc_dens_tbl}
 \begin{tabular}{@{\extracolsep{\fill}}cccccc}
  \toprule
\multirow{3}*{Magnitude Range} & \multirow{3}*{Sp-T Range} & \multirow{3}{1.75cm}{Completeness Limits$^a$ (pc)} & \multirow{3}*{$N_{\rm comp}$$^b$} & \multicolumn{2}{c}{$(10^{-3}\ \mbox{UCDs}\ {\rm pc}^{-3}\ {\rm mag}^{-1})$}\\\cmidrule(l{.55cm}r{.55cm}){5-6}
 \multicolumn{4}{c}{ } & Total $\rho_{\rm (comp)}$$^c$ & $\rho_{\rm (20pc)}$$^d$\\
  \midrule
$10.5\leqslant\mbox{\Mj}<11.5$         & $\simeq$M7V--M9V     & 68          & $79\pm8.9$ & $0.77\pm0.09$ & [$4.0$]\\
$11.5\leqslant\mbox{\Mj}<12.5$         & $\simeq$M9.5V--L2.5  & 42          & $25\pm5.0$ & $1.0\pm0.2$   & [$2.0$]\\
$12.5\leqslant\mbox{\Mj}<14.0$         & $\simeq$L3--L6       & 34          & $20\pm4.5$ & $1.6\pm0.3$   & [$1.6$]\\
$14.0\leqslant\mbox{\Mj}\leqslant14.9$ & $\simeq$L6.5--L9     & \textemdash & $ 1$       & $>3.0$        & [$>1.0$]\\ 
  \midrule
{\bf Total $\rho$} & M7--L9 & \textemdash & 125 & {\bf $6.41\pm3.01$} & $\simeq8.6$\\
 \bottomrule
 \end{tabular}
 \end{center}
{($^a$) The distance limits over which our UCD catalogue is considered to be complete in each \Mj~magnitude range.\\
($^b$) The cumulative number of objects within the complete space volumes of the specified \Mj~magnitude ranges.\\
($^c$) The space densities for the UCD catalogue derived using the distance limits from column three (see text for details). Uncertainties for each \Mj~magnitude range are calculated based on the Poisson error in the number of objects within each \Mj~range and each corresponding completeness derived distance estimates, then propagated through the space density calculations. The uncertainties quoted in our total space densities were found by adding the individual \Mj~range uncertainties in quadrature.\\
($^d$) The space densities found for the 20\,pc 2M2U sample of \citet{cruz07} given for comparison.}
 \end{minipage}
\end{table*}

\section{Conclusions}
In this paper we have demonstrated that it is possible to systematically search the `zone of avoidance' of the southern Galactic plane for UCDs, using near-IR and optical photometric constraints, as well as utilising proper motion data, down to the 2MASS and SSA limiting magnitudes of ($\mbox{\Ks} = 14.5\,\mbox{\rm mag}$, and $\mbox{\In}\simeq18.5\,\mbox{\rm mag}$). The more well characterised colours and two-colour planes one can work with, the better the efficiency in rejecting contaminant objects encountered at low Galactic latitudes (seven optical/near-IR colours used here). We also find reduced proper-motion is an effective tool in this respect. In our analysis we highlight and quantify difficulties and limitations working at low Galactic latitude using these existing survey data sets, in particular the cross-correlation with the digitised optical photographic plate material, that can lead to issues of incompleteness.

We have presented a candidate UCD catalogue compiled at southern low Galactic latitudes with a sky coverage of 5042~deg$^2$ (completeness corrected areal coverage of \catarea~deg$^2$), within $|\mbox{{\em b}}|\leqslant 15^\circ$. Our catalogue contains \catnum~members, sixteen of which have spectroscopic confirmation (94 per cent of the spectroscopic targets). From an analysis of the characterised photometric and spectroscopic UCD samples, we derive completeness corrected space densities over three UCD \Mj~ranges, and a lower limit to the space density on our forth lowest luminosity range. We find a total space density of $(6.41\pm3.01)\times10^{-3}\,{\rm pc}^{-3}$ (for $10.5\leqslant\mbox{\Mj}\la 14.9$) in agreement with the value derived at higher Galactic latitudes by \citet{cruz07}. We also compare our space density results with \citet{phan_bao_gp07} and find a value of $\rho=(1.61\pm0.20)\times10^{-3}\,{\rm pc}^{-3}$ when re-binned to the same \Mj~range used by these authors (M8--L3.5), a result which is in almost exact agreement. 

Our UCD catalogue also contains some potentially very interesting objects, which include possible companions to known {\em Hipparcos} stars, and a nearby $<20$\,pc sample based on spectro-photometric, and photometric distance estimates. This catalogue may also contain some interesting examples of UCDs with unusual metallicity/gravity. Due to the locations of these objects in the Galactic plane they will also make ideal candidates for future high-strehl adaptive optics imaging (high stellar densities for reference tip-tilt correction stars), to search for low-mass sub-stellar/planetary mass companions.

\section*{Acknowledgements}

This publication makes use of data products from the Two Micron All Sky Survey, which is a joint project of the University of Massachusetts and the Infrared Processing and Analysis Center/California Institute of Technology, funded by the National Aeronautics and Space Administration and the National Science Foundation.Our research has also made use of data obtained from the SuperCOSMOS Science Archive, prepared and hosted by the Wide Field Astronomy Unit, Institute for Astronomy, University of Edinburgh, which is funded by the UK Science and Technology Facilities Council. This research has also made use of the SIMBAD database, operated at CDS, Strasbourg, France, as well as of NASA's Astrophysics Data System Bibliographic Services. We acknowledge that our research has benefited extensively from the M, L, and T dwarf compendium housed at {\sc http://dwarfarchives.org} and maintained by Chris Gelino, Davy Kirkpatrick, and Adam Burgasser, and also from the IRTF Spectral Library: http://irtfweb.ifa.hawaii.edu/$\sim$spex/IRTF\_Spectral\_Library. The authors would also like to acknowledge the use of David Fanning's very useful 'Coyote's Guide to IDL Programming' resource located at http://www.dfanning.com/index.html\#toc.

S.L.F. acknowledges funding support from the ESO-Government of Chile Mixed Committee 2009, and from the GEMINI--CONICYT grant \# 32090014/2009. Support for S.L.F. is also provided by the Ministry for the Economy, Development, and Tourism's Programa Iniciativa Cient\'{i}fica Milenio through grant P07-021-F, awarded to The Milky Way Millennium Nucleus. Support for R.K. is provided by Proyecto FONDECYT Regular \# 1080154, by Proyecto DIUV23/2009, and the Centro de Astrof\'isica de Valpara\'iso. A.D.J. is supported by a FONDECYT postdoctorado fellowship under project number \# 3100098.

\bibliography{./sf_gp_ucd_cat}

\newcounter{chapter}{0}

\appendix

\section{Supplementary Materials}

The following supplementary material is available on-line for the article:

\subsection{On-line Complete Photometric Candidate Data Table}\label{online_phot}

Spectral type and distance estimates are presented for the southern Galactic plane photometric UCD catalogue candidate members.

\subsection{On-line complete SGPUCD Catalogue Table}\label{online_cat}

The full Southern Galactic Plane UCD catalogue (SGPUCD) containing \catnum~members. Astrometric, photometric, and proper-motion data are presented, as well as spectral types and distances derived for both the spectroscopic and photometric samples.

\bsp						  
 						  
 \label{lastpage}				  
 						  
 \end{document}